\documentclass[fleqn,10pt]{wlscirep}
\title{Learning behavior accounts for background-related advantage in AI-assisted education}

\author[1$*$]{Jingwei Yi}
\author[2,3$\dagger$]{Yueqi Xie}
\author[1$*$]{Jiyan He}
\author[4]{Rui Ye}
\author[3$\dagger$]{Junming Huang}
\author[5]{Bin Zhu}
\author[6]{Sean Rintel}
\author[3,7]{Yu Xie}
\author[8]{Xing Xie}
\author[8$\dagger$]{Fangzhao Wu}
\affil[1]{University of Science and Technology of China, Hefei 230026, China}
\affil[2]{Hong Kong University of Science and Technology, Hong Kong}
\affil[3]{Paul and Marcia Center on Contemporary China, Princeton University, Princeton, NJ 08544, United States}
\affil[4]{Shanghai Jiao Tong University, Shanghai 200240, China}
\affil[5]{Microsoft AI Asia, Beijing 100080, China}
\affil[6]{Microsoft Research Cambridge, United Kingdom}
\affil[7]{Center for Social Research, Guanghua School of Management, Peking University, Beijing 100871, China}
\affil[8]{Microsoft Research Asia, Beijing 100080, China}
\affil[$\dagger$]{Correspondence: yxieay@connect.ust.hk, pub@junminghuang.com, fangzwu@microsoft.com}

\affil[$*$]{Jingwei Yi and Jiyan He are MSRA Intern students.}

\usepackage{color}
\usepackage{colortbl}
 \usepackage{CJKutf8}
\usepackage{bm}
\usepackage{hyperref}
\usepackage{xcolor}
\usepackage{lineno}
\usepackage{verbatim}
\usepackage{booktabs} 
\usepackage{algorithm}
\usepackage{algorithmic}
\usepackage{epsfig}
\usepackage{amsmath}
\usepackage{amsthm}

\usepackage{color}
\usepackage{amsmath}
\usepackage{graphicx,subfigure}
\usepackage{multirow}
\usepackage{makecell}
\usepackage{amssymb}
\usepackage[most]{tcolorbox} 
\tcbuselibrary{theorems} 
\tcbuselibrary{breakable}
\tcbuselibrary{skins}

\usepackage{threeparttablex}
\usepackage{tabularx}
\usepackage{longtable}
\usepackage{enumitem}

\usepackage{subfigure}

\newtcolorbox{questionbox}{
    colback=white,
    colframe=black!50,
    boxrule=0.5pt,
    arc=0pt,
    outer arc=0pt,
    left=10pt,
    right=10pt,
    top=6pt,
    bottom=6pt,
    margin bottom=3mm
}
\AtBeginDocument{%
}
\newcommand{\supplementtitlepage}{%
  \clearpage
  {\raggedright\sffamily\bfseries\fontsize{20}{25}\selectfont
  Supplementary Information for ``Learning behavior accounts for background-related advantage in AI-assisted education''\par}
  \vskip10pt
  {\raggedright\sffamily\fontsize{12}{16}\selectfont
  Jingwei Yi\textsuperscript{1$*$}, Yueqi Xie\textsuperscript{2,3$\dagger$}, Jiyan He\textsuperscript{1$*$}, Rui Ye\textsuperscript{4}, Junming Huang\textsuperscript{3$\dagger$}, Bin Zhu\textsuperscript{5}, Sean Rintel\textsuperscript{6}, Yu Xie\textsuperscript{3,7}, Xing Xie\textsuperscript{8}, and Fangzhao Wu\textsuperscript{8$\dagger$}\par}
  \vskip18pt
  {\raggedright\sffamily\fontsize{10}{12}\selectfont
  \textsuperscript{1}University of Science and Technology of China, Hefei 230026, China\par
  \textsuperscript{2}Hong Kong University of Science and Technology, Hong Kong\par
  \textsuperscript{3}Paul and Marcia Center on Contemporary China, Princeton University, Princeton, NJ 08544, United States\par
  \textsuperscript{4}Shanghai Jiao Tong University, Shanghai 200240, China\par
  \textsuperscript{5}Microsoft AI Asia, Beijing 100080, China\par
  \textsuperscript{6}Microsoft Research Cambridge, United Kingdom\par
  \textsuperscript{7}Center for Social Research, Guanghua School of Management, Peking University, Beijing 100871, China\par
  \textsuperscript{8}Microsoft Research Asia, Beijing 100080, China\par
  \textsuperscript{$\dagger$}Correspondence: yxieay@connect.ust.hk, pub@junminghuang.com, fangzwu@microsoft.com\par
  \textsuperscript{$*$}Jingwei Yi and Jiyan He are MSRA Intern students.\par}
  \vskip18pt
}
\newcommand{\startsupplement}{%
  \clearpage
  \setcounter{section}{0}
  \setcounter{subsection}{0}
  \setcounter{subsubsection}{0}
  \setcounter{figure}{0}
  \setcounter{table}{0}
  \setcounter{equation}{0}
  \renewcommand{\theHsection}{supp.\arabic{section}}
  \renewcommand{\theHsubsection}{supp.\arabic{section}.\arabic{subsection}}
  \renewcommand{\theHsubsubsection}{supp.\arabic{section}.\arabic{subsection}.\arabic{subsubsection}}
  \renewcommand{\theHfigure}{supp.\arabic{figure}}
  \renewcommand{\theHtable}{supp.\arabic{table}}
  \renewcommand{\theHequation}{supp.\arabic{equation}}
  \startcontents[si]
  \supplementtitlepage
  \section*{Contents}
  \printcontents[si]{}{1}{}
  \clearpage
}
\begin{abstract}
    Generative AI has been found, and will likely be found increasingly, useful in education.
    However, existing AI-for-education studies provide inconsistent evidence on its average effects.
    More broadly, research on prior educational technologies shows that average effects often mask substantial heterogeneity across student populations.
    Motivated by this evidence, this study examines heterogeneity in students’ learning behavior with AI, which students benefit from AI assistance, and how learner profiles and learning behavior shape these patterns.
    To this end, we recruited 318 university students to participate in structured learning experiments lasting up to 125 minutes. Our findings indicate that students’ learning behavior is strongly associated with learning outcomes, with behaviors characterized by proactive and critical engagement, rather than limited engagement, associated with significantly better performance.
    These behavioral differences are related to learner profiles, with students from higher-ranking universities and those with greater prior knowledge tending to benefit more, consistent with their greater likelihood of adopting proactive interaction strategies.
    Accounting for learning behavior substantially weakens or eliminates the associations between learner profiles and learning outcomes, suggesting that how students use AI is a key pathway through which background differences are linked to learning gains.
    Overall, this work provides a deeper understanding of AI assistance in education by showing how differences in learner profiles and learning behavior shape who benefits from AI-supported learning. These insights can help educators and students better navigate and integrate AI into educational practices.
\end{abstract}

\begin{document}
\begin{CJK*}{UTF8}{gbsn}
  \flushbottom
  \maketitle
  %
  \clearpage
\begin{figure*}[!t]
  \centering
  \includegraphics[width=1.0\textwidth]{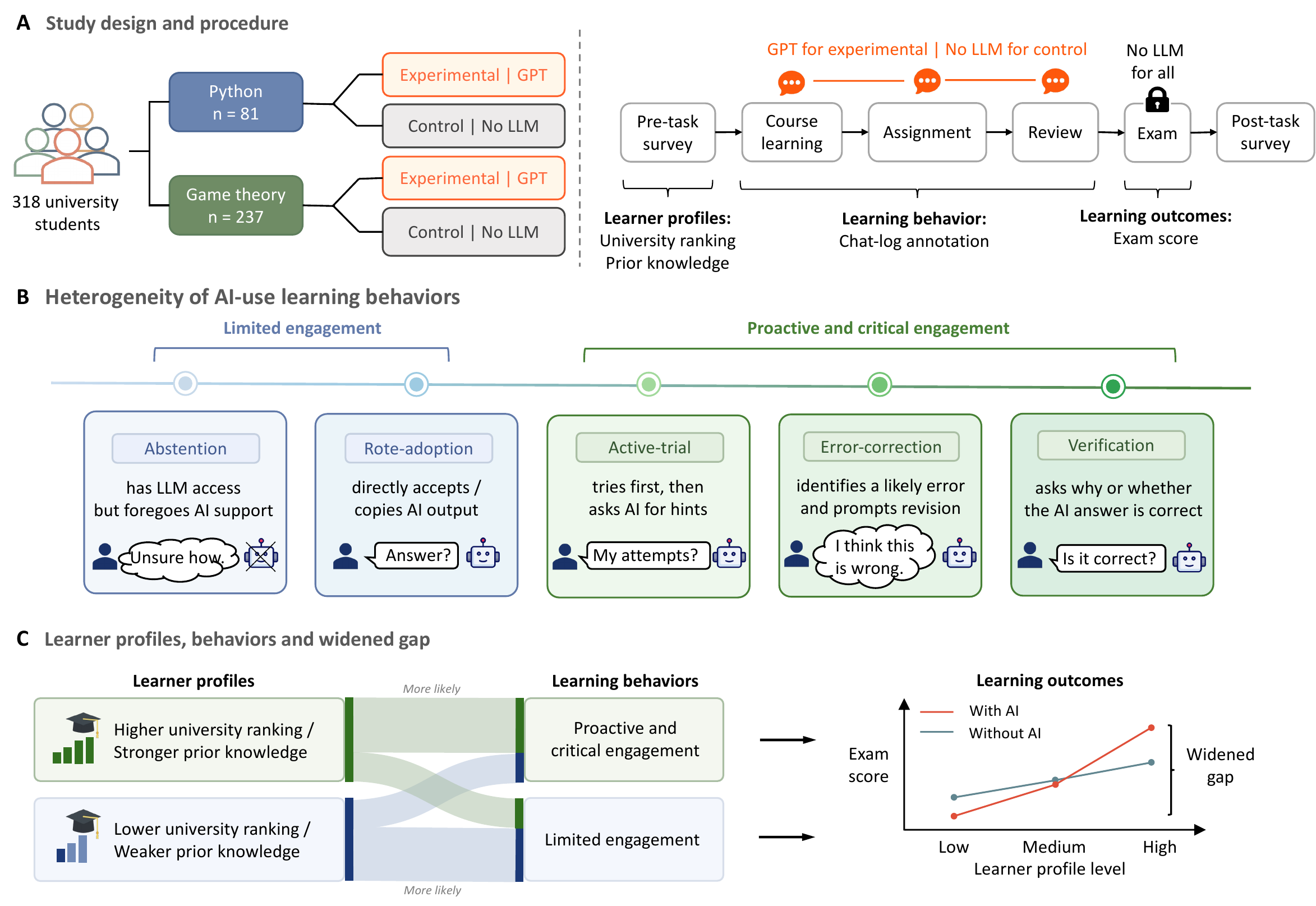}
  \caption{
    \textbf{Study design and conceptual framework linking AI-use learning behavior to heterogeneous learning outcomes.}
    \textbf{a}, Study design and procedure. A total of 318 university students were assigned to one of two course contexts, Python or game theory, and randomly allocated within each course to either the experimental group with GPT access or the control group without LLM-based tool access. The procedure includes a pre-task survey, course learning, an assignment, review, an exam without GPT access for all participants, and a post-task survey.
    \textbf{b}, Conceptual overview of heterogeneous learning behavior during AI-assisted study. Student interactions with AI range from limited engagement, including abstention and rote-adoption of AI output, to proactive and critical engagement, including active-trial, error-correction, and verification.
    \textbf{c}, Conceptual link between learner profiles, learning behavior, and learning outcomes. Students with higher university ranking or stronger prior knowledge are more likely to adopt proactive and critical engagement strategies, whereas students with weaker learner profiles are more likely to show limited engagement with AI. These differences in learning behavior may translate AI assistance into larger exam score gains for advantaged students, thereby widening pre-existing learning disparities.
  }
  \label{fig:overall}
\end{figure*}

Large language models (LLMs)\cite{openai2023gpt4,singh2025openai,meta2025llama,guo2025deepseek} possess a strong capability to streamline the retrieval, processing, and delivery of well-structured information and solutions across various domains~\cite{white2025livebench,wang2024mmlu}.
This makes them potentially beneficial at different stages of the educational process, including knowledge acquisition and summarization, review, and problem-solving\cite{futterer2023chatgpt,cooper2023examining}.
In fact, students’ spontaneous usages already reflect this influence:
A College Board survey (Jun 2024–Jun 2025) found that 84\% of U.S. high school students used generative AI for schoolwork~\cite{adair2025ushighschool}, while a HEPI survey (Dec 2024) found that 88\% of full-time undergraduate students in the UK reported using generative AI for their studies~\cite{hepi2025studentgenai}.

Because the primary goal of education is for students to acquire knowledge rather than simply complete assignments, the involvement of LLMs in learning has sparked widespread debate.
LLMs can provide individualized support during study, and this potential has driven rapid commercial development of AI-powered tutoring features such as Khan Academy's Khanmigo\cite{khan2023khanmigo} and a growing range of domain-specific AI tutors\cite{sonkar2024pedagogical, lee2024llava, chen2023gptutor}.
Yet the educational value of AI assistance may depend on how students engage with the assignment.
Students may use AI-generated responses to support understanding and review, but they may also rely on them to complete the assignment with limited engagement.
This possibility has led educators to worry that easy access to AI-generated responses may reduce the active thinking essential for genuine knowledge acquisition and create significant issues of academic integrity\cite{VOA2023, NYT2023,simkute2025new}.
Existing empirical findings further underscore this ambiguity. Studies report positive average effects in some settings\cite{chen2026learning}, no statistically significant benefit in others\cite{tarning2025more,lehmann2024ai}, and possible learning losses when AI assistance is not sufficiently scaffolded to support independent reasoning~\cite{bastani2025generative,stromberg2026generative}.
One reason for these divergent conclusions may be that average effects obscure variation across students and forms of AI use.

Research on earlier educational technologies, including online courses, lecture videos, flipped classrooms, adaptive tutoring software, and handheld calculators, has shown that average effects often mask substantial heterogeneity across student subgroups~\cite{xu2014performance,cacault2021distance,setren2019effects,roschelle2016online,muralidharan2019disrupting,hembree1986effects}.
These studies indicate that learner profiles and learning behavior can shape both the direction and the magnitude of learning effects, with some tools disproportionately benefiting higher-ability students and others narrowing pre-existing gaps.
This perspective suggests that the effects of generative AI in education should also be expected to vary across students.
Yet our knowledge of heterogeneous AI-assisted learning outcomes and how learning behavior shapes these patterns remains limited.

To address this gap, we conducted a randomized controlled experiment involving 318 university students with diverse learner profiles to examine who benefits from AI assistance and how learning behavior is associated with these benefits.
As illustrated in \autoref{fig:overall}\textbf{a}, participants were assigned to one of two course contexts, Python or game theory, and within each course were randomly allocated to either the experimental group with GPT access or the control group without LLM-based tool access.
The procedure includes a pre-task survey, course learning, an assignment, a structured review, an exam without GPT access for all participants, and a post-task survey.
Because learning has both near-term and long-term components~\cite{soderstrom2015learning}, this study focuses on proximal, near-term transfer as reflected in the immediate exam score following a controlled study session, a common and informative index of instructional effectiveness in prior educational research~\cite{vanlehn2011relative}.

Our findings indicate that heterogeneity in the effects of AI assistance on learning outcomes is associated with how students interact with the AI assistant.
As summarized in \autoref{fig:overall}\textbf{b}, learning behavior ranges from limited engagement with AI output to proactive and critical engagement, including independent attempts, verification of AI responses, and correction of AI-generated errors.
Among students in the experimental group, those whose interactions are characterized by proactive and critical engagement achieve higher exam scores relative to the control group.
In contrast, students who show limited engagement with AI show no significant improvement and in some cases perform worse than their control counterparts.
Further analysis shows that learner profiles are associated with the likelihood of adopting these more productive interaction practices.
As conceptualized in \autoref{fig:overall}\textbf{c}, students from higher-ranked universities or with stronger prior knowledge are more likely to engage with AI in a proactive and critical manner, and consequently tend to derive greater benefits from AI assistance.
Accounting for learning behavior substantially weakens the associations between learner profiles and exam score, suggesting that proactive and critical AI use is an important pathway through which university ranking and prior knowledge are linked to AI-assisted learning outcomes.
These findings suggest that AI assistance may amplify pre-existing differences in academic preparation rather than automatically serving as an equalizing force, raising important concerns about educational equity as AI becomes increasingly adopted in education.
They also inform the design and evaluation of AI-supported learning systems, as well as instructional and monitoring practices that guide students toward more productive AI-use learning behavior.

\section*{Results}

\begin{table*}[!h]
  \caption{Participant characteristics and study duration by course and group.}
  \centering
  \scalebox{0.85}{
    \begin{tabular}{lllcc}
      \Xhline{1.0pt}
      \textbf{Course}              & \textbf{Condition} & \textbf{Intervention} & \textbf{Sample size} & \textbf{Avg. duration
      (min)}                                                                                                                   \\ \midrule
      \multirow{3}{*}{game theory} & experimental       & GPT-assisted          & 158                  & 102.339 $\pm$ 16.949
      \\
                                   & control            & no LLM allowed        & 79                   & 96.711 $\pm$ 20.694
      \\
                                   & subtotal           & -                     & 237                  & 100.463 $\pm$ 18.431
      \\ \midrule
      \multirow{3}{*}{Python}      & experimental       & GPT-assisted          & 54                   & 103.537 $\pm$ 21.418
      \\
                                   & control            & no LLM allowed        & 27                   & 104.959 $\pm$ 17.320
      \\
                                   & subtotal           & -                     & 81                   & 104.011 $\pm$ 20.046
      \\ \midrule
      overall                      & all                & -                     & 318                  & 101.367 $\pm$ 18.887  \\ \Xhline{1.0pt}
    \end{tabular}
  }
  \label{tab:stat}
\end{table*}

\begin{figure*}[!t]
  \centering
  \includegraphics[width=1.0\textwidth]{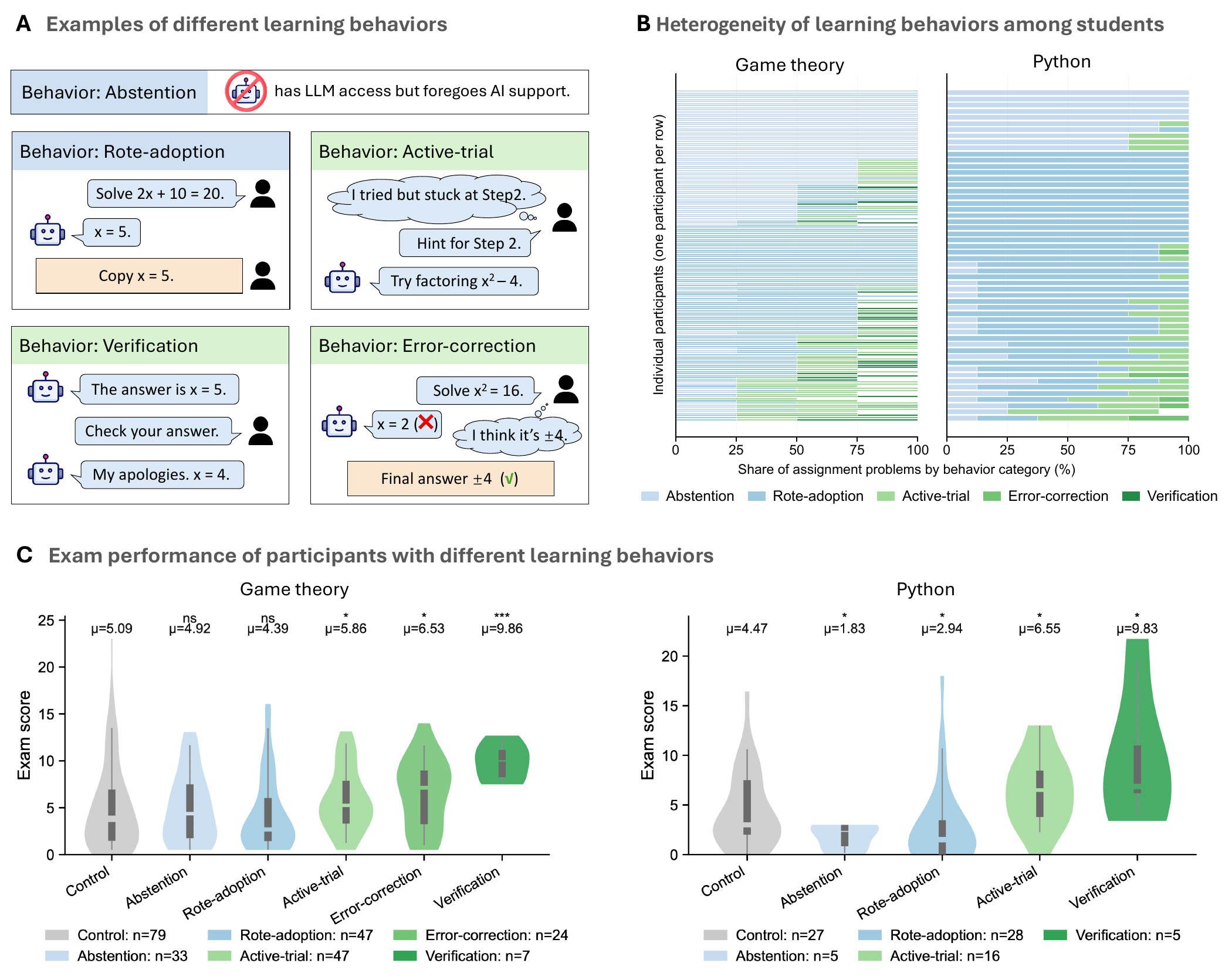}
  \caption{
    \textbf{Students adopt distinct AI-use learning behavior patterns, and proactive and critical engagement is associated with higher exam score.}
    \textbf{a}, Schematic examples of five annotated learning behavior patterns among experimental participants during assignment work: abstention, rote-adoption, active-trial, verification, and error-correction.
    These patterns distinguish participants who forgo available AI support, directly adopt AI output, attempt independently before seeking help, verify AI responses, or identify and correct AI-generated errors.
    \textbf{b}, Student-level composition of learning behavior across assignment problems in game theory and Python.
    Each horizontal bar represents one experimental participant, and colored segments indicate the share of assignment problems classified into each behavior pattern.
    \textbf{c}, Exam score distributions by learning behavior pattern in game theory and Python.
    Violin plots show score distributions with overlaid box plots.
    Legends report group sizes ($n$), and $\mu$ denotes mean exam score.
    Significance annotations are based on one-sided Brunner--Munzel tests for selected pairwise comparisons, adjusted using the Benjamini--Hochberg procedure.
  }
  \label{fig:behavior}
\end{figure*}

\subsection*{Experimental design}

We recruited 318 adult university students from institutions spanning a wide range of academic tiers to participate in a randomized controlled trial involving two courses: Python and game theory.
Participants were pre-screened to ensure that they had not previously learned the target subject, Python or game theory, while possessing prerequisite knowledge relevant to the experiment, including foundational mathematics for game theory and basic programming competence for Python.
Within each course, participants were randomly allocated to the experimental group or the control group in a 2:1 ratio, yielding the group sizes reported in \autoref{tab:stat}. This unequal allocation was adopted because the larger experimental group was necessary to support a comprehensive characterization of participants' generative AI interaction patterns, while the control group provided the no-LLM comparison group.
Welch's $t$-tests confirmed that prior knowledge scores did not differ significantly between the experimental and control groups in either course ($P > 0.05$), validating the comparability of the two groups at baseline.

Both groups engaged in the same structured process, which included studying course materials with tailored readings and video tutorials, completing a 20-minute assignment, reviewing with predefined questions, and taking a 20-minute exam.
Participants received performance-based compensation tied to their scores on both the assignment and the exam.
The experimental group had GPT access during the learning, assignment, and review stages, while the control group was prohibited from using any LLM-based tools.
Both groups completed the exam without GPT access.
Exam score served as the primary measure of learning outcomes.
Additionally, we conducted surveys before and after the study to collect demographic and academic information, assess prior knowledge, and gather participant feedback.
Full details are provided in the method section and Supplementary Information.

\subsection*{Heterogeneity in learning behavior shapes learning outcomes}

We analyze interaction logs from experimental participants during the assignment phase to characterize how students use the AI assistant in practice.
For each participant's chat events, we manually annotated whether the participant had attempted the problem before the chat, the purpose of each chat request, and how the participant modified the answer after the chat.
We then aggregated these chat-level annotations across assignment problems to derive student-level learning behavior patterns.
We next examine the association between these behavior patterns and subsequent exam score.

\paragraph{Heterogeneity in learning behavior.}
We characterize learning behavior along a broad distinction between limited engagement and proactive and critical engagement.
Within this framework, we identify five main learning behavior patterns among experimental participants.
Abstention denotes completing the assignment without using the assistant despite having GPT access.
Rote-adoption denotes direct use of AI-generated answers with minimal independent reasoning.
Active-trial denotes independent attempts before consulting the AI.
Verification denotes explicit checking or follow-up questioning of AI responses against one's own reasoning.
Error-correction denotes identifying and revising incorrect AI outputs.
As shown in \autoref{fig:behavior}\textbf{a}, abstention and rote-adoption are grouped as limited engagement, while active-trial, verification, and error-correction are grouped as proactive and critical engagement.

As shown in \autoref{fig:behavior}\textbf{b}, learning behavior varies substantially across students.
Some participants primarily show limited engagement, whereas others more frequently show proactive and critical engagement across assignment problems.
The error-correction group is absent in Python because GPT-4o answers on Python assignment problems are near ceiling in accuracy, leaving few observable cases in which students could identify and revise incorrect AI outputs.
Detailed analyses of model answer accuracy are provided in the Supplementary Information.

\begin{figure*}[!t]
  \centering
  \includegraphics[width=1.0\textwidth]{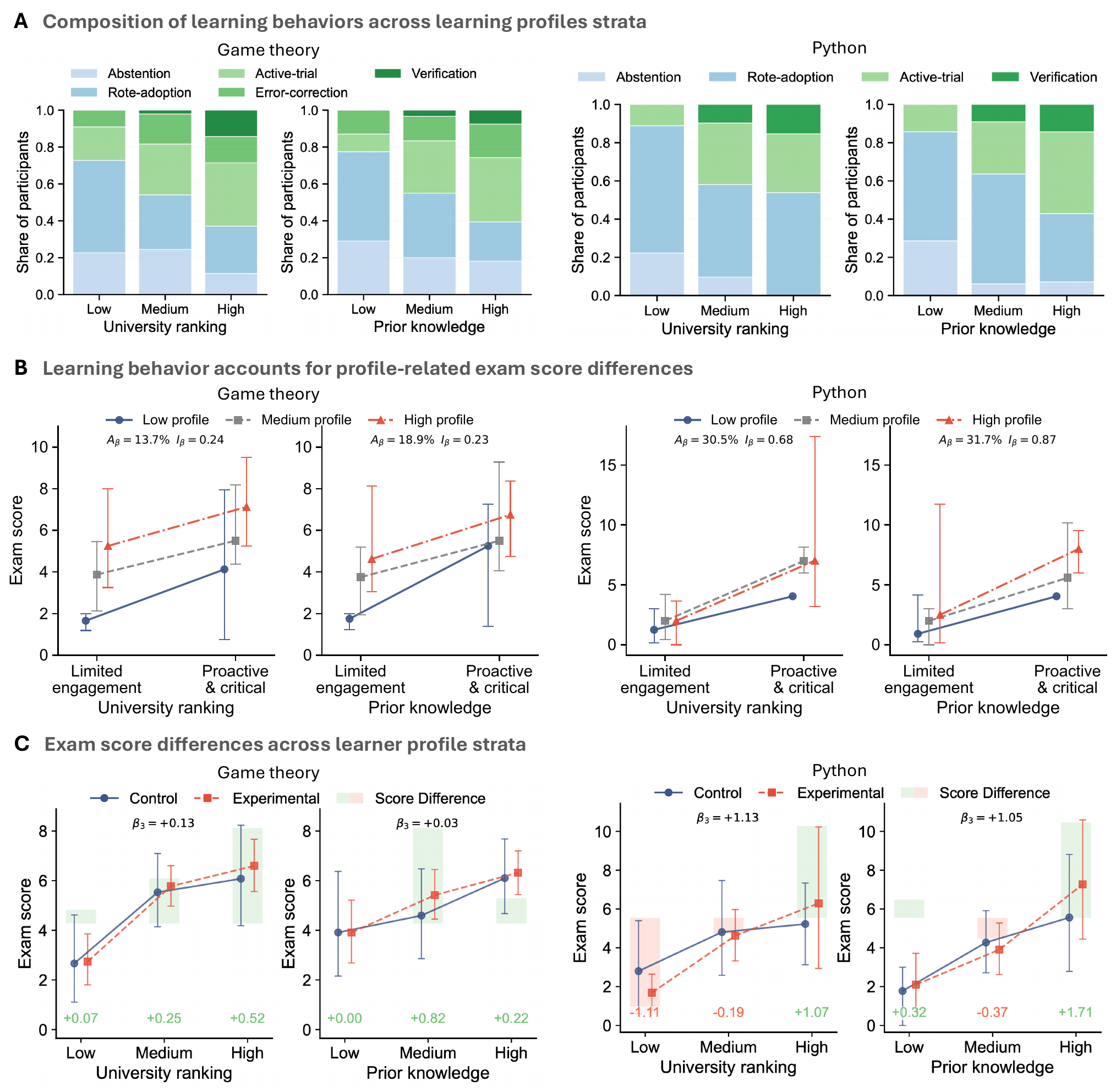}
  \caption{
    \textbf{Students with stronger learner profiles show more proactive-critical AI use and larger gains in exam score.}
    \textbf{a}, Learning behavior composition across university ranking and prior knowledge strata in game theory and Python.
    Stacked bars show the share of experimental participants in each behavior pattern.
    \textbf{b}, Exam score by learning behavior group, with separate lines for learner profile strata based on university ranking and prior knowledge.
    Points show median exam scores with bootstrap 95\% confidence intervals.
    Line overlap indicates smaller exam score differences between learner profile strata after learning behavior is fixed.
    $A_{\beta}$ denotes the proportion of the profile association accounted for by learning behavior, and $I_{\beta}$ denotes the indirect association through learning behavior.
    \textbf{c}, Exam score differences between the experimental and control groups across learner profile strata.
    Points show mean exam scores with bootstrap 95\% confidence intervals.
    Shaded bars show differences computed as experimental minus control.
    $\beta_{3}$ denotes the experimental-group-by-resource interaction coefficient from an OLS regression.
  }
  \label{fig:inequal}
  \vspace*{0.03\textheight}
\end{figure*}

\paragraph{Impact of learning behavior on learning outcomes.}
We then compare exam score across these behavior patterns.
As shown in \autoref{fig:behavior}\textbf{c}, abstention is close to the control group, indicating that unused GPT access provides little observable learning advantage.
Rote-adoption also shows limited benefit relative to the control group, even though it may help students complete assignment problems more directly.
By contrast, active-trial, verification, and error-correction are associated with higher exam scores.
These patterns involve independent attempts before asking, explicit evaluation of AI responses, or correction of AI-generated errors, which may help students use AI assistance as support for reasoning rather than as a substitute for engagement.
Together, these results suggest that GPT access alone does not determine learning outcomes.
Instead, how students incorporate AI into assignment work is associated with subsequent exam score.

\subsection*{AI-assisted learning behavior varies across learner profiles}
We next examine whether learning behavior differs across learner profiles.
We focus on two background dimensions, university ranking and prior knowledge, and ask whether these dimensions are associated with the adoption of proactive and critical engagement during AI-assisted assignment work.
In this analysis, proactive and critical engagement refers to active-trial, verification, and error-correction, while limited engagement refers to abstention and rote-adoption.
We then assess whether accounting for learning behavior patterns attenuates background-related differences in exam score, and whether exam score differences between the experimental and control groups vary across learner profile strata.
Details of the regression specifications and proportion tests in this subsection are provided in the method section.

\paragraph{Distribution of learning behavior across learner profiles.}
As shown in \autoref{fig:inequal}\textbf{a}, the composition of learning behavior differs across university ranking and prior knowledge strata in both courses.
Proactive and critical engagement is more common among students from higher-ranked universities and among those with stronger prior knowledge.
We further test these associations using OLS linear probability models that predict proactive and critical engagement from learner profiles.
In game theory, both university ranking and prior knowledge are positively associated with proactive and critical engagement ($a_{\text{univ}} = +0.171$, $P = 0.010$ and $a_{\text{prior}} = +0.161$, $P = 0.002$).
In Python, prior knowledge is also positively associated with proactive and critical engagement ($a_{\text{prior}} = +0.217$, $P = 0.046$).
Although the overall association with university ranking is not significant ($a_{\text{univ}} = +0.161$, $P = 0.127$), a follow-up one-sided two-proportion $z$-test shows that students from mid- and high-ranking universities are more likely than students from low-ranking universities to adopt proactive and critical engagement (proportion difference $= +0.32$, $P = 0.035$).

\paragraph{Learning behavior attenuates profile-related exam score differences.}
We next examine whether learning behavior explains part of the profile-related differences in exam score by asking whether these differences persist after accounting for learning behavior.
We estimate OLS models of exam score with learner profiles alone and after adding the binary learning behavior indicator.
In game theory, adding the learning behavior indicator decreases the university ranking coefficient from 1.765 to 1.524 and the prior knowledge coefficient from 1.212 to 0.983, while both behavior-adjusted learner profile associations remain significant.
Learning behavior accounts for 13.7\% and 18.9\% of the total associations, respectively.
The $R^{2}$ increases by 0.032 in both models.
In Python, the university ranking coefficient decreases from 2.235 to 1.552 and the prior knowledge coefficient decreases from 2.753 to 1.881, and both behavior-adjusted learner profile associations become non-significant.
Learning behavior accounts for 30.5\% and 31.7\% of the total associations, respectively.
The $R^{2}$ increases by 0.179 and 0.156 in the two models.
These reductions indicate that accounting for learning behavior attenuates profile-related differences in exam score, especially in Python.

The corresponding indirect associations are 0.24 (95\% CI, 0.017--0.556) for university ranking and 0.23 (95\% CI, 0.017--0.526) for prior knowledge in game theory.
In Python, the indirect associations are 0.68 (95\% CI, $-0.125$--1.714) for university ranking and 0.87 (95\% CI, 0.032--2.105) for prior knowledge.
The confidence intervals exclude zero for both learner profiles in game theory and for prior knowledge in Python, indicating significant indirect associations.
Although the confidence interval for university ranking in Python includes zero, its positive estimate is directionally consistent with the same pattern.
As shown in \autoref{fig:inequal}\textbf{b}, Python shows near-overlapping profile lines within each learning behavior category, whereas game theory shows more separated lines.
Taken together, coefficient attenuation, significant indirect associations for three of the four profile-course combinations, and within-behavior profile-line convergence indicate that learning behavior partly explains profile-related differences in exam score.

\paragraph{Exploratory analysis of AI-assisted learning gains across learner profiles.}
We next examine whether AI-assisted learning gains vary across learner profile strata.
We compare exam score between the control and experimental groups using an interaction model that includes learner profiles, experimental group assignment, and their interaction:
\begin{equation}
  \mathrm{ExamScore}_{i} \sim \alpha + \beta_{1}\, \mathrm{Background}_{i}
  + \beta_{2}\, \mathrm{Experimental}_{i}
  + \beta_{3}\, (\mathrm{Background}_{i} \times \mathrm{Experimental}_{i}) + \varepsilon_{i},
  \label{eq:interaction}
\end{equation}
where $\mathrm{Background}_{i}$ denotes either university ranking or prior knowledge, and $\mathrm{Experimental}_{i} \in \{0,1\}$ is an indicator equal to one for membership in the experimental group.
The interaction coefficient $\beta_{3}$ captures whether AI-assisted learning gains vary across learner profile strata.

As shown in \autoref{fig:inequal}\textbf{c}, all four interaction coefficients are positive, suggesting a directionally consistent pattern in which students with stronger learner profiles show larger AI-assisted gains relative to the control group.
This directional pattern is more pronounced in Python, where the interaction coefficients are $\beta_3 = +1.13$ for university ranking and $\beta_3 = +1.05$ for prior knowledge, compared with $\beta_3 = +0.13$ and $\beta_3 = +0.03$ in game theory.
However, these interaction estimates do not reach statistical significance.
The confidence intervals in \autoref{fig:inequal}\textbf{c} show substantial within-stratum variation relative to the estimated learning gains.
This uncertainty is expected in small samples, as detecting interaction effects typically requires larger samples than detecting main effects~\cite{mcclelland1993statistical}.
We therefore treat this pattern as exploratory evidence that AI-assisted learning gains may vary across learner profiles, a possibility that becomes increasingly important as AI assistance is adopted by more learners across broader educational contexts.

\begin{figure*}[!t]
  \centering
  \includegraphics[width=1.0\textwidth]{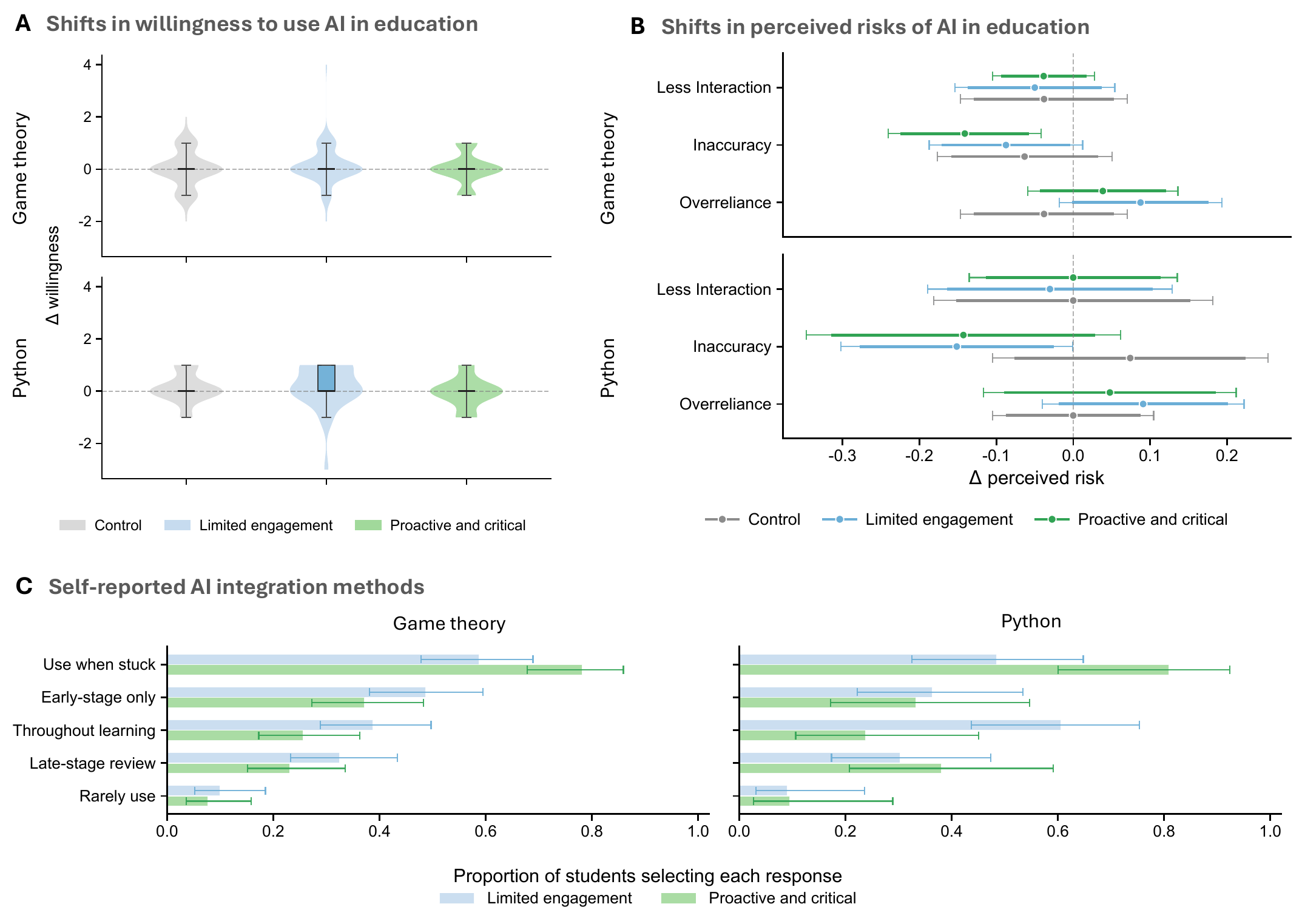}
  \caption{
    \textbf{Participant reports align with learning behavior patterns in AI attitudes and integration practices.}
    \textbf{a}, Changes from before to after the study in willingness to use AI in education, by course and group.
    Violin plots show individual-level changes with overlaid box plots.
    Whiskers span the 5th to 95th percentiles.
    \textbf{b}, Changes from before to after the study in perceived risks of AI use in education, by course, group, and risk dimension.
    Points indicate means.
    Thick and thin error bars indicate 90\% and 95\% confidence intervals.
    \textbf{c}, Self-reported AI integration methods among experimental participants by learning behavior pattern.
    Bars show the proportion of students selecting each response, with Wilson 95\% confidence intervals.
  }
  \label{fig:feedback}
\end{figure*}

\subsection*{Participant feedback analysis}

We finally examine participant feedback to assess whether the learning behavior patterns identified from interaction logs are consistent with students' subjective accounts of AI-assisted learning.
This analysis focuses on three complementary outcomes: changes in willingness to use AI in education, changes in perceived risks of AI use, and self-reported practices for integrating AI assistance into learning.
These survey responses help interpret why variation in learning behavior may be associated with different learning outcomes.
Additional post-task ratings of AI assistant quality and the stage-based study design are reported in the Supplementary Information.

\paragraph{Attitudinal shifts in willingness and risk perception.}
Changes from before to after the study in willingness to use AI in education are generally small across both courses and groups, as shown in \autoref{fig:feedback}\textbf{a}.
The clearest increase appears among limited engagement experimental participants in Python, whereas proactive and critical participants and participants in game theory show little change.
This pattern suggests that, when AI assistance provides relatively accurate and immediate feedback, limited engagement may make students more willing to use AI by reducing the perceived difficulty of assignment work and increasing dependence on the tool.

Risk perceptions vary across risk dimensions, as shown in \autoref{fig:feedback}\textbf{b}.
Concerns that AI use reduces interpersonal interaction show little change between experimental participants and the control group across both learning behavior patterns.
Concerns about inaccurate AI responses decline more clearly among both limited engagement and proactive and critical experimental participants relative to the control group.
This decline suggests that direct experience with AI assistance may increase confidence in response accuracy or usefulness.
Concerns about overreliance increase more among limited engagement participants than among proactive and critical participants.
This pattern is consistent with the behavioral annotation results, because direct answer generation can make AI assistance feel useful while also increasing students' awareness of their dependence on the tool.

\paragraph{Self-reported AI integration methods.}
\autoref{fig:feedback}\textbf{c} summarizes how experimental participants report integrating AI assistance into their learning process.
Across both courses, proactive and critical participants more often report using AI when stuck, suggesting that AI assistance is used as targeted support after independent effort.
Limited engagement participants more often report using AI throughout learning or only at early stages, suggesting less selective use of AI assistance during assignment work.
These self-reports are consistent with the behavior-based annotation results and help explain why proactive and critical engagement is associated with better learning outcomes.

\section*{Discussion}
LLMs are becoming part of everyday educational practice, but their role in learning remains contested.
This debate reflects a central tension between an enormous expansion of access to information, explanation, review, and problem-solving support, and the possibility that the expansion may reduce the active cognitive engagement required for durable knowledge acquisition.
Prior empirical studies report mixed effects of AI assistance on academic performance, suggesting that a simple average effect is either poorly defined, because it depends on context, or socially uninformative 
for understanding its educational consequences.
Our results support this view by showing that AI assistance depends on learner profiles and learning behavior.

This study identifies learning behavior as a key pathway through which GPT access is associated with heterogeneous learning outcomes.
Students in the experimental group do not use the assistant in a uniform way.
Some show limited engagement by either not using the assistant or adopting AI-generated answers with little independent reasoning, whereas others use the assistant in a more proactive and critical manner through active-trial, verification, and error-correction.
These behavioral differences are closely associated with exam score.
Proactive and critical engagement is associated with higher exam score, while limited engagement shows limited benefit relative to the control group.
Productive AI use is also unevenly distributed across learner profiles.
Students with higher university ranking or stronger prior knowledge more often adopt proactive and critical learning behavior.
Accounting for learning behavior attenuates these exam score differences, especially in Python.

Together, these findings suggest that GPT access alone does not determine who benefits from AI-assisted learning.
The educational value of AI assistance depends on whether students use it to support reasoning, verification, and reflection rather than to substitute for effort.
This has broader implications for AI use in education.
Broad access to AI assistance may not reduce inequality if students with stronger learner profiles are better able to use it productively, while less prepared students are more likely to abstain from use or rely on direct answers.
Equal access to the same tool can therefore produce unequal learning returns.
Policy makers and educators should treat AI literacy as educational support, teaching students to question AI responses, attempt problems before consultation, and avoid substituting AI output for active engagement.
Such guidance may shape whether AI reinforces existing advantages or broadens effective learning support.

We acknowledge several limitations of this study.
First, the study examines adult university students in two controlled short-course settings, and future work should test whether similar behavioral pathways appear across other educational stages, subjects, and institutional contexts.
Second, the experiment captures a single structured learning session and focuses on proximal learning outcomes measured by an immediate exam.
This design allows detailed observation of AI-assisted learning behavior, but it does not establish longer-term learning effects.
Third, the analysis of learning behavior and exam score within the experimental group is observational.
Although random assignment identifies the group-level effect of GPT access, behavioral pathways  reflect the interaction effects of the access with unmeasured learner characteristics.
Longitudinal and intervention studies with a much larger sample, ideally across more diverse 
learning contexts, are needed to address whether and how targeted guidance can improve learning behavior and help more students benefit from AI assistance.

\section*{Method}
\subsection*{Related work}
The application of large language models (LLMs) has shown significant promise across various domains. Numerous studies have demonstrated that LLMs can enhance productivity in tasks ranging from knowledge work to software development. For instance, generative AI has been shown to improve efficiency and quality in knowledge-intensive tasks~\cite{doi:10.1126/science.adh2586}, while tools like GitHub Copilot have been highlighted for their ability to assist developers in coding and debugging~\cite{peng2023impact}. Additionally, research has emphasized ChatGPT’s potential to foster creativity and support complex problem-solving, albeit with task-dependent effectiveness~\cite{lee2024empirical}.

In the context of education, recent research presents polarized views. On the theoretical front, many studies have explored the potential applications of generative AI in pedagogy, envisioning both its benefits and challenges. Frameworks for integrating AI into education have been proposed, offering structured methods for leveraging generative AI in teaching~\cite{su2023unlocking, wardat2023chatgpt, UnlockingChatGPT2023,liu2024socraticlm,kreijkes2025effects}.
These works often emphasize the transformative potential of tools like ChatGPT in enhancing engagement and facilitating personalized learning experiences~\cite{lim2023generative, baidoo2023education}. Beyond theoretical insights, practical tools have been developed to support specific educational goals, such as GPTutor for programming education~\cite{chen2023gptutor} and LLaVA-Docent for multimodal art appreciation~\cite{lee2024llava}. Such innovations underscore the optimism surrounding generative AI’s role in reshaping educational paradigms.

Conversely, concerns about generative AI’s application in education persist~\cite{simkute2025new}. Surveys have revealed significant apprehension among educators, particularly regarding academic dishonesty and plagiarism facilitated by AI tools~\cite{ibrahim2023perception}. Experimental studies in various educational contexts have further examined these issues. For example, Michalewicz et al.\cite{pardos2023learning} identified limitations in the pedagogical quality of ChatGPT-generated hints compared to human tutors, while Zhou et al.\cite{pankiewicz2023large} noted mixed results in automating feedback on programming assignments. Additionally, studies on ChatGPT’s role in solving programming bugs and explaining concepts have highlighted its dependency risks and limitations in fostering critical thinking~\cite{surameery2023use, biswas2023role}. These findings illustrate the dual-edged nature of generative AI in education, emphasizing the need for cautious and guided integration.

\subsection*{Experimental setup}
This study was approved by the Microsoft Research Ethics Review Program and the Princeton Institutional Review Board (IRB).
Informed consent was obtained from all participants before the main study.

\paragraph{Recruitment and consent.}
We conducted a randomized controlled trial with adult university students in two course contexts, Python and game theory.
Participants were recruited through two partner companies, DataOcean and Abaka AI.
During recruitment, candidates submitted academic resumes and completed a course-specific screening test.
Eligible Python candidates were required to have basic programming experience but no prior exposure to Python programming.
Eligible game theory candidates were required to have experience with equations, optimization, and calculus concepts such as derivatives, but no previous formal education in game theory.
After candidates entered the study platform, we further checked course eligibility using the pre-task survey.
Python participants were excluded if they reported having learned Python or having no prior programming background.
Game theory participants were excluded if they reported having learned game-theory concepts such as Nash equilibrium or static games.
Eligible participants received individual platform credentials and completed the main-study consent form before entering the experiment.
The consent form described the study workflow, survey and behavioral data collection, platform usage policies, and screen-recording requirement.
Screen recordings were required during the main study to support compliance checks and verify that participants did not use unauthorized external resources.

\paragraph{Main study workflow.}
Eligible participants were randomly allocated in a 2:1 ratio to the experimental group or the control group.
The unequal allocation provided a larger experimental sample for characterizing heterogeneous AI-use behavior, while preserving a no-LLM comparison group.
All participants completed the same fixed stage-based workflow on the study platform.
The workflow included a 10-minute pre-task phase with a baseline survey and timed prior knowledge assessment, a 40-minute course learning phase, a 20-minute assignment phase, a 20-minute review phase, a 20-minute exam phase, and a 5-minute post-task survey.
Participants who failed to advance to the next phase within a 20-minute grace period after the allotted time were treated as withdrawn.
The experimental group had GPT access during the learning, assignment, and review phases.
The control group did not have GPT access, and neither group had GPT access during the exam phase.
During the assignment and exam phases, participants could refer to the course materials, while the exam used problems distinct from the assignment and review phases.
Exam score from this restricted final assessment served as the primary learning outcomes measure.

\paragraph{Compensation and incentives.}
Participants received a base compensation of 150 RMB.
Additional performance-based incentives were calculated from assignment score, exam score, and completion speed.
The incentive score $S$ was defined as
\begin{equation}
    S = 0.4 \times S_a \times \left(1 + \max\left(0, 1 - \frac{T_a}{20}\right)\right)
    + 0.6 \times S_e \times \left(1 + \max\left(0, 1 - \frac{T_e}{20}\right)\right),
\end{equation}
where $S_a$ and $S_e$ denote assignment score and exam score, and $T_a$ and $T_e$ denote assignment and exam completion time in minutes.
The formula assigned 40\% weight to the assignment and 60\% weight to the exam, with a speed-based adjustment within each 20-minute assessment phase.
Participants ranked in the top 10\% by the incentive score received an additional 200 RMB, and those ranked in the top 10\%--30\% received an additional 100 RMB.
The average total compensation was 210 RMB.
Full details of stage-specific task materials, platform logging, data filtering, and scoring procedures are provided in the Supplementary Information.

\subsection*{Measures}
This subsection describes the primary measures of learning outcomes, learner profiles, and learning behavior.

\paragraph{Learner profiles.}
We operationalized learner profiles using two pre-task measures.
University ranking was self-reported in ordinal categories based on the BCUR ranking and collapsed into three tiers for stratified analyses.
The tiers were high, corresponding to top 50, medium, corresponding to top 50--500, and low, corresponding to outside top 500.
Prior knowledge was measured using a timed prerequisite quiz administered before the learning tasks.
Based on the course-specific score distributions and score granularity, we stratified prior knowledge using cut points of 3 and 6.5 in Python and 3.5 and 6.5 in game theory.

\paragraph{Learning outcomes.}
The primary learning outcomes measure was exam score, measured during a 20-minute exam phase in which GPT access was disabled for both the experimental and control groups.
Exam score was computed from participants' submissions within the stage time window.
Python responses were graded automatically using problem-specific unit tests, whereas game theory responses were graded by human reviewers using a predefined rubric.
Full scoring procedures are provided in the Supplementary Information.

\paragraph{Learning behavior.}
We characterized learning behavior among experimental participants during the assignment phase using platform interaction logs and chat records.
For each participant's chat events, annotators manually labeled whether the participant had attempted the problem before the chat, the purpose of the chat request, and how the participant modified the answer after the chat.
We aggregated these chat-level annotations across assignment problems and assigned each participant to one predominant learning behavior pattern using the rule described in the Supplementary Information.

We identified five learning behavior patterns among experimental participants.
Abstention denotes completing the assignment without using the assistant despite having GPT access.
Rote-adoption denotes direct use of AI-generated answers with minimal independent reasoning.
Active-trial denotes independent attempts before consulting the assistant.
Verification denotes explicit checking of AI responses or follow-up requests to understand, confirm, or evaluate the solution.
Error-correction denotes identifying and revising incorrect AI outputs.
We group abstention and rote-adoption as limited engagement.
We group active-trial, verification, and error-correction as proactive and critical engagement, reflecting independent attempts before asking, explicit evaluation of AI responses, or correction of AI-generated errors.
Full operational definitions and grouping procedures are provided in the Supplementary Information.

\subsection*{Statistical analysis}
This subsection describes the hypothesis tests and regression specifications used throughout the results section.
Unless otherwise noted, all analyses were conducted separately for Python and game theory.

\paragraph{Hypothesis tests.}
Baseline balance on prior knowledge between the experimental and control groups was assessed using two-sided Welch's $t$-tests.
To compare exam score distributions between each learning behavior pattern and the control group, we used one-sided Brunner--Munzel tests, with the direction of the alternative hypothesis specified a priori.
Proactive and critical engagement profiles were tested against the alternative that their scores exceeded those of the control group.
Limited-engagement profiles were tested against the alternative that their scores fell below those of the control group.
For each course, the resulting $P$-values were adjusted for multiple comparisons using the Benjamini--Hochberg procedure.
For the Python follow-up analysis of university ranking, we compared the prevalence of proactive and critical engagement between students from low-ranking universities and those from mid- or high-ranking universities using a one-sided two-proportion $z$-test.
The alternative hypothesis specified that the proportion in the mid- and high-ranking group exceeded that in the low-ranking group.
Because learning behavior was observed after random assignment rather than randomized, comparisons across learning behavior patterns were interpreted as observational associations rather than causal estimates of behavior-specific effects.

\paragraph{Regression models.}
To assess whether student background characteristics predicted the adoption of proactive and critical engagement within the experimental group, we estimated a linear probability model using ordinary least squares (OLS).
The outcome variable was a binary indicator of proactive and critical engagement, coded as one for active-trial, verification, or error-correction and zero for abstention or rote-adoption.
The model took the form
\begin{equation}
    \mathrm{ProactiveCritical}_{i} \sim a_{0} + a\, \mathrm{Background}_{i} + \eta_{i},
    \label{eq:lpm}
\end{equation}
where $\mathrm{ProactiveCritical}_{i} \in \{0,1\}$ denotes this proactive and critical engagement indicator, and $\mathrm{Background}_{i}$ denotes either university ranking tier ($\mathrm{UnivRank}_{i}$) or prior knowledge score ($\mathrm{PriorKnowledge}_{i}$).
Both background variables were encoded as ordered numeric variables with levels Low, Mid, and High mapped to 1, 2, and 3, respectively.
\autoref{eq:lpm} was estimated separately for each background variable.
The resulting coefficients are reported as $a_{\mathrm{univ}}$ and $a_{\mathrm{prior}}$ in the results, representing the estimated change in the probability of proactive and critical engagement associated with a one-level increase in university ranking and prior knowledge, respectively.

To quantify the extent to which adjustment for learning behavior pattern attenuated background-associated differences in exam score within the experimental group, we estimated two nested OLS models.
The background-only specification took the form
\begin{equation}
    \mathrm{ExamScore}_{i} \sim \alpha + \beta\, \mathrm{Background}_{i},
    \label{eq:bg_only}
\end{equation}
and the behavior-augmented specification took the form
\begin{equation}
    \mathrm{ExamScore}_{i} \sim \alpha^{\prime} + \beta^{\prime}\, \mathrm{Background}_{i}
    + \gamma\, \mathrm{ProactiveCritical}_{i},
    \label{eq:bg_behavior}
\end{equation}
where $\mathrm{Background}_{i}$ is defined as in \autoref{eq:lpm}, and $\mathrm{ProactiveCritical}_{i}$ is the binary learning behavior indicator defined above.
The coefficient $\beta$ in \autoref{eq:bg_only} represents the total association between the learner profile and exam score.
The coefficient $\beta^{\prime}$ in \autoref{eq:bg_behavior} represents the corresponding behavior-adjusted association, and $\gamma$ represents the association between proactive and critical engagement and exam score after accounting for the learner profile.
The proportion of the total association accounted for by learning behavior was denoted by $A_{\beta}$ and calculated as
\begin{equation}
    A_{\beta} = \frac{\beta - \beta^{\prime}}{\beta} \times 100\%.
    \label{eq:accounted}
\end{equation}
The gain in explained variance was summarized by $\Delta R^{2} = R^{2}_{\text{aug}} - R^{2}_{\text{base}}$, where $R^{2}_{\text{aug}}$ and $R^{2}_{\text{base}}$ denote the coefficients of determination from \autoref{eq:bg_behavior} and \autoref{eq:bg_only}, respectively.
We denoted the indirect association through learning behavior by $I_{\beta} = a\gamma$, where $a$ was obtained from \autoref{eq:lpm} and $\gamma$ from \autoref{eq:bg_behavior}.
We used 10,000 participant-level nonparametric bootstrap resamples to obtain percentile 95\% confidence intervals for $I_{\beta}$.
Because learner profiles and learning behavior were not randomized, this mediation-style decomposition was interpreted as associational rather than as evidence of causal direct or indirect effects.

To test whether AI assistance modified the association between student background and exam score, we estimated the OLS interaction model introduced in the results section in \autoref{eq:interaction}, pooling both experimental and control participants.
In this model, $\mathrm{Experimental}_{i} \in \{0,1\}$ is a binary indicator equal to one for experimental group membership.
The interaction coefficient $\hat{\beta}_{3}$ captures the difference in the slope relating student background to exam score between the experimental and control groups.
Additional implementation details are provided in the Supplementary Information.

  \section*{Data availability}
  The content of pre-task survey, post-task survey, assignment questions, and exam questions are available in the Supplementary Information.
  The analysis-result datasets supporting the findings, together with the course materials and other materials provided through the study platform, are publicly available at \url{https://github.com/yjw1029/GenAI-Learning-Outcomes}.
  Raw participant-level experimental data are not publicly available because participant consent and ethics approvals restrict open release.
  Qualified research collaborators approved by the corresponding authors may request controlled access to the data, subject to institutional approval and conditions prohibiting re-identification, redistribution, and unauthorized secondary use.

  \section*{Code availability}
  The code applied in the experiments is publicly available at \url{https://github.com/yjw1029/GenAI-Learning-Outcomes}.

  \bibliography{main}

  \section*{Acknowledgments}
  We thank Aidi Li for assistance with experimental preparations.

  \section*{Author contributions}

  J.Y. conceived the idea of this work, designed, prepared and organized experiments, set up the experimental platform, analyzed the results, and contributed to the writing of this manuscript.
  Yueqi Xie (Y.X.) conceived the idea of this work, designed and prepared experiments, analyzed the results, and contributed to the writing of this manuscript.
  Jiyan He (J.H.) designed and prepared experiments, set up the experimental platform.
  R.Y. designed and prepared experiments.
  Junming Huang (J.H.) prepared and organized the experiment, analyzed the results and contributed to the writing of this manuscript.
  B.Z. contributed to the writing of this manuscript.
  S.R. contributed to the writing of this manuscript.
  Yu Xie (Y.X.) contributed to the writing of this manuscript and coordinated the research project.
  X.X. contributed to the writing of this manuscript and coordinated the research project.
  F.W. conceived the idea of this work, analyzed the results, and contributed to the writing of this manuscript and coordinated the research project.

  \section*{Additional information}

  \textbf{Supplementary Information} accompanies this manuscript in the attached supplementary information file.

  \textbf{Competing interests:}
  The authors declare no competing interests.

  \startsupplement

\section{Platform and data collection}
\label{sec:platform}

We developed a web-based platform to standardize the delivery of study materials and tasks while capturing behavioral traces throughout the main study. These platform logs formed the final behavioral dataset used in our analyses.

\begin{figure*}[!t]
    \centering
    \includegraphics[width=\textwidth]{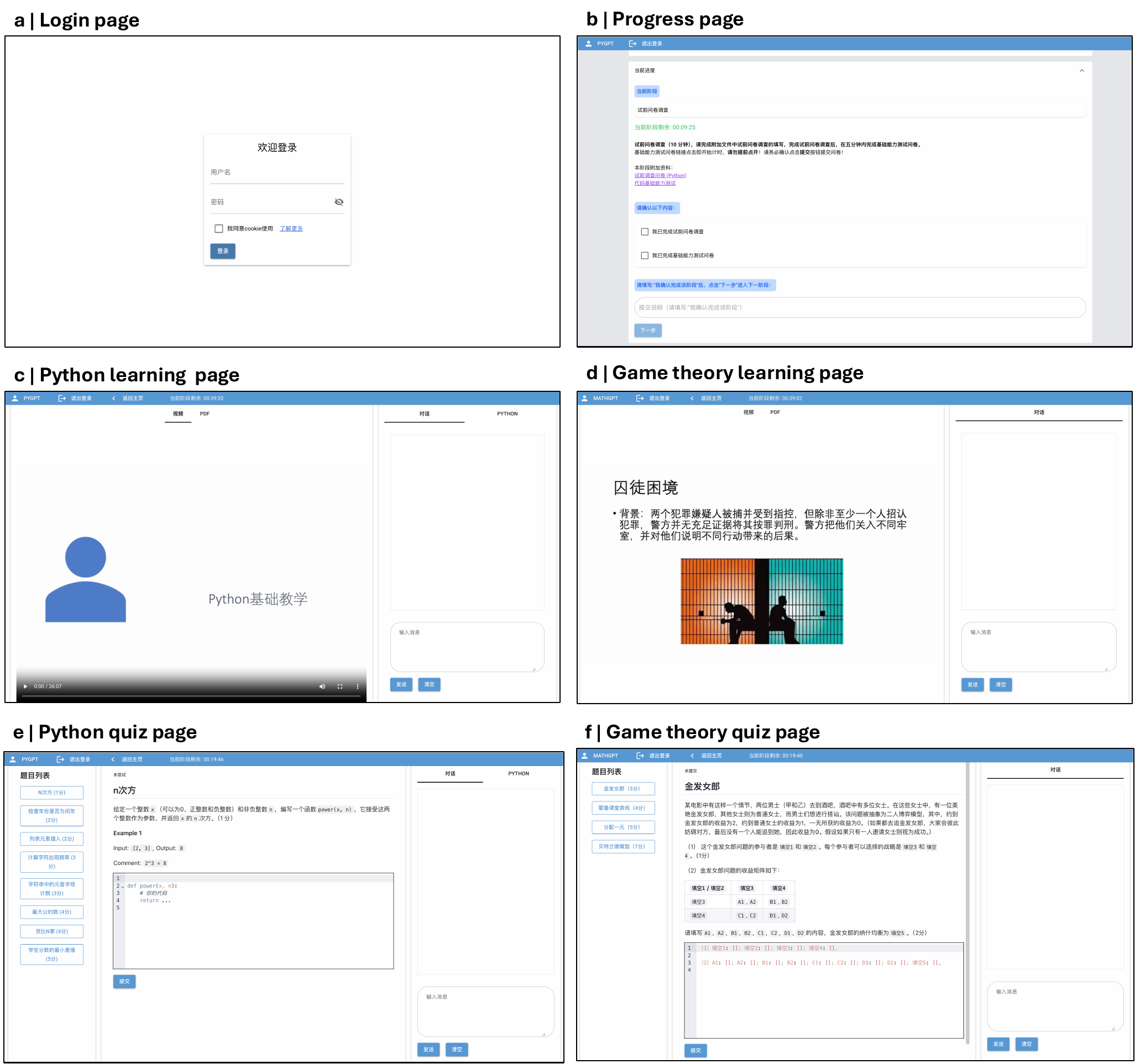}
    \caption{\textbf{Representative platform pages.} \textbf{a}, Account login. \textbf{b}, Stage progress page that introduces the current stage and provides navigation to stage-specific pages. \textbf{c}, Python learning interface. \textbf{d}, Game theory learning interface. \textbf{e}, Python assignment interface. \textbf{f}, Game theory assignment interface. Experimental participants additionally had access to an embedded GPT-4o chat panel during eligible stages. Control participants did not.}
    \label{fig:platform}
\end{figure*}

\subsection{Recruitment and screening (company-managed)}
Participant recruitment and the initial eligibility screening were conducted in collaboration with two partner companies. During this recruitment phase, the companies (i) distributed a screening notice, (ii) collected candidates' schedules and resumes, and (iii) administered a screening survey. Eligible Python candidates were required to have basic programming experience but no prior exposure to Python programming. Eligible game theory candidates were required to have experience with equations, optimization, and calculus concepts such as derivatives, but no previous formal education in game theory. Importantly, these recruitment-phase materials were managed by the companies and were \emph{not} transferred into the research data repository. The only information passed to the research team for study execution was an eligibility outcome indicating whether a candidate was invited to the main study.

\subsection{Main study workflow (platform-completed)}
The main study was completed on the platform. Invited participants received individual credentials and logged in through the authentication page (\autoref{fig:platform}\textbf{a}). Prior to beginning the study workflow, participants were required to complete the main-study informed consent form (full text in Supplementary Information \autoref{sec:exp-info}).

After consent, the platform enforced a fixed, stage-based workflow via a progress page (\autoref{fig:platform}\textbf{b}) and stage-specific access controls. The study comprised a pre-task survey, a timed prior knowledge assessment, a brief platform familiarization stage, timed course learning (\autoref{fig:platform}\textbf{c}--\textbf{d}), a timed assignment (Python or game theory; \autoref{fig:platform}\textbf{e}--\textbf{f}), a review stage during which reference answers were available, a timed exam without GPT access, and a post-task survey.

Stage timers were tracked on the server and displayed to participants. Upon the end of a stage's allotted time, the platform allowed a 20-minute grace period during which participants could remain on the progress page and enter the next stage; if the grace period elapsed without entering the next stage, the session was treated as abandoned and the participant was logged out. The prior knowledge assessment pages additionally enforced a hard time limit and auto-submitted responses upon timeout.

The experimental group had GPT access through an embedded chat interface powered by GPT-4o during the course learning, assignment, and review phases. The control group did not have GPT access. The platform configuration enforced stage-specific availability, and tool calling was not enabled.

The study was primarily completed remotely. Participants could use their own computers; participants without reliable computer access could instead complete the study at company-designated locations where computers were provided. The study procedure and platform constraints were identical across settings. Throughout the main study, participants were required to record their screen for compliance verification.

\subsection{Stage timing, materials, and incentives}
Unless otherwise specified, each stage of the main study was time-bound. The pre-task phase lasted 10 minutes and included a pre-task survey and a 5-minute prior knowledge assessment. The course learning phase lasted 40 minutes. Participants were provided with written materials and course-specific video tutorials. The Python course and the game theory course each included a 26-minute instructional video. Python participants also had access to an integrated Python interpreter.

The assignment, review, and exam phases each lasted 20 minutes. During the assignment phase, participants completed problems designed to reinforce the learning objectives and could refer to the learning materials. Python participants received 8 progressively challenging programming problems, while game theory participants solved 4 problems. The review phase provided practice questions and answer keys. Python participants received 32 review problems, and game theory participants received 10 review problems. During the exam phase, participants completed problems distinct from those in the assignment and review phases. Python participants received 8 programming problems, and game theory participants solved 4 problems. Participants could refer to the learning materials and assignment solutions during the exam, but GPT access was disabled for both the experimental and control groups.

After the exam, participants completed a 5-minute post-task survey that collected feedback on GPT access, perceptions of GPT's utility in education, motivations for study participation, and attitudes toward large language models in education. Participants received a base compensation of 150 RMB. Additional performance-based incentives were calculated from assignment score, exam score, and completion speed. The incentive score $S$ was defined as
\begin{equation}
    S = 0.4 \times S_a \times \left(1 + \max\left(0, 1 - \frac{T_a}{20}\right)\right)
    + 0.6 \times S_e \times \left(1 + \max\left(0, 1 - \frac{T_e}{20}\right)\right),
\end{equation}
where $S_a$ and $S_e$ denote assignment score and exam score, and $T_a$ and $T_e$ denote assignment and exam completion time in minutes. Participants ranked in the top 10\% by the incentive score received an additional 200 RMB, and those ranked in the top 10\%--30\% received an additional 100 RMB. The average total compensation was 210 RMB.

\subsection{Server-side behavioral logging}
All behavioral traces were recorded on the server as an append-only event log. Concretely, the platform persisted each interaction as a \texttt{UserAction} record in a SQLite database (\texttt{user\_actions.db}). Each record contains a participant identifier (\texttt{username}), an event name (\texttt{action}), a server-side timestamp (\texttt{timestamp}; Unix timestamp, in seconds), and a JSON payload (\texttt{value}) containing stage- and page-specific metadata.

The logging schema was designed to support both compliance auditing (e.g., consent, stage progression) and fine-grained behavioral analysis (e.g., drafting trajectories, clickstreams). The platform recorded the following primary event categories:
\begin{itemize}[leftmargin=*]
    \item \textbf{Consent.} A consent submission event (\texttt{action=consent}) indicating the participant's decision and submission time.
    \item \textbf{Stage progression.} Progress events (\texttt{action=progress}) indicating the current workflow stage, enabling the reconstruction of stage entry times, stage durations, and completion status.
    \item \textbf{Surveys.} For each survey instrument, the platform stored (i) per-page navigation events, (ii) saved answers (\texttt{::answers}), and (iii) the final submission event (\texttt{::submit}). For the timed prior knowledge assessment, the platform additionally logged a start time and enforced auto-submission upon timeout.
    \item \textbf{Problem navigation and clicks.} When participants navigated between questions, the platform logged a per-problem click event (\texttt{\{problem\}::click}) with stage and problem identifiers.
    \item \textbf{Answer drafting trajectories.} While participants worked on a question, the platform periodically saved the content of the answer editor as \texttt{\{problem\}::answer}. To reduce redundant writes, a snapshot was stored only when the content differed from the previously saved version. The snapshot timer ran every 5 seconds, enabling reconstruction of fine-grained drafting trajectories.
    \item \textbf{Submissions and assessment outcomes.} For game theory (free-text/fill-in) questions, the platform logged explicit submission events (\texttt{\{problem\}::submit}). For Python questions, the platform provided a built-in judge and logged judge events (\texttt{\{problem\}::judge}) containing pass/fail outcomes and diagnostic information.
    \item \textbf{LLM chat (experimental group only).} The platform logged the full user-visible chat message sequence (\texttt{::chat}) with timestamps and its associated task context. Tool calling was not enabled. System prompts were used internally to generate model responses but were not stored in the behavioral log; only user-visible messages were retained.
    \item \textbf{Auxiliary Python execution.} In Python stages where an execution sandbox was available, execution requests and outputs were logged (e.g., \texttt{learning::python} and \texttt{\{problem\}::python}) to capture debugging and experimentation behavior.
    \item \textbf{Review and answer access.} During the review stage, the platform exposed reference answers and logged answer-view events (\texttt{\{problem\}::show\_answer}) indicating when participants opened or closed the answer panel.
\end{itemize}

\autoref{tab:platform-events} summarizes the main event types and their semantics.
\begin{table*}[ht]
    \centering
    \caption{\textbf{Platform event taxonomy (server-side logs).} Each log entry is stored as a tuple \texttt{(username, action, timestamp, value)}; \texttt{value} is a JSON payload that includes raw stage identifiers (e.g., \texttt{learning/a1/a2/review}) and task identifiers when applicable.}
    \label{tab:platform-events}
    \footnotesize
    \setlength{\tabcolsep}{6pt}
    \renewcommand{\arraystretch}{1.12}
    \begin{tabularx}{\textwidth}{>{\raggedright\arraybackslash\ttfamily}p{0.28\textwidth}X}
        \toprule
        \textbf{Action (pattern)} & \textbf{Semantics (selected examples)}                                                            \\
        \midrule
        consent                   & Main-study consent submission and decision.                                                       \\
        progress                  & Stage progression events (stage label), enabling reconstruction of stage timing.                  \\
        \{survey\}::answers       & Saved survey answers (per-page saves and intermediate saves).                                     \\
        \{survey\}::submit        & Final submission of a survey instrument.                                                          \\
        \{survey\}::start\_time   & Start-time marker for the timed prior knowledge assessment (used to enforce a fixed time budget). \\
        \{survey\}::page\_change  & Survey page navigation with page index.                                                           \\
        \{problem\}::click        & Question-navigation events with stage and problem identifiers.                                    \\
        \{problem\}::answer       & Periodic answer-editor snapshots (every 5 s, recorded only when content changes).                 \\
        \{problem\}::submit       & Submission events for game theory free-text/fill-in questions.                                    \\
        \{problem\}::judge        & Python judge submissions (pass/fail) with diagnostic outputs.                                     \\
        \makecell[l]{\{problem\}::python                                                                                              \\learning::python} & Python execution-sandbox events (executed code and execution results). \\
        \makecell[l]{\{problem\}::chat                                                                                                \\learning::chat} & LLM chat histories with timestamps and task context (user-visible messages only; tool calling disabled; system prompts not stored). \\
        \{problem\}::show\_answer & Review-phase answer viewing events (open/close state changes).                                    \\
        \bottomrule
    \end{tabularx}
\end{table*}

\subsection{Initial dataset and main-stage eligibility screening}
Course eligibility was screened in two steps. During recruitment, the partner companies administered the course-specific screening questions shown in Supplementary Information \autoref{tab:screening-questions-python} and \autoref{tab:screening-questions-game-theory} and invited eligible candidates to the main study. After invited participants entered the platform, we applied an additional course-eligibility check using the pre-task survey. In the Python course, participants were excluded if they indicated in Q8-P (\autoref{tab:pre-task-survey-questions}) that they had not learned any programming language or that they had already learned Python. In the game theory course, participants were excluded if they indicated in Q10-G (\autoref{tab:pre-task-survey-questions}) that they had learned game-theory concepts such as Nash equilibrium or static games. These main-stage survey exclusions are distinct from both the company-managed recruitment screening and the post-hoc filtering described in Supplementary Information \autoref{sec:data-processing}.

After participants completed all required stages, screen recordings were manually reviewed to verify compliance with study rules. Specifically, we checked that participants did not access outside resources beyond the tools provided within the platform during the study session. Participants who passed this screen-recording audit were included in the candidate dataset constructed from platform logs.
In total, we collected platform logs and audited screen recordings for 346 participants; this set served as the input to the data-processing pipeline (Supplementary Information \autoref{sec:data-processing}).

\section{Data processing}
\label{sec:data-processing}

\subsection{Filtering and exclusions}
Starting from the 346 platform-completed study sessions described in Supplementary Information \autoref{sec:platform}, we applied post-hoc data-quality filters based on platform records to construct the final analytic dataset.
Two exclusion filters are applied:
\begin{itemize}[leftmargin=*]
    \item \textbf{Extremely disengaged participation.} Participants were excluded if they (i) obtained a total score of 0 and (ii) exhibited no meaningful attempt behaviors on the easiest items. We operationalized ``no meaningful attempt'' as the absence of substantive drafting and submission activity on a basic Python item involving \texttt{len} (exam problem~1) and on the first four blanks of a basic game theory item (exam problem~2, the ``big pig/small pig'' sub-questions).
    \item \textbf{Cheating signals not captured by screen recordings (Python only).} Participants were excluded if their Python solutions systematically relied on syntax that was (i) not covered by the course materials, (ii) not introduced in their recorded LLM chat history, and (iii) inconsistent with their self-reported prior programming-language experience. We did not apply an analogous automated syntax-based screen to game theory because responses were fill-in-the-blank and did not admit a comparable programmatic check.
\end{itemize}

In total, these filters excluded 13 participants in Python and 4 participants in game theory for extremely disengaged participation, and excluded 11 additional participants in Python for syntax-based cheating signals (28 excluded in total). After filtering, the final dataset used in our analyses includes 318 participants.

\subsection{Score computation}
\label{sec:score-computation}
We computed task-performance scores for the assignment, the exam, and the prior knowledge assessment from platform logs and survey records.

\paragraph{Prior knowledge assessment.}
Prior knowledge scores were computed from the merged survey responses.
Each question response was checked against a predefined answer key; we then computed a weighted sum of correct items and reported both the raw weighted score and a normalized score on a 0--1 scale (dividing by the total weight).
For Python items, correctness was based on direct string matching.
For game theory items, we applied lightweight normalization (e.g., trimming whitespace and normalizing minus signs) and accepted simple numeric equivalences (e.g., decimals or fractions) when applicable.

\paragraph{Python assignment and exam.}
Python assignment and exam scores were computed programmatically from the platform behavior database.
For each participant and each assessment stage, we identified the stage start time from platform ``progress'' events and retained only submissions within a 20-minute window after stage entry.
Each code submission was evaluated against problem-specific unit tests, and the submission score was computed as the fraction of test cases passed multiplied by the problem's maximum points.
For each problem, we used the maximum score achieved across eligible submissions; overall scores were then computed by summing problem scores within the assignment and within the exam, respectively.

\paragraph{Game theory assignment and exam.}
Game theory assignment and exam responses were fill-in-the-blank and were scored via human review.
We used a standardized per-blank annotation format, in which each blank was marked as correct/incorrect and stored as a dictionary of binary correctness flags along with the extracted final submission metadata (answer text, submission timestamp, and action identifier).
To define the final submission for review, we selected the last answer submission within 20 minutes of entering the corresponding assessment stage, based on the platform action log.
Scores were then computed by mapping each blank's correctness flag to a predefined point value and summing across blanks within each problem, and then across problems within the assignment and within the exam, respectively.

\subsection{Annotation of chat actions}
\label{sec:chat-action-annotation}
For participants in the experimental group, we annotated chat actions to characterize how participants solicited and used assistance from a large language model while solving problems. Each annotation corresponds to a single platform chat event (a unique chat action identifier) and was performed within the same problem context.

\paragraph{Action-level labels.}
For each chat action, annotators labeled a short window of participant activity immediately before and after the chat. The schema has three components, which were recorded per chat action:
\begin{itemize}[leftmargin=*]
    \item \textbf{Pre-chat attempt} (\texttt{pre.tried}): whether the participant made a substantive attempt on the problem before asking the model (yes, no, or unknown).
    \item \textbf{Request content and intent} (\texttt{ask.type} and \texttt{ask.goal}): the request type (final-answer request, concept-level question, debugging question, or challenge) and the intended goal (final answer, hint, explanation, correction, verification, or rechecking), with an ``unknown'' option when the intent could not be inferred.
    \item \textbf{Post-chat incorporation} (\texttt{post.mode}): how the participant incorporated the model response (copy and paste, near-verbatim typing, adaptation, a mixture of these, or no observable follow-up work), with an ``unknown'' option when incorporation could not be inferred.
\end{itemize}
For Python, we additionally recorded whether the model response directly provided a correct solution to the participant's request (\texttt{ask.correct}). For game theory, we optionally recorded per-blank answer states before the chat, in the model's response, and after the chat (stored as per-blank dictionaries with binary correctness flags). This representation supports downstream checks of whether a participant adopted, rejected, or corrected an incorrect model suggestion at the blank level.

\paragraph{Annotation windows.}
To support consistent interpretation, each chat action was labeled with a pre-chat window and a post-chat window. The pre-chat window typically spanned from the first action after entering the problem (or the end of the previous post-chat window within the same problem) up to the chat action. The post-chat window typically spanned from the first action after the chat until the action immediately preceding the next chat within the same problem; if there was no subsequent chat, the window extended to the final action in that problem. Annotators adjusted these boundaries only when the default window did not reflect the relevant behavioral context.

\subsection{Deriving participant-level learning behavior patterns}
\label{sec:chat-behavior-groups}
We aggregated action-level labels during the assignment phase and assigned each participant to a single predominant learning behavior pattern, separately for Python and game theory. Participants who completed the assignment without using the assistant despite having GPT access were labeled \textbf{Abstention}. For participants who used chat, we derived problem-level indicators and then aggregated these indicators across problems to obtain participant-level rates and flags.

\paragraph{Python groups.}
For each Python problem, we marked whether a participant attempted the problem before asking the model and whether the participant requested a final answer. We then computed the fraction of problems showing evidence of attempting before asking and the fraction of problems containing at least one final-answer request without a prior attempt.
We defined \textbf{Rote-adoption} as requesting final answers without a prior attempt for at least 75\% of problems. We defined \textbf{Active-trial} as exhibiting evidence of attempting before asking on at least 25\% of problems. We defined \textbf{Verification} as explicit checking of an AI response or a follow-up request to understand, confirm, or evaluate the solution within the same problem. To ensure that each participant was assigned to exactly one group, we used a fixed precedence rule with verification resolved before active-trial, rote-adoption, and other chat users.

\paragraph{Game theory groups.}
For each game theory problem, we defined an attempt indicator using either \texttt{pre.tried} or, when available, the fraction of blanks that were attempted before the first chat. We defined \textbf{Rote-adoption} as requesting final answers without a prior attempt and subsequently adopting responses verbatim (as indicated by post-chat incorporation modes), for at least 50\% of eligible problems. Problems were considered eligible when there was sufficient post-chat evidence to assess incorporation, excluding cases with no observable post-chat work.
We defined \textbf{Active-trial} as exhibiting evidence of attempting before asking on at least 50\% of eligible problems. We defined \textbf{Verification} analogously to Python and included cases in which a participant challenged a model response that was judged incorrect for at least one blank. We defined \textbf{Error-correction} when a participant subsequently produced correct blank values that corrected model-suggested errors for blanks that were not already correct before the chat. Final assignment used a fixed precedence rule with error-correction and verification resolved before active-trial, rote-adoption, and other chat users.

For analyses using broader behavior families, we grouped abstention and rote-adoption as \textbf{limited engagement}. We grouped active-trial, verification, and error-correction as \textbf{proactive and critical engagement}. These aggregate categories match the definitions used in the Main Paper.

\begin{figure*}[ht]
    \centering
    \includegraphics[width=\textwidth]{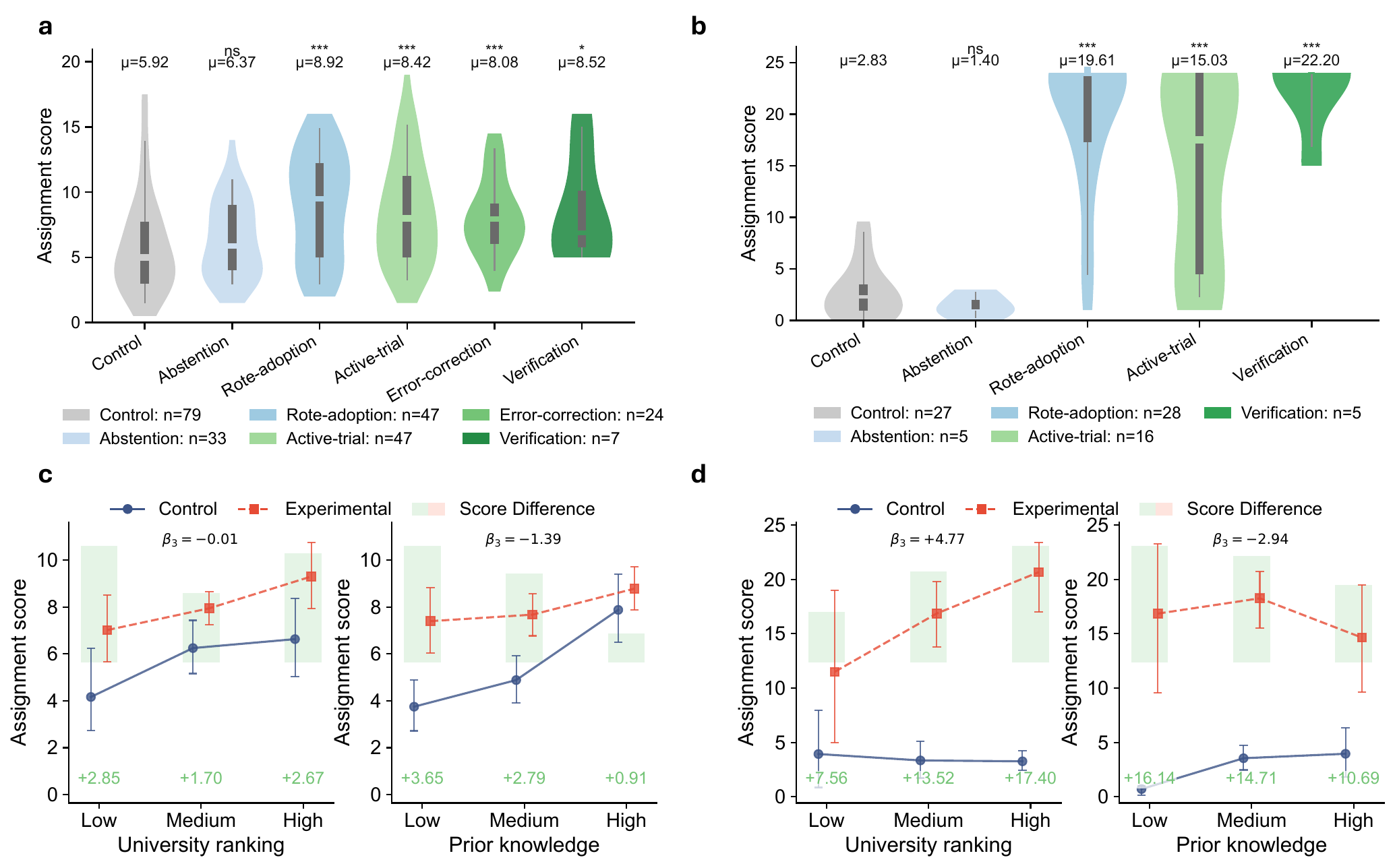}
    \caption{\textbf{AI assistance is associated with higher assignment scores across learning behavior patterns and learner profiles.} \textbf{a} and \textbf{b}, Assignment score distributions by learning behavior pattern and randomized group for game theory (\textbf{a}) and Python (\textbf{b}). Significance annotations are based on two-sided Wilcoxon rank-sum tests comparing each learning behavior pattern with the control group. $\mathrm{ns}$ denotes $P \ge 0.05$, $^{*}P<0.05$, and $^{***}P<0.001$. $n$ and $\mu$ denote group size and mean score. \textbf{c} and \textbf{d}, Assignment scores across university ranking and prior knowledge strata for game theory (\textbf{c}) and Python (\textbf{d}). Points show mean assignment scores with error bars, and shaded bars show differences computed as experimental minus control. $\beta_{3}$ denotes the experimental-group-by-learner-profile interaction coefficient from an OLS regression.}
    \label{fig:assignment}
\end{figure*}

\section{Additional experimental results}
\label{sec:additional-exp-results}

\subsection{Impact on assignment scores}
\label{sec:a1-score-analysis}

This subsection reports additional analyses of how learning behavior patterns and learner profiles are associated with assignment scores in the experimental and control groups.

\textbf{Learning behavior patterns are associated with substantial differences in assignment score.}
In both courses, experimental participants showing abstention have assignment scores that do not differ significantly from those of control participants, indicating that GPT access alone is not associated with higher assignment score when the assistant is not used.
By contrast, participants showing rote-adoption, active-trial, verification, or error-correction generally achieve higher assignment scores than the control group, as shown in Figure~\ref{fig:assignment}\textbf{a,b}.
These differences are particularly pronounced in Python, where assistant-using participants show a clear separation from both abstention and the control group.
Together, these results suggest that assistant use can facilitate immediate assignment completion across distinct learning behavior patterns, including patterns focused primarily on answer acquisition.

\textbf{Assignment score differences across learner profiles are not consistent across courses.}
Experimental participants have higher mean assignment scores than control participants in every learner profile stratum, as shown in \autoref{fig:assignment}\textbf{c,d}.
Unweighted OLS interaction tests provide limited evidence of systematic heterogeneity.
In game theory, the experimental-control difference decreases with prior knowledge ($\beta_3=-1.39$, $P=0.032$), whereas the interaction with university ranking is not statistically significant ($\beta_3=-0.01$, $P=0.993$).
In Python, the experimental-control difference tends to increase with university ranking, but the interaction is not statistically significant ($\beta_3=+4.77$, $P=0.087$).
The interaction with prior knowledge is also not statistically significant ($\beta_3=-2.94$, $P=0.319$).
Thus, the assignment score advantage of the experimental group is present across learner profile strata, but its variation with learner profiles is not consistent across courses or background measures.

\subsection{Post-task participant evaluations}
\label{sec:user-study-eval}

This subsection summarizes participants' evaluations of the AI assistant and the stage-based study design, including course learning, assignment, review, and exam phases.

\begin{figure*}[ht]
    \centering
    \includegraphics[width=\textwidth]{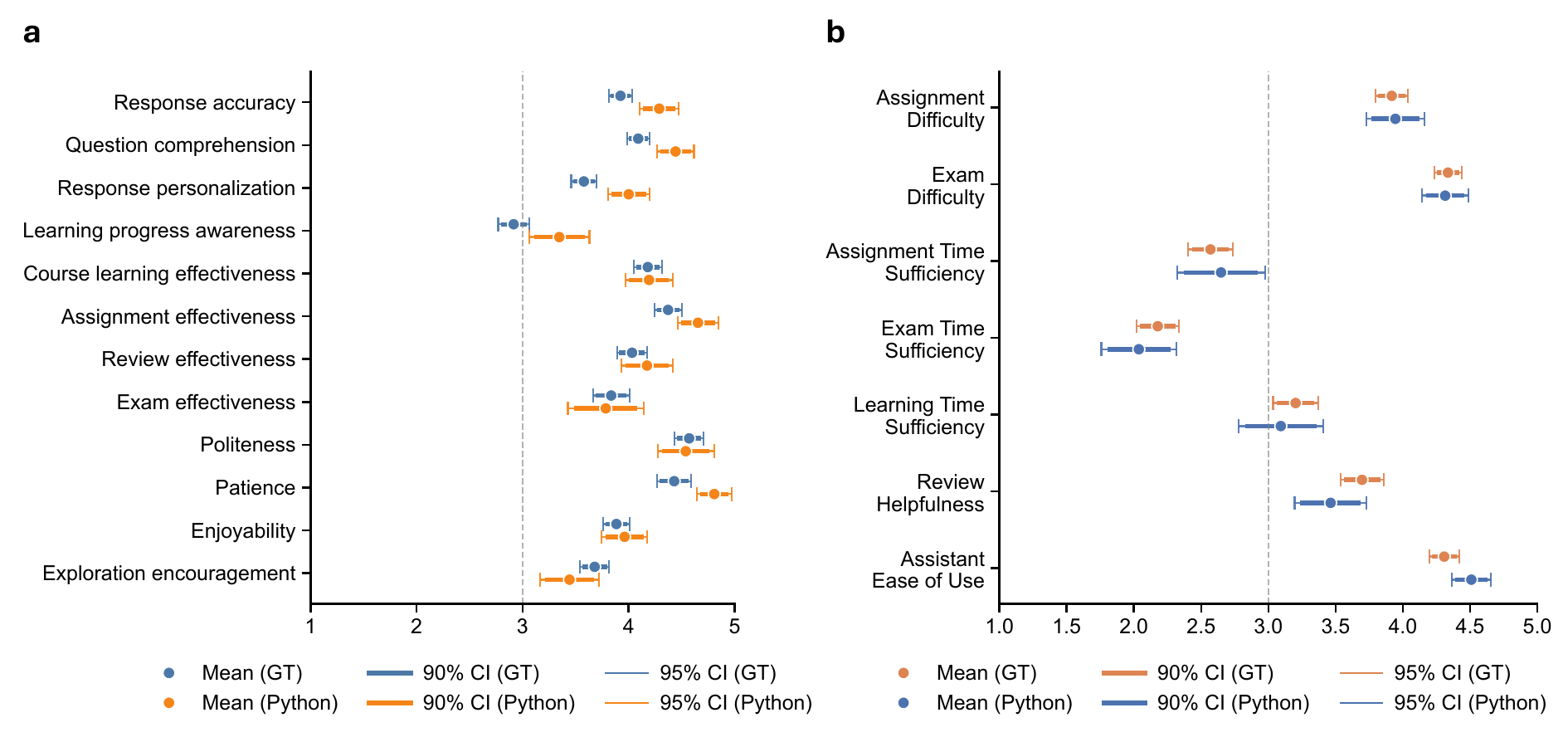}
    \caption{\textbf{Post-task evaluations cover participants' experiences with the AI assistant and the stage-based study design.} \textbf{a}, Experimental participants' ratings of the AI assistant across response quality, personalization, learning effectiveness, interaction quality, enjoyability, and exploration encouragement. \textbf{b}, Participants' ratings of assignment and exam difficulty, time sufficiency during the assignment, exam, and learning phases, review helpfulness, and assistant ease of use. Points show mean ratings for game theory and Python. Thick and thin error bars indicate 90\% and 95\% confidence intervals. The dashed line marks the scale midpoint.}
    \label{fig:user-study-eval}
\end{figure*}

\textbf{Participants distinguish immediate task support from personalized learning support.}
Experimental participants generally rate the assistant positively, as shown in \autoref{fig:user-study-eval}\textbf{a}.
Politeness, patience, and assignment effectiveness receive high ratings, while learning progress awareness, response personalization, and exploration encouragement receive comparatively lower ratings.
Exam effectiveness is also rated lower than assignment effectiveness.
This pattern suggests that participants perceive the assistant as effective for responsive task-level support, but less effective at tracking individual learning trajectories or supporting later exam preparation.
Python participants rate response accuracy and question comprehension more highly than game theory participants, consistent with the higher assistant accuracy observed for Python assignment problems.

\textbf{Participants report difficult tasks and limited time for the assignment and exam phases.}
Perceived assignment and exam difficulty are high, while time sufficiency for the assignment and exam phases is consistently below the scale midpoint, indicating that participants felt time-constrained during the study, as shown in \autoref{fig:user-study-eval}\textbf{b}.
These ratings reflect the same design choice.
We included questions spanning simple to challenging levels and enough items to avoid ceiling effects, which helped preserve score variation across participants but also tightened the available time.
Ratings for learning time sufficiency are closer to neutral, and review helpfulness is positive but more moderate.
Assistant ease of use receives the strongest ratings overall.

\begin{figure*}[ht]
    \centering
    \includegraphics[width=\textwidth]{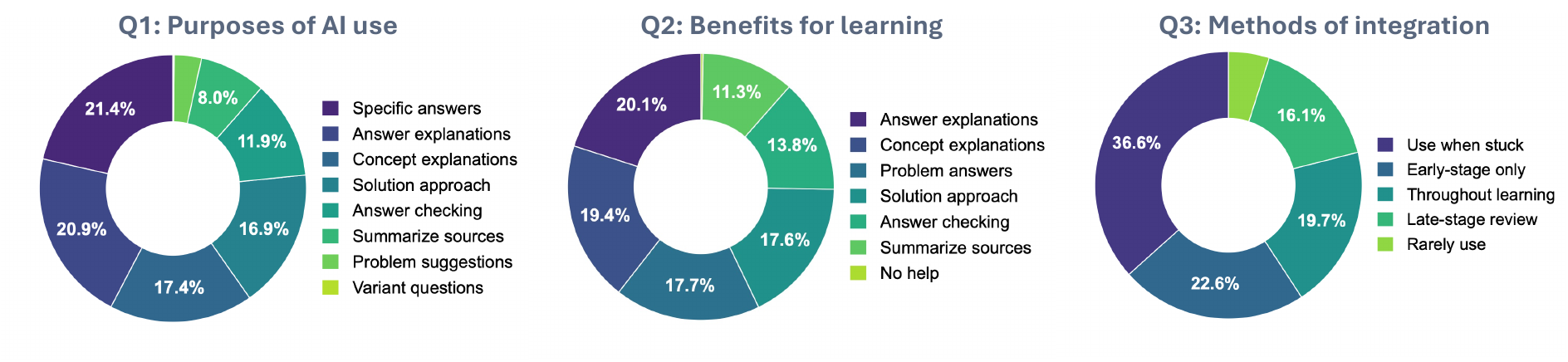}
    \caption{\textbf{Self-reported purposes, perceived learning benefits, and methods of integrating AI assistance.} \textbf{Q1}, Purposes for which experimental participants used AI assistance. \textbf{Q2}, Aspects of AI assistance that participants perceived as beneficial for learning. \textbf{Q3}, Methods through which participants integrated AI assistance into learning. Percentages represent shares of all selected responses within each multiple-response question rather than shares of participants.}
    \label{fig:self-report-pattern}
\end{figure*}

\textbf{Self-reported purposes and methods of integrating AI assistance.}
Participants report using AI assistance for both direct task support and broader learning support, as shown in \autoref{fig:self-report-pattern}.
Requests for specific answers and answer explanations account for substantial shares of the purposes reported in \textbf{Q1}, while concept explanations, solution approaches, and answer checking are also common.
In \textbf{Q2}, participants frequently identify answer explanations, concept explanations, problem answers, and solution approaches as benefits for learning.
For integration methods in \textbf{Q3}, using AI assistance when stuck is the most frequently selected response.
Other participants report using it primarily at an early stage or throughout the learning process.
These aggregate self-reports show that AI assistance serves multiple roles, ranging from targeted support after difficulty arises to more continuous involvement in learning.

\subsection{Feature effectiveness}
\label{sec:feature-effectiveness}

\begin{figure*}[t]
    \centering
    \includegraphics[width=\textwidth]{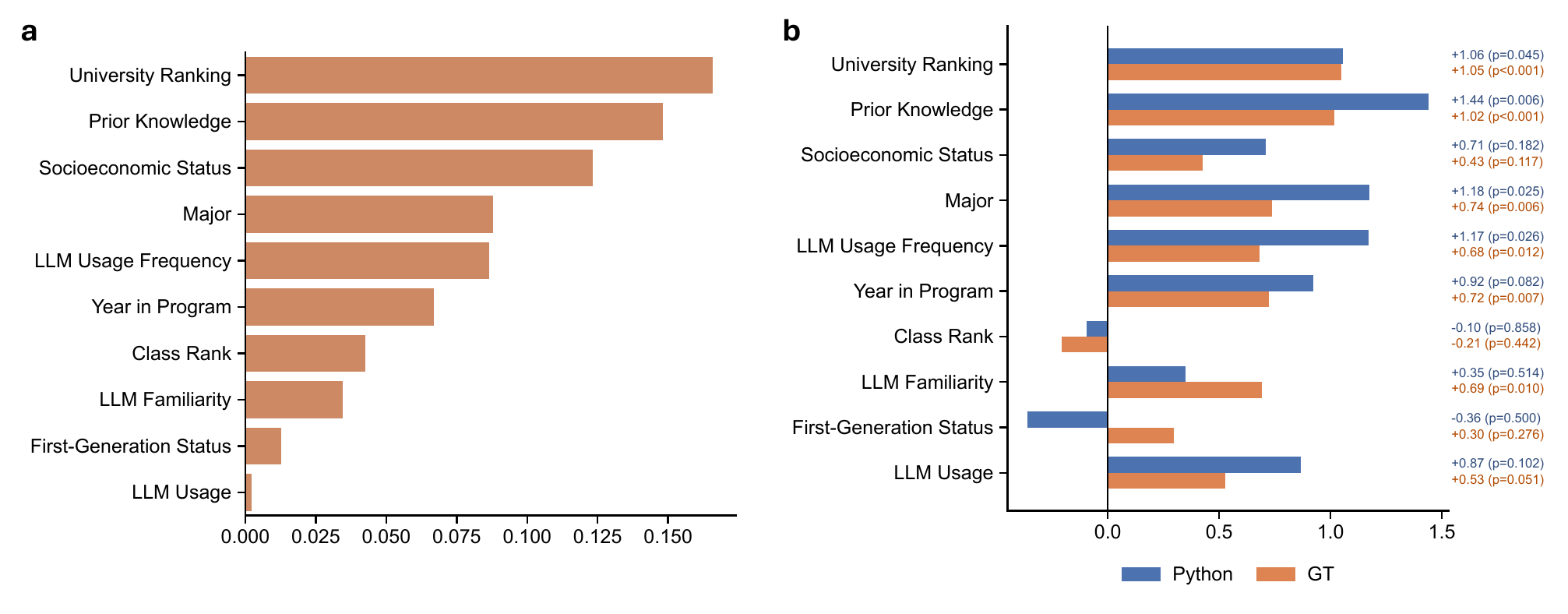}
    \caption{\textbf{University ranking and prior knowledge are the most consistent learner-profile predictors of exam score across courses.} \textbf{a}, SHAP feature importance from a random-forest model predicting exam score from pre-task learner-profile and prior LLM-experience features in a combined-course analysis. Exam score is z-scored within course, and higher values indicate greater predictive contribution. \textbf{b}, Associations between the same features and exam score estimated using course-specific univariate linear regressions with standardized predictors. Labels show the standardized coefficient and two-sided $P$ value for Python and game theory.}
    \label{fig:feature-effectiveness}
\end{figure*}

This subsection explains why the Main Paper focuses on university ranking and prior knowledge.
We began with a broader set of pre-task learner-profile and prior LLM-experience features.
We then selected features that are highly important for predicting exam score, significantly associated with exam score, and consistent across both Python and game theory.
\autoref{fig:feature-effectiveness} summarizes this screening and shows that university ranking and prior knowledge are the most robust features under these criteria.

Accordingly, the Main Paper uses university ranking and prior knowledge as the primary learner profiles.
Both features are highly important, statistically significant, and consistent across Python and game theory, as shown in \autoref{fig:feature-effectiveness}\textbf{a,b}.
Other candidate features are weaker, less statistically robust, or less consistent across courses, so they are not emphasized in the Main Paper.

\subsection{Direct adoption and assignment LLM accuracy}
\label{sec:direct-adoption-accuracy}

This subsection provides additional diagnostics for interpreting learning behavior during the assignment phase.
Direct adoption captures answer seeking in which a participant requests or uses answer-like AI output before making a meaningful independent attempt.
This measure complements the Main Paper analysis of rote-adoption and limited engagement.
We analyze direct adoption at participant and problem levels and then examine the accuracy of assistant responses during assignment work across difficulty tiers.

\begin{figure*}[t]
    \centering
    \includegraphics[width=\textwidth]{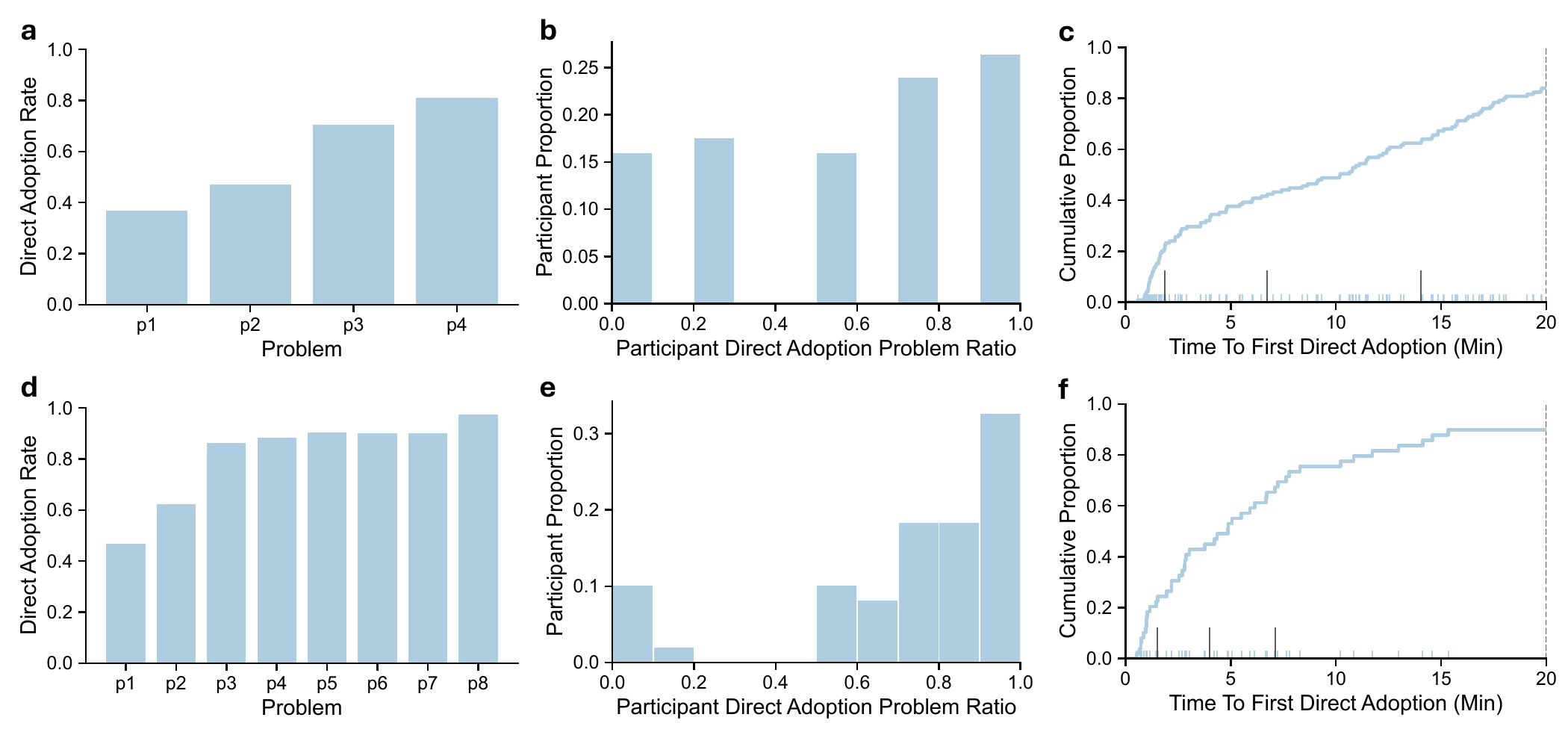}
    \caption{\textbf{Direct adoption increases across assignment problems and is concentrated among high-adoption participants.} \textbf{a}, game theory problem level direct adoption rate. \textbf{b}, game theory participant level distribution of direct adoption problem ratio. \textbf{c}, game theory cumulative distribution of time to first direct adoption. \textbf{d}, Python problem level direct adoption rate. \textbf{e}, Python participant level distribution of direct adoption problem ratio. \textbf{f}, Python cumulative distribution of time to first direct adoption.}
    \label{fig:direct-adoption}
\end{figure*}

\autoref{fig:direct-adoption} shows that direct adoption is not uniformly distributed across participants.
In game theory, problem level direct adoption rises from roughly one-third of participants on the first problem to more than 80\% on the final problem, as shown in \autoref{fig:direct-adoption}\textbf{a}.
In Python, the same pattern is sharper.
Direct adoption is already moderate on the first two problems and exceeds 80\% for most later problems, as shown in \autoref{fig:direct-adoption}\textbf{d}.
The participant level distributions show that direct adoption is concentrated among students with high direct adoption problem ratios, although both courses also include students who rarely adopt answer-like outputs directly, as shown in \autoref{fig:direct-adoption}\textbf{b,e}.
The timing distributions further show that direct adoption often appears early in the assignment phase, with many first direct adoption events occurring within the first several minutes, while some participants first turn to this behavior later in the session, as shown in \autoref{fig:direct-adoption}\textbf{c,f}.
These patterns support the Main Paper's distinction between limited engagement and proactive and critical engagement.

\begin{figure*}[t]
    \centering
    \includegraphics[width=\textwidth]{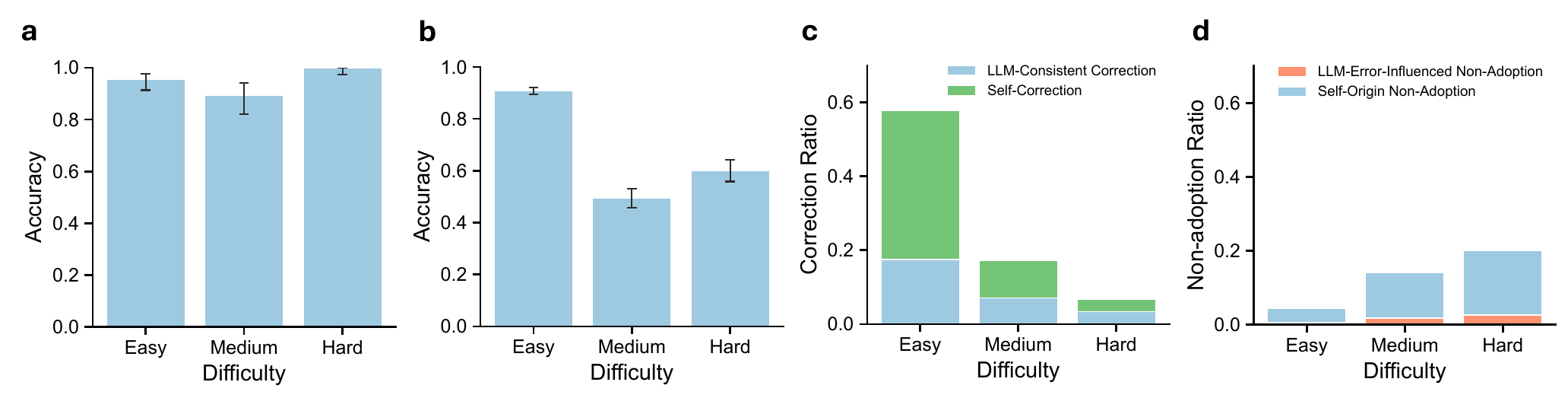}
    \caption{\textbf{Assignment LLM accuracy is high in Python but lower and more difficulty-sensitive in game theory.} \textbf{a}, Python LLM answer accuracy during assignment work by difficulty tier. \textbf{b}, game theory LLM answer accuracy during assignment work by difficulty tier. Error bars show Wilson 95\% confidence intervals. \textbf{c}, game theory correction ratio after incorrect LLM answers, partitioned into LLM-consistent correction and self-correction. \textbf{d}, game theory non-adoption ratio after correct LLM answers, partitioned into LLM-error-influenced non-adoption and self-origin non-adoption.}
    \label{fig:in-task-llm-accuracy}
\end{figure*}

Assignment LLM accuracy is near ceiling in Python but more variable in game theory, as shown in \autoref{fig:in-task-llm-accuracy}\textbf{a,b}.
For Python, accuracy remains high across all difficulty tiers, with only a modest dip on medium items and near-perfect performance on hard items, as shown in \autoref{fig:in-task-llm-accuracy}\textbf{a}.
For game theory, easy items also show high accuracy, but accuracy drops sharply on medium items and only partially recovers on hard items, as shown in \autoref{fig:in-task-llm-accuracy}\textbf{b}.
This lower reliability increases the need for students to verify and correct model outputs.
When the game theory assistant gives an incorrect answer, correction is most common on easy items and becomes much less common on medium and hard items, as shown in \autoref{fig:in-task-llm-accuracy}\textbf{c}.
Most correction after incorrect answers is aligned with a later correct LLM response on easy items, while self-correction accounts for a smaller share and nearly disappears for hard items.
Conversely, non-adoption after correct game theory answers is lowest on easy items and higher on medium and hard items, as shown in \autoref{fig:in-task-llm-accuracy}\textbf{d}.
Most of this non-adoption is self-origin rather than directly attributable to exposure to a conflicting wrong LLM answer.
These diagnostics are consistent with the Main Paper conclusion that proactive and critical engagement is associated with higher exam score, while limited engagement can show limited benefit when model reliability is lower.

\subsection{Accuracy of different LLMs on assignment questions}
\label{sec:llm-accuracy}

We additionally evaluate the accuracy of different LLMs on the assignment questions used in the study.
For each model in \autoref{tab:llm-models} and each item, we sample 10 independent responses
at temperature 0.7 using a structured output format.
For both Python and game theory, we evaluate assignment items only.
Python responses are code-only and automatically judged with test cases.
Game theory responses fill the required blanks and are scored by human annotators using the official rubric
under the same participant-facing constraints.
We summarize accuracy by easy, medium, and hard difficulty tiers as in the Main Paper.

\begin{table}[ht]
    \centering
    \caption{\textbf{Tested LLMs and release details.} LLM Arena scores are from the leaderboard as of February 27, 2026.}
    \label{tab:llm-models}
    \footnotesize
    \setlength{\tabcolsep}{6pt}
    \renewcommand{\arraystretch}{1.12}
    \begin{tabular}{lccc}
        \toprule
        \textbf{Model Name} & \textbf{Vendor} & \textbf{Release Date} & \textbf{LLM Arena Score} \\
        \midrule
        GPT-4o              & OpenAI          & May 13, 2024          & 1346                     \\
        DeepSeek-R1         & DeepSeek        & January 20, 2025      & 1398                     \\
        Claude 3.7 Sonnet   & Anthropic       & February 24, 2025     & 1372                     \\
        Llama 4 Maverick    & Meta            & April 5, 2025         & 1328                     \\
        GPT-4.1             & OpenAI          & April 14, 2025        & 1413                     \\
        Gemini 2.5 Flash    & Google          & June 17, 2025         & 1411                     \\
        GPT-5               & OpenAI          & August 7, 2025        & 1426                     \\
        Gemini 3 Pro        & Google          & November 18, 2025     & 1486                     \\
        Gemini 3 Flash      & Google          & December 17, 2025     & 1473                     \\
        Claude Sonnet 4.6   & Anthropic       & February 17, 2026     & 1458                     \\
        \bottomrule
    \end{tabular}
\end{table}

\begin{figure}[ht]
    \centering
    \includegraphics[width=\linewidth]{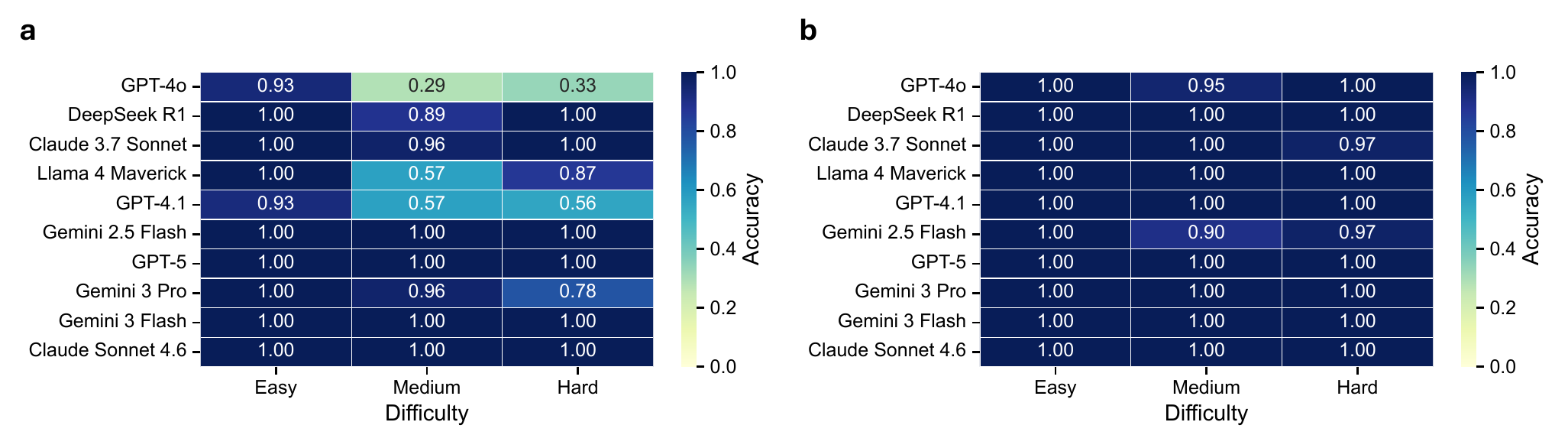}
    \caption{\textbf{LLM accuracy varies by difficulty and course.} \textbf{a}, game theory accuracy by difficulty tier across all evaluated models. \textbf{b}, Python accuracy by difficulty tier across the same model set, showing high performance with small drops on harder items for some models.}
    \label{fig:llm-accuracy}
\end{figure}

\autoref{fig:llm-accuracy} reveals distinct patterns across courses.
For Python, accuracy is near ceiling across difficulty tiers, as shown in \autoref{fig:llm-accuracy}\textbf{b}.
This suggests that LLMs can generally solve these assignment questions, consistent with students'
reported experience in the Main Paper.
For game theory, easy items also approach ceiling, whereas
medium and hard items show lower accuracy for earlier models and a clear upward trend as model capability
increases, as shown in \autoref{fig:llm-accuracy}\textbf{a}.
This improvement is consistent with the Main Paper discussion that confidently incorrect or hallucinated
answers can mislead students.
As models become stronger, such failure modes are mitigated, reducing but not eliminating the risk of misleading guidance on more challenging items.

\subsection{Case study}
\label{sec:case-study}

This subsection presents representative cases that illustrate the qualitative observations above.
For ease of understanding, we translate participants' original Chinese chat messages and inputs into English
for presentation.

\begin{figure}[ht]
    \centering
    \includegraphics[width=\linewidth]{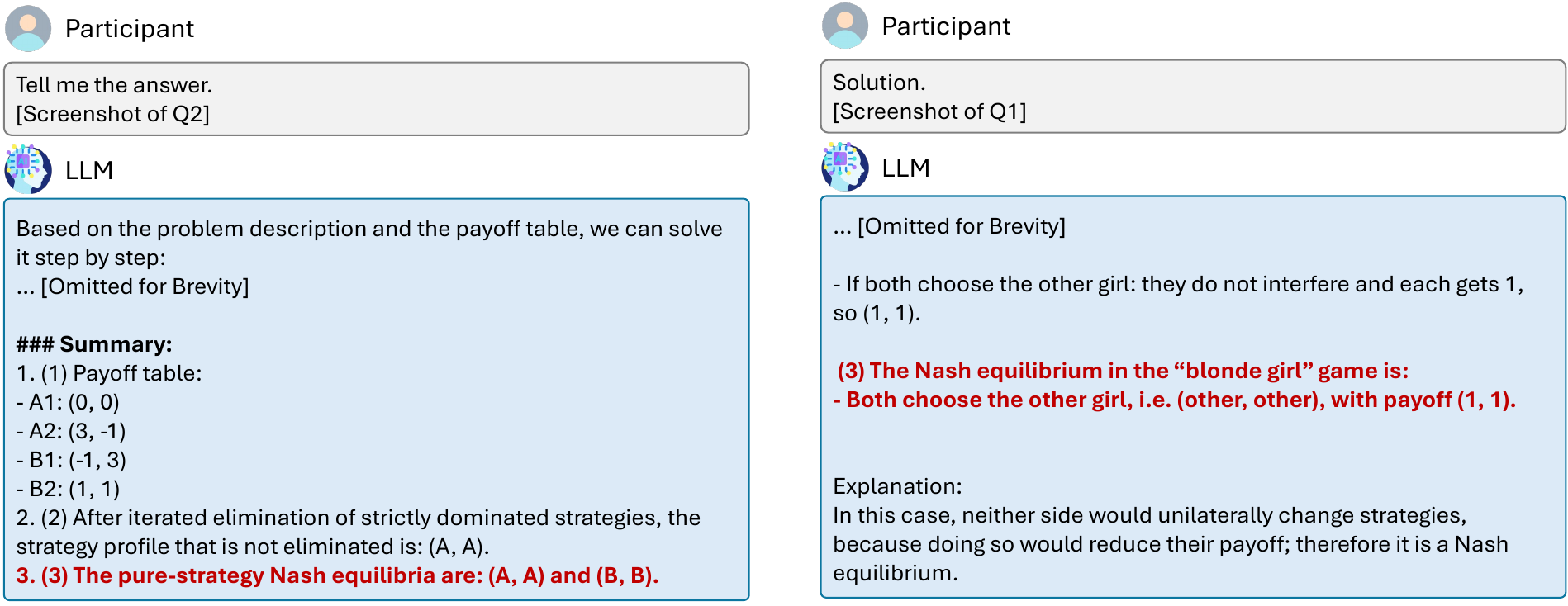}
    \caption{\textbf{Game theory examples show plausible but incorrect LLM responses.} Two representative LLM responses contain incorrect final answers driven by hallucination or reasoning errors. Gray boxes denote participant chat turns, and blue boxes denote LLM chat turns.}
    \label{fig:gt-incorrect}
\end{figure}

\begin{figure}[ht]
    \centering
    \includegraphics[width=\linewidth]{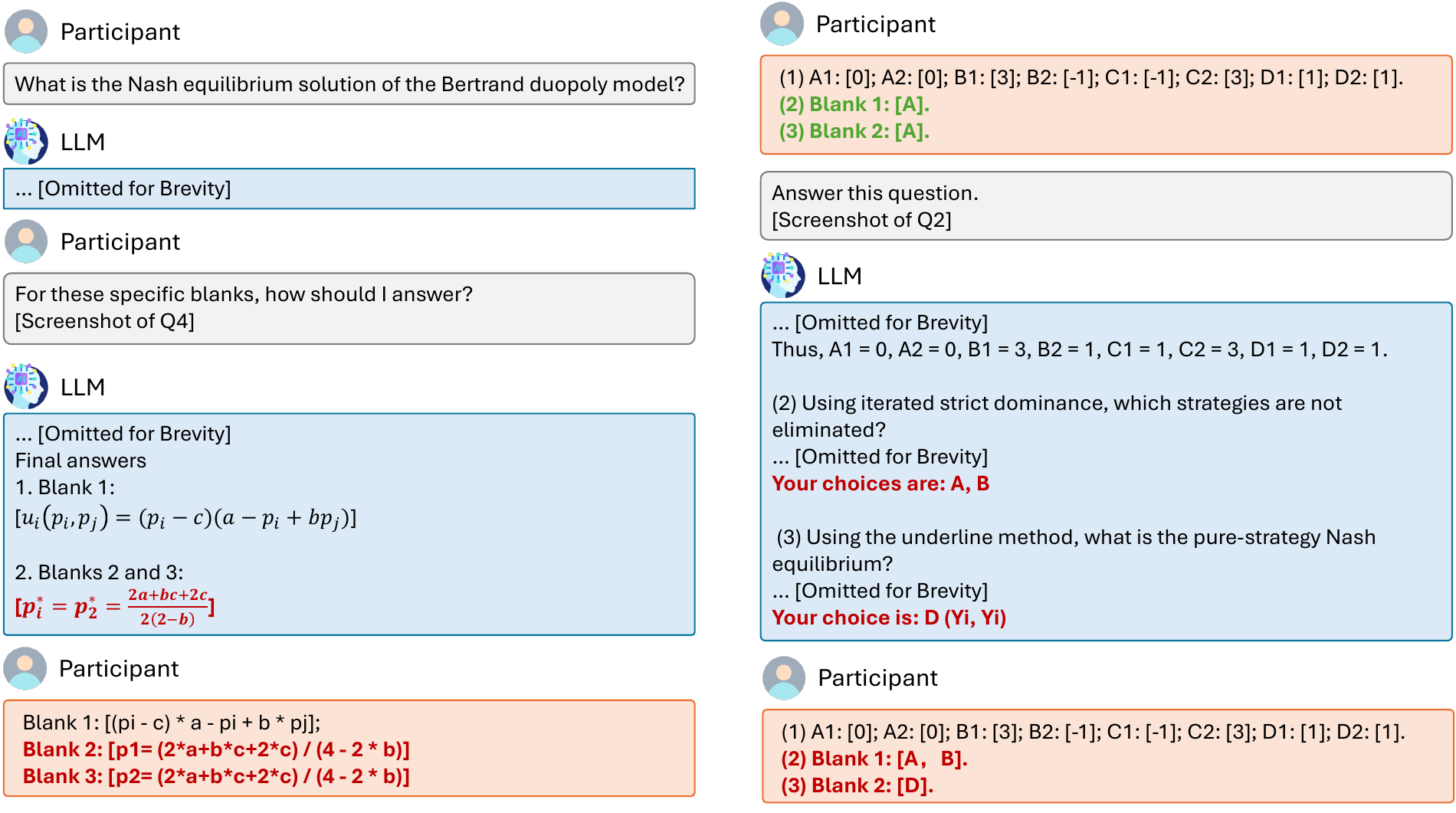}
    \caption{\textbf{Incorrect LLM guidance can change participants' final answers in game theory.} \textbf{a}, A participant adopts an incorrect LLM answer. \textbf{b}, A participant initially gives the correct answer but changes it to an incorrect one after receiving an incorrect LLM suggestion. Gray boxes denote participant chat turns, blue boxes denote LLM chat turns, and orange boxes denote participant answer-editing actions.}
    \label{fig:gt-incorrect-effect}
\end{figure}

\begin{figure}[ht]
    \centering
    \includegraphics[width=\linewidth]{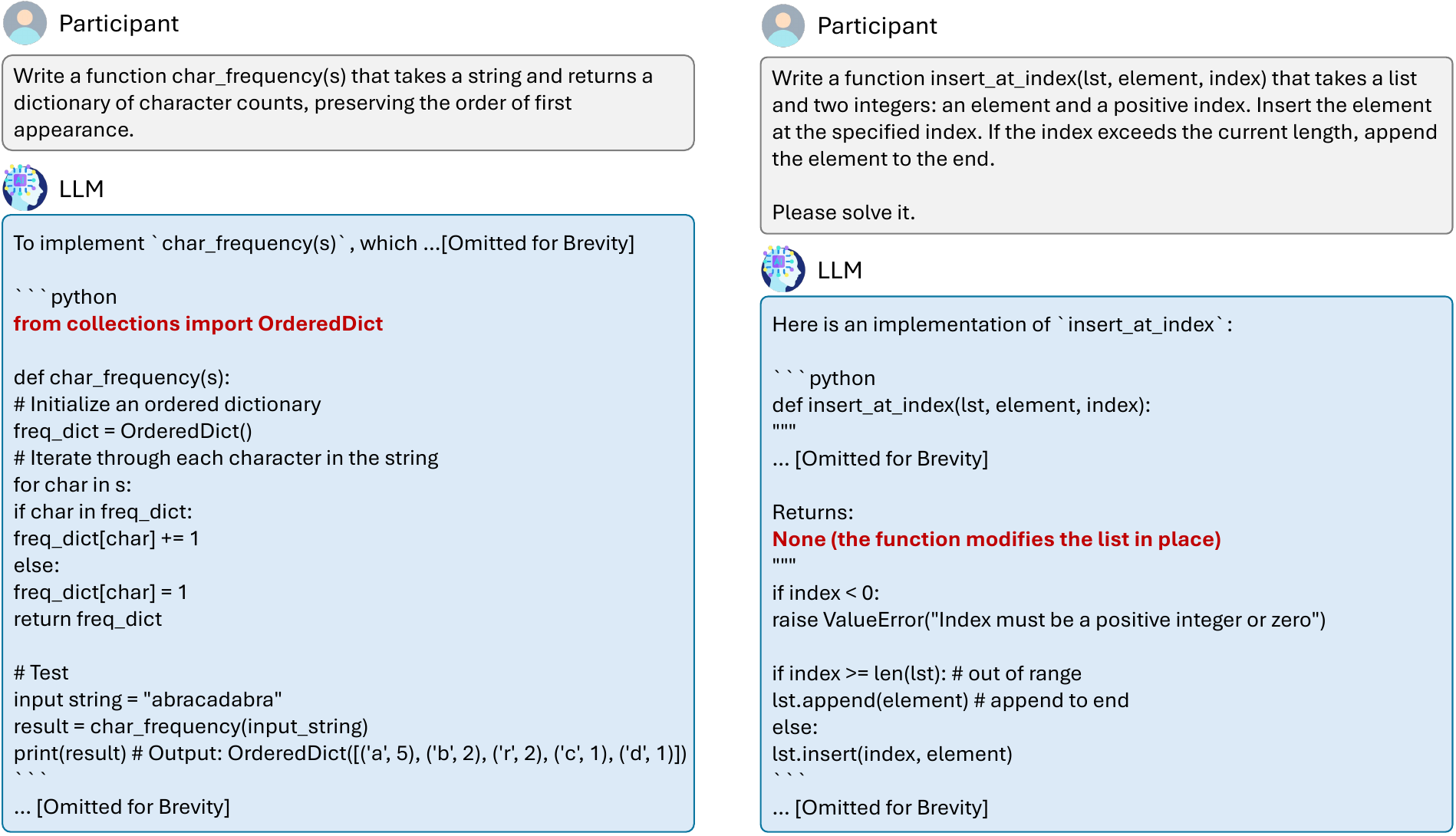}
    \caption{\textbf{Python failures often reflect specification mismatches rather than incorrect core logic.} Two representative LLM responses fail the platform judge because of constraint violations, including disallowed imports and in-place list mutation without the required return value. Gray boxes denote participant chat turns, and blue boxes denote LLM chat turns.}
    \label{fig:python-incorrect}
\end{figure}

\autoref{fig:gt-incorrect} and \autoref{fig:python-incorrect} show representative incorrect LLM
responses observed in our offline evaluation.
For Python, near-ceiling accuracy is consistent with the observation
that most model outputs are correct; the remaining failures often reflect judge-facing specification
mismatches rather than incorrect core logic, as shown in \autoref{fig:llm-accuracy}\textbf{b} and \autoref{fig:python-incorrect}.
By contrast, lower game theory accuracy is consistent with a
hallucination-driven failure mode, in which models produce plausible-looking but incorrect reasoning and
final answers, as shown in \autoref{fig:llm-accuracy}\textbf{a} and \autoref{fig:gt-incorrect}.
\autoref{fig:gt-incorrect-effect} further shows how such incorrect guidance can affect participants'
final answers in game theory.

\begin{figure*}[!t]
    \centering
    \includegraphics[width=\textwidth]{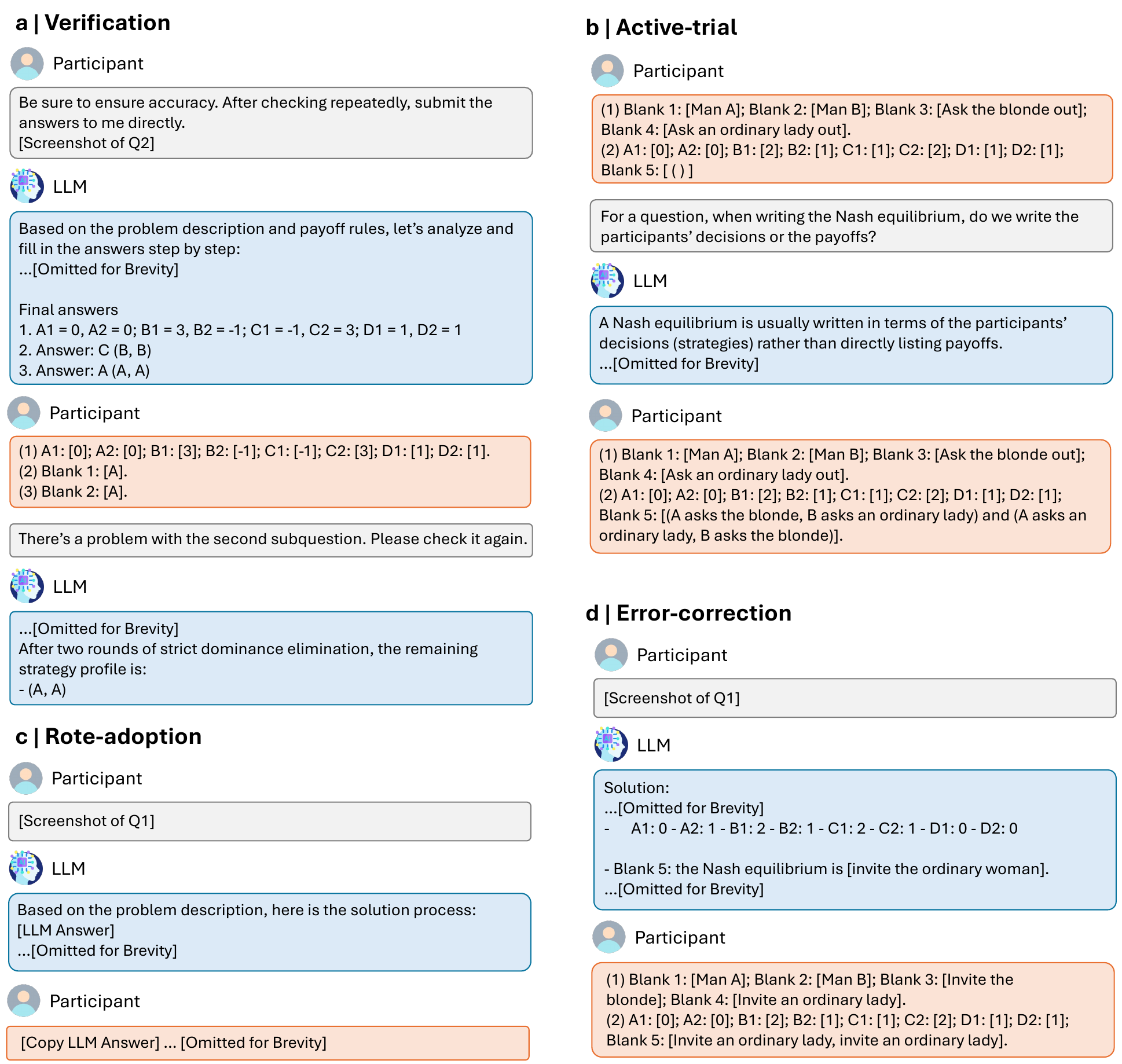}
    \caption{\textbf{Game theory examples illustrate distinct learning behavior patterns.} Representative game theory problem-solving examples show different learning behavior patterns. Gray boxes denote participant chat turns, blue boxes denote LLM chat turns, and orange boxes denote participant answer-editing actions.}
    \label{fig:gt-behavior-cases}
\end{figure*}

\begin{figure*}[!t]
    \centering
    \includegraphics[width=\textwidth]{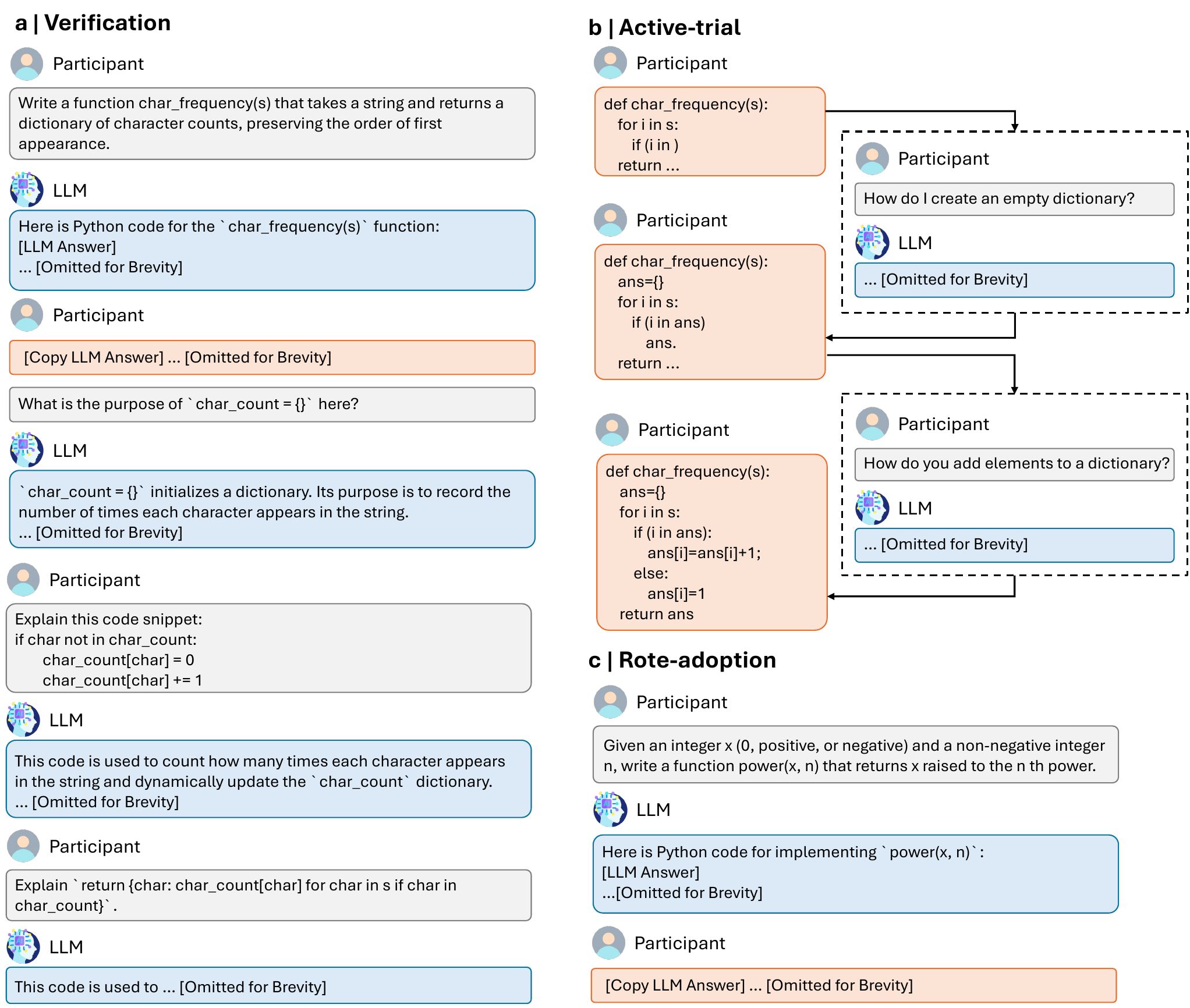}
    \caption{\textbf{Python examples illustrate distinct learning behavior patterns.} Representative Python problem-solving examples show different learning behavior patterns. Gray boxes denote participant chat turns, blue boxes denote LLM chat turns, and orange boxes denote participant answer-editing actions.}
    \label{fig:python-behavior-cases}
\end{figure*}

\autoref{fig:gt-behavior-cases} and \autoref{fig:python-behavior-cases} provide additional qualitative
context for the Main Paper results on heterogeneity across learners.
They show representative problem-solving artifacts illustrating different learning behavior patterns in game theory and Python.

\clearpage


\section{Additional experimental material}
\label{sec:exp-info}

\subsection{Consent forms}

In our study, candidates were required to submit their resumes and complete a simple screening test during the recruitment phase. Those who passed the screening were then invited to participate in the main study. Before both the recruitment phase and the formal experiment, participants were shown and required to sign separate informed consent forms before proceeding with subsequent procedures. This subsection presents the details of these two consent forms. For privacy protection purposes, all personal identifiable information (including names and email addresses) in this consent form has been anonymized. The original consent form contained actual contact information of the research team members and was provided to all study participants.

\begin{nolinenumbers}
    \begin{tcolorbox}[
            breakable,        
            enhanced,         
            colback=white,
            colframe=black!50,
            boxrule=0.5pt,
            arc=0pt,
            title={\textbf{Microsoft Research – Participant Consent Form (Recruit)}},
            fonttitle=\large,
            before skip=12pt, 
            after skip=12pt,  
            pad at break*=3mm 
        ]

        \textbf{Project Title:} LLMs in Education

        \textbf{Principal Investigator:} [Senior Researcher] (Microsoft Research), [Professor] (Princeton University)

        \textbf{Invitation:} Thank you for taking the time to consider volunteering in the screening phase of a research project conducted by Microsoft Research in collaboration with Princeton University. The purpose of this research is to provide insights into leveraging AI for creating a more interactive and tailored learning environment.

        \textbf{Procedures:} You participation is voluntary. If you choose to participate, you will need to first submit your resume to [Data Collection Partner]. If your resume meets eligibility criteria, you will be invited to complete a 10-minute math or coding test for screening purposes.

        If your test results meet the participation requirements, we will invite you to take part in our user study.

        You will have another opportunity to agree or decline to participate in the user study. At that time, you will be able to choose to participate either through an online meeting or by visiting the [Data Collection Partner] lab in person. In our user study, you will spend 2 hours participating in a course study (there may be LLM assistance during the learning process), and provide some personal information and usage feedback.

        \textbf{NOTE:} Not all participants will receive an invitation for math or code test, nor will all participants receive an invitation to participate in the user study.

        During the screening phase, we plan to collect the following information about you:
        \begin{itemize}
            \item Resume
            \item Answers to the code or math screening test
        \end{itemize}

        \textbf{Compensation:} The screening process, which includes submitting a resume and completing a screening math or coding test, does not offer any compensation.

        If you are invited to participate in the user study, you will receive a basic compensation of 150 RMB. Additional incentives will be provided based on the participants' performance at learning the course material. The top 10\% will receive an additional incentive of 200 RMB. Those ranked in the 10\% to 40\% range will receive an incentive of 100 RMB. The average total compensation is anticipated to be around 210 RMB.

        \textbf{Benefits:} There will be no direct benefit to you as a result of participating in this study.

        We hope that findings from this research will help better understand whether and how large models can aid in student learning, and also reduce inequalities in education.

        \textbf{Risks:} The risks of participating are similar to what you might experience while performing everyday tasks. Risks include fatigue or test-taking anxiety. You are free to skip any questions you prefer not to answer.

        \textbf{Privacy \& Confidentiality:} Researchers will keep your participation and the information you share as confidential as possible.

        For participants who decline future contact from [Data Collection Partner] for participation in other studies, the resumes and math/coding test responses from the screening stage will be deleted within 30 days from submission (if you are not invited to continue in the user study) or within 30 days after the completion of compensation payment (if you go on to complete the user study).

        If you would like to withdraw your resume or test submission before 30 days, you may contact the research team at [MSR-Contact@microsoft.com] and we will remove your study information from our database within 3 work days.

        Data generated during the user study will be transferred abroad.

        For questions about how Microsoft manages your privacy, please see the Microsoft Privacy Statement (\url{http://go.microsoft.com/fwlink/?LinkId=521839}).

        \textbf{Conflict of Interest Disclosure:} This research project receives sole funding from Microsoft Research. [University Researchers], employees of Princeton University, do not have any financial interests or affiliations with Microsoft Research.

        \textbf{Participation is your choice:} Whether or not you participate is entirely up to you. You can decide to participate now and stop participating later. Your decision of whether or not to participate will have no impact on any other services or agreements you have with Microsoft outside of this research.

        \textbf{Questions or Concerns:} If you have any questions or concerns about this study at any time, you may contact the research team at [MSR-Contact@microsoft.com] and [University-Contact@princeton.edu]. If you have any questions about your rights as a research participant, please contact the Microsoft Research Ethics Review Program at [MSR-Ethics@microsoft.com] and the Princeton Institutional Review Board (IRB) at [IRB@princeton.edu].

        \subsection*{Consent}
        Would you like to participate in the screening phase of this study as described above?
        \begin{itemize}
            \item[$\square$] Yes, I would like to participate.
            \item[$\square$] No, thanks.
        \end{itemize}

        Can [Data Collection Partner] contact you in the future to invite you to participate in other studies? If you answer yes, [Data Collection Partner] will retain your resume and Prior Knowledge assessment results indefinitely for recruitment purposes.
        \begin{itemize}
            \item[$\square$] Yes
            \item[$\square$] No
        \end{itemize}

        \smallskip
        \emph{If you would like to keep a copy of this consent form, please print or save one now.}

    \end{tcolorbox}
\end{nolinenumbers}

\begin{nolinenumbers}
    \begin{tcolorbox}[
            breakable,
            enhanced,
            colback=white,
            colframe=black!50,
            boxrule=0.5pt,
            arc=0pt,
            title={\textbf{Microsoft Research – Participant Consent Form (Main Study)}},
            fonttitle=\large,
            before skip=12pt,
            after skip=12pt,
            pad at break*=3mm
        ]

        \textbf{Project Title:} LLMs in Education

        \textbf{Principal Investigator:} [Senior Researcher] (Microsoft Research), [Professor] (Princeton University)

        \textbf{Invitation:} Thank you for taking the time to consider volunteering in this research project conducted by Microsoft Research in collaboration with Princeton University. The purpose of this research is to provide insights into leveraging AI for creating a more interactive and tailored learning environment.

        \textbf{Procedures:} You participation is voluntary. If you choose to participate, the study will take around 2 hours and can be completed either online or in person. You will complete a pre-course survey online, which includes a test of your coding or math skill, then proceed to participate in an online course on either Python code or game theory (researchers will assign courses). Depending on the arrangement, you may or may not have access to an AI digital learning assistant during the course. After the course, you will complete a post-course survey online. In total, your participation will take around 2 hours.

        \textbf{NOTE:} Participants are not allowed to exit once they enter each stage of the course; they must complete it within the allotted time and return to the progress page. However, participants can stay on the progress page for up to 20 minutes before proceeding to the next stage. (These 20 minutes are not included in the total experiment duration.)

        During the study, we plan to collect the following information about you:
        \begin{itemize}
            \item Academic rank of university
            \item Major
            \item Grade
            \item Class ranking
            \item Income range
            \item Perceptions and usage of large language models
            \item Math and programming skills
            \item A log of what you type and click while interacting with our online assignment platform, including course assignments and review
            \item A subjective account of your experience on our platform and evaluation of its features
            \item Your perspective on applying large language models in education
        \end{itemize}

        Screen recording is required to ensure that participants are not using external tools to complete course assignments. Screen recordings do not show your face. \textbf{Participants taking part remotely will be reminded at the start of the session to close any apps showing identifiable or private information and silence any notifications (e.g. email pop-ups).}

        \textbf{Compensation:} All participants will receive a basic compensation of 150 RMB. Additional incentives will be provided based on the participants' performance at learning the course material, where the performance metric is defined as: 20\% \textit{score of assignment 1} (1 + speed of assignment 1) + 30\% \textit{score of assignment 2} (1 + speed of assignment 2). Based on the performance in the two main tasks (Assignment 1 and Assignment 2), we will conduct a ranking. The top 10\% will receive an additional incentive of 200 RMB. Those ranked in the 10\% to 40\% range will receive an incentive of 100 RMB. The average total compensation is anticipated to be around 210 RMB.

        \textbf{NOTE:} Participants are not permitted to use any outside resources to complete course assignments. If you use any outside resources (e.g., phone, internet), you will not be compensated. Any data you provided will be destroyed within 30 days.

        \textbf{Benefits:} There will be no direct benefit to you as a result of participating in this study.

        We hope that findings from this research will help better understand whether and how large models can aid in student learning, and also reduce inequalities in education.

        \textbf{Risks:} The risks of participating are similar to what you might experience while performing everyday tasks. Risks include fatigue, test-taking anxiety, or discomfort answering sensitive questions. You are free to skip any questions you prefer not to answer. After your information is deidentified, there is a potential for individuals to be reidentified based on unique characteristics in the data, but we are taking several precautions to prevent this, including the following: We will not share screen recordings outside of the study team, we will report findings in aggregate (individual responses are grouped together to show patterns), and we will label records with a code instead of your name.

        \textbf{Privacy \& Confidentiality:} Researchers will keep your participation and the information you share as confidential as possible.

        The information you share, including screen recordings, will be labeled in our records with a code instead of your name or other direct identifier. The key to this code will be stored separately and destroyed after data collection is complete.

        Screen recordings will be retained until we have verified that no outside resources were used to complete study tasks, after which they will be deleted. This will be completed by the time the final writeup of the study is concluded, or within two years, whichever is sooner.

        Researchers may share the results of this study publicly, such as in journal articles or conference presentations, but your identity will not be disclosed.

        Information collected during this study may be used for future research studies or to improve products or services at Microsoft.

        If you decide to withdraw from the study, and want researchers to remove your study information, you can contact the team at [MSR-Contact@microsoft.com], we will delete your screen recordings, responses to the questionnaire (including details such as university rankings, major, grade, class ranking, income range, perceptions and usage of large language models, answers from math and programming skills tests, a subjective account of your experience on our platform and evaluation of its features, and your perspective on applying large language models in education), and records of interactions with the platform within 3 work days. However, once we publish the research findings in journal articles or conference presentations, it will no longer be possible to remove your information from the displayed results. The research outcomes are reported as aggregate results (combining individual responses to showcase outcomes), ensuring your personal information is not disclosed.

        Data generated during the experimental processes will be transferred abroad.

        For questions about how Microsoft manages your privacy, please see the Microsoft Privacy Statement (\url{http://go.microsoft.com/fwlink/?LinkId=521839}).

        \textbf{Conflict of Interest Disclosure:} This research project receives sole funding from Microsoft Research. [University Researchers], employees of Princeton University, do not have any financial interests or affiliations with Microsoft Research.

        \textbf{Participation is your choice:} You participation is voluntary. Whether or not you participate is entirely up to you. You can decide to participate now and stop participating later. Your decision of whether or not to participate will have no impact on any other services or agreements you have with Microsoft outside of this research.

        \textbf{Questions or Concerns:} If you have any questions or concerns about this study at any time, you may contact the research team at [MSR-Contact@microsoft.com] and [University-Contact@princeton.edu]. If you have any questions about your rights as a research participant, please contact the Microsoft Research Ethics Review Program at [Ethics-Contact@microsoft.com] and the Princeton Institutional Review Board (IRB) at [IRB@princeton.edu].

        \textbf{Consent:} Would you like to participate in this study as described above?
        \begin{itemize}
            \item[$\square$] Yes, I would like to participate.
            \item[$\square$] No, thanks.
        \end{itemize}

        If you would like to keep a copy of this consent form, please print or save one now.

    \end{tcolorbox}
\end{nolinenumbers}

\subsection{Screening test questions}
During recruitment, candidates were required to submit their resumes and complete a screening test to assess course-specific eligibility. Eligible Python candidates needed basic programming experience but no prior exposure to Python programming. Eligible game theory candidates needed experience with equations and optimization problems but no previous formal education in game theory. All participants were adult university students. To preserve the validity of the screening process, candidates were not informed about the purpose of the resume submission and screening questions while completing them. The original participant-facing wording is reproduced below.
The screening questions for Python and game theory are presented in \autoref{tab:screening-questions-python} and \autoref{tab:screening-questions-game-theory}, respectively.

\begin{longtable}{|l|p{0.80\textwidth}|}
    \caption{Screening questions of Python course.} \label{tab:screening-questions-python}                                                                                                                                                    \\
    \toprule
    \textbf{Question No.} & \textbf{Question Description}                                                                                                                                                                                     \\
    \midrule
    \endfirsthead
    \multicolumn{2}{c}%
    {{\tablename\ \thetable{} -- continued from previous page}}                                                                                                                                                                               \\
    \toprule
    \textbf{Question No.} & \textbf{Question Description}                                                                                                                                                                                     \\
    \midrule
    \endhead
    \midrule \multicolumn{2}{r}{{Continued on next page}}                                                                                                                                                                                     \\
    \endfoot
    \bottomrule
    \endlastfoot
    Q1                    & \textbf{Personal Information}                                                                                                                                                                                     \\
                          & What is your name (enter in Pinyin, e.g., xiaoming)? \underline{\hspace{1cm}}                                                                                                                                     \\
    \midrule
    Q2                    & \textbf{Academic Background}                                                                                                                                                                                      \\
                          & What is your major? \underline{\hspace{1cm}}                                                                                                                                                                      \\
    \midrule
    Q3                    & \textbf{Programming Experience}                                                                                                                                                                                   \\
                          & Which programming languages do you have a basic understanding of (multiple choices allowed)?                                                                                                                      \\
                          & $\square$ C $\square$ PHP $\square$ Matlab $\square$ Rust $\square$ R $\square$ Java $\square$ C\# $\square$ JavaScript $\square$ Python $\square$ Fortran $\square$ Go $\square$ Other: \underline{\hspace{1cm}} \\
    \midrule
    Q4                    & \textbf{Coding Exercise}                                                                                                                                                                                          \\
                          & Please use all the languages you know to complete the following problem:                                                                                                                                          \\
                          & \textbf{Problem:} Write a function SUMA that takes three numbers and returns the sum of all positive integers among them.                                                                                         \\
                          & \textbf{Requirements:} Please use all the programming languages you are familiar with to solve this problem. Make sure your code can correctly handle input containing positive integers.                         \\
                          & \textbf{Example:} Input: 1, -2, 3 Output: 4 (because 1 + 3 = 4).                                                                                                                                                  \\
                          & Your solution: \underline{\hspace{1cm}}                                                                                                                                                                           \\
\end{longtable}

\begin{longtable}{|l|p{0.80\textwidth}|}
    \caption{Screening questions of Game Theory course.} \label{tab:screening-questions-game-theory}                                                   \\
    \toprule
    \textbf{Question No.} & \textbf{Problem Description}                                                                                               \\
    \midrule
    \endfirsthead
    \multicolumn{2}{c}%
    {{\tablename\ \thetable{} -- continued from previous page}}                                                                                        \\
    \toprule
    \textbf{Problem No.}  & \textbf{Problem Description}                                                                                               \\
    \midrule
    \endhead
    \midrule \multicolumn{2}{r}{{Continued on next page}}                                                                                              \\
    \endfoot
    \bottomrule
    \endlastfoot
    Q1                    & \textbf{Personal Information}                                                                                              \\
                          & What is your name (enter in Pinyin, e.g., xiaoming)? \underline{\hspace{1cm}}                                              \\
    \midrule
    Q2                    & \textbf{Academic Background}                                                                                               \\
                          & What is your major? \underline{\hspace{1cm}}                                                                               \\
    \midrule
    Q3                    & \textbf{Prerequisite Knowledge Assessment}                                                                                 \\
                          & Have you studied derivatives?                                                                                              \\
                          & $\square$ Yes \quad $\square$ No                                                                                           \\
    \midrule
    Q4                    & \textbf{Additional Background}                                                                                             \\
                          & Have you studied game theory?                                                                                              \\
                          & $\square$ Yes \quad $\square$ No                                                                                           \\
    \midrule
    Q5                    & \textbf{Mathematical Skills Assessment}                                                                                    \\
                          & Please answer the following question:                                                                                      \\
                          & Find the derivative of the function $f(x)=3x^2-2x+5+e^x$. And find the value of the derivative of the function when $x=1$. \\
                          & Please show the calculation process and give the final answer.                                                             \\
                          & Answer: \underline{\hspace{8cm}}                                                                                           \\
                          & \underline{\hspace{8cm}}                                                                                                   \\
                          & \underline{\hspace{8cm}}                                                                                                   \\
\end{longtable}

\subsection{Pre-task surveys}

Participants completed a pre-task survey at the beginning of the main study. The survey collected demographic information, student background measures, and prior experience with and attitudes toward LLMs. Participants also completed a timed prior knowledge assessment measuring prerequisite programming or mathematical knowledge. The original participant-facing wording is reproduced below. The pre-task survey is presented in \autoref{tab:pre-task-survey-questions}, and the prior knowledge assessments for Python and game theory are presented in \autoref{tab:foundation-python} and \autoref{tab:foundation-game-theory}, respectively.

\begin{longtable}{|l|p{0.80\textwidth}|}
    \caption{Pre-task survey questions for Python and Game Theory courses.}
    \label{tab:pre-task-survey-questions}
    \\
    \toprule
    \textbf{Question No.} & \textbf{Question Content}                                                                                                                                                                                                                \\
    \midrule
    \endfirsthead
    \multicolumn{2}{c}%
    {{\tablename\ \thetable{} -- continued from previous page}}                                                                                                                                                                                                      \\
    \toprule
    \textbf{Question No.} & \textbf{Question Content}                                                                                                                                                                                                                \\
    \midrule
    \endhead
    \midrule \multicolumn{2}{r}{{Continued on next page}}                                                                                                                                                                                                            \\
    \endfoot
    \bottomrule
    \endlastfoot
    Q1                    & University: According to the BCUR ranking, where is your university? (If not listed, choose Other and specify.)                                                                                                                          \\
                          & \url{https://www.shanghairanking.cn/rankings/bcur/2024}                                                                                                                                                                                  \\
                          & $\bigcirc$ Top 10 \quad $\bigcirc$ Top 10-50 \quad $\bigcirc$ Top 50-100 \quad $\bigcirc$ Top 100-500 \quad $\bigcirc$ Outside top 500 \quad $\bigcirc$ Not sure / prefer not to say \quad $\bigcirc$ Other                              \\
    \midrule
    Q2                    & Major: What is your major?                                                                                                                                                                                                               \\
                          & \underline{\hspace{6cm}}                                                                                                                                                                                                                 \\
    \midrule
    Q3                    & Year: What year are you in?                                                                                                                                                                                                              \\
                          & $\bigcirc$ Year 1 \quad $\bigcirc$ Year 2 \quad $\bigcirc$ Year 3 \quad $\bigcirc$ Year 4                                                                                                                                                \\
                          & $\bigcirc$ Master's Year 1 (PhD-track Y1) \quad $\bigcirc$ Master's Year 2 (PhD-track Y2)                                                                                                                                                \\
                          & $\bigcirc$ Master's Year 3 (PhD-track Y3) \quad $\bigcirc$ PhD Year 1 (PhD-track Y4)                                                                                                                                                     \\
                          & $\bigcirc$ PhD Year 2 (PhD-track Y5) \quad $\bigcirc$ Not sure / prefer not to say \quad $\bigcirc$ Other                                                                                                                                \\
    \midrule
    Q4                    & What is your approximate rank in your cohort/major?                                                                                                                                                                                      \\
                          & $\bigcirc$ Top 10\% \quad $\bigcirc$ Top 20\% \quad $\bigcirc$ Top 30\% \quad $\bigcirc$ Below 30\% \quad $\bigcirc$ Not sure / prefer not to say                                                                                        \\
    \midrule
    Q5                    & First-generation student: Are you the first in your family to attend college?                                                                                                                                                            \\
                          & $\bigcirc$ Yes \quad $\bigcirc$ No \quad $\bigcirc$ Not sure / prefer not to say                                                                                                                                                         \\
    \midrule
    Q6                    & Socioeconomic status: What is your household annual income range?                                                                                                                                                                        \\
                          & $\bigcirc$ Below 50,000 RMB \quad $\bigcirc$ 50,000 - 100,000 RMB \quad $\bigcirc$ 100,000 - 500,000 RMB                                                                                                                                 \\
                          & $\bigcirc$ 500,000 - 1,000,000 RMB \quad $\bigcirc$ 1,000,000 - 10,000,000 RMB \quad $\bigcirc$ Above 10,000,000 RMB \quad $\bigcirc$ Not sure / prefer not to say                                                                       \\
    \midrule
    Q7                    & How do you find solutions when you face learning problems? (Select all that apply)                                                                                                                                                       \\
                          & $\square$ Try on my own \quad $\square$ Consult documentation \quad $\square$ Ask classmates or teachers                                                                                                                                 \\
                          & $\square$ Online search (e.g., Baidu, Google) \quad $\square$ Technical Q\&A platforms (e.g., Stack Overflow, GitHub)                                                                                                                    \\
                          & $\square$ Watch learning videos or online courses (e.g., YouTube, Bilibili) \quad $\square$ Use LLMs for help (e.g., ChatGPT) \quad $\square$ Other                                                                                      \\
    \midrule
    Q8-P                  & \textbf{[Python Course]} Programming background: If you have taken programming courses, list all languages you learned. (Select all that apply)                                                                                          \\
                          & $\square$ Java \quad $\square$ C++ \quad $\square$ C \quad $\square$ Python \quad $\square$ JavaScript \quad $\square$ Matlab \quad $\square$ R \quad $\square$ SQL \quad $\square$ I have not learned programming \quad $\square$ Other \\
    \midrule
    Q8-G                  & \textbf{[Game Theory Course]} Math background: Describe your experience with derivatives.                                                                                                                                                \\
                          & $\bigcirc$ I am very familiar with derivatives, including partial derivatives.                                                                                                                                                           \\
                          & $\bigcirc$ I learned derivatives, but I am unclear or do not remember how to compute derivatives or partials.                                                                                                                            \\
                          & $\bigcirc$ I only know basic concepts of derivatives and am not very familiar.                                                                                                                                                           \\
                          & $\bigcirc$ I have not learned derivatives or related concepts.                                                                                                                                                                           \\
                          & $\bigcirc$ Other                                                                                                                                                                                                                         \\
    \midrule
    Q9-P                  & \textbf{[Python Course]} Programming frequency: How often do you program?                                                                                                                                                                \\
                          & $\bigcirc$ Almost daily \quad $\bigcirc$ Almost weekly \quad $\bigcirc$ Occasionally \quad $\bigcirc$ Rarely, almost never                                                                                                               \\
    \midrule
    Q9-G                  & \textbf{[Game Theory Course]} Math familiarity: When did you last study or apply calculus (e.g., derivatives)?                                                                                                                           \\
                          & $\bigcirc$ Within the past week \quad $\bigcirc$ Within the past month \quad $\bigcirc$ Within the past six months \quad $\bigcirc$ Within the past year \quad $\bigcirc$ More than a year ago \quad $\bigcirc$ Other                    \\
    \midrule
    Q10-G                 & \textbf{[Game Theory Course]} Math background: Check all math concepts you have learned. (Select all that apply)                                                                                                                         \\
                          & $\square$ Derivatives \quad $\square$ Matrices \quad $\square$ Nash equilibrium \quad $\square$ Complex analysis \quad $\square$ Partial derivatives                                                                                     \\
                          & $\square$ Static games \quad $\square$ Hypothesis testing \quad $\square$ Bellman equation \quad $\square$ Groups/Rings/Fields \quad $\square$ I have not learned any of the above                                                       \\
    \midrule
    Q11                   & Usage: Have you used LLM chat models (e.g., ChatGPT) before?                                                                                                                                                                             \\
                          & $\bigcirc$ Yes \quad $\bigcirc$ No                                                                                                                                                                                                       \\
    \midrule
    Q12                   & Frequency: If yes, how often do you use LLM chat models (e.g., ChatGPT) to support learning?                                                                                                                                             \\
                          & $\bigcirc$ Never \quad $\bigcirc$ Rarely \quad $\bigcirc$ Sometimes \quad $\bigcirc$ Often \quad $\bigcirc$ Always                                                                                                                       \\
    \midrule
    Q13                   & Familiarity: On a 1-5 scale, how familiar are you with LLM chat models (e.g., ChatGPT)?                                                                                                                                                  \\
                          & $\bigcirc$ 1. Not familiar \quad $\bigcirc$ 2. Slightly familiar \quad $\bigcirc$ 3. Moderately familiar \quad $\bigcirc$ 4. Quite familiar \quad $\bigcirc$ 5. Extremely familiar                                                       \\
    \midrule
    Q14                   & Impact: How do you rate their impact on learning (1-5)?                                                                                                                                                                                  \\
                          & $\bigcirc$ 1: Negative \quad $\bigcirc$ 2: Slightly negative \quad $\bigcirc$ 3: No impact \quad $\bigcirc$ 4: Positive \quad $\bigcirc$ 5: Very positive                                                                                \\
    \midrule
    Q15                   & Task usage: If yes, what types of tasks do you use LLM chat models (e.g., ChatGPT) for? (Select all that apply)                                                                                                                          \\
                          & $\square$ Concept explanations \quad $\square$ Answers and explanations for specific questions \quad $\square$ Help translate or summarize literature                                                                                    \\
                          & $\square$ Help write reports or papers \quad $\square$ Research inspiration through dialogue \quad $\square$ Leisure \quad $\square$ Other                                                                                               \\
    \midrule
    Q16                   & Reasons for not using: If you have not used LLM-based platforms (e.g., ChatGPT), what are the main reasons? (Select all that apply)                                                                                                      \\
                          & $\square$ I have used LLM-based platforms (e.g., ChatGPT) \quad $\square$ I am not aware of LLM-based platforms                                                                                                                          \\
                          & $\square$ Concerns about answer quality or truthfulness \quad $\square$ Accessibility or connectivity issues                                                                                                                             \\
                          & $\square$ May weaken independent thinking \quad $\square$ Risk of plagiarism accusations \quad $\square$ Other                                                                                                                           \\
    \midrule
    Q17                   & Potential risks: What risks do you think exist when using LLM chat models (e.g., ChatGPT) for learning? (Select all that apply)                                                                                                          \\
                          & $\square$ I worry about over-reliance, which may weaken my critical thinking and problem-solving.                                                                                                                                        \\
                          & $\square$ I worry the information from these tools may be inaccurate or unreliable.                                                                                                                                                      \\
                          & $\square$ I worry these tools may reduce deep learning.                                                                                                                                                                                  \\
                          & $\square$ I worry this may reduce interaction and collaboration with peers/teachers.                                                                                                                                                     \\
                          & $\square$ I think there are no risks in using LLM chat models for learning \quad $\square$ I am not aware of LLM-based platforms \quad $\square$ Other                                                                                   \\
    \midrule
    Q18                   & Expectations: What do you hope to gain from learning with LLM chat models (e.g., ChatGPT)? (Select all that apply)                                                                                                                       \\
                          & $\square$ Improve learning outcomes for related knowledge \quad $\square$ Gain deeper understanding of related knowledge                                                                                                                 \\
                          & $\square$ Higher motivation and engagement during learning \quad $\square$ I am not aware of LLM-based platforms \quad $\square$ Other                                                                                                   \\
    \midrule
    Q19                   & Willingness: How willing are you to use LLM chat models (e.g., ChatGPT) to support learning in the future?                                                                                                                               \\
                          & $\bigcirc$ Very willing \quad $\bigcirc$ Somewhat willing \quad $\bigcirc$ Neutral \quad $\bigcirc$ Not very willing \quad $\bigcirc$ Not willing at all \quad $\bigcirc$ I am not aware of LLM-based platforms                          \\
    \midrule
    Q20                   & Assessment: How should educators design assessments to avoid misuse of LLM chat models (e.g., ChatGPT)? (Select all that apply)                                                                                                          \\
                          & $\square$ Design open-ended questions \quad $\square$ Ban LLM chat models or any AI tools during learning (including homework)                                                                                                           \\
                          & $\square$ Ban LLM chat models or any AI tools during quizzes \quad $\square$ Provide guidance on using LLM chat models                                                                                                                   \\
                          & $\square$ Provide dedicated restricted LLM tools (e.g., no direct solutions) \quad $\square$ I am not aware of LLM-based platforms \quad $\square$ Other                                                                                 \\
\end{longtable}

\begin{longtable}{|l|p{0.80\textwidth}|}
    \caption{Prior Knowledge assessment questionnaire for Python course.}
    \label{tab:foundation-python}                                                                                                                                                                                              \\
    \toprule
    \textbf{Question No.} & \textbf{Question Content}                                                                                                                                                                          \\
    \midrule
    \endfirsthead
    \multicolumn{2}{c}%
    {{\tablename\ \thetable{} -- continued from previous page}}                                                                                                                                                                \\
    \toprule
    \textbf{Question No.} & \textbf{Question Content}                                                                                                                                                                          \\
    \midrule
    \endhead
    \midrule \multicolumn{2}{r}{{Continued on next page}}                                                                                                                                                                      \\
    \endfoot
    \bottomrule
    \endlastfoot
    Q1                    & What is the role of loops in programming?                                                                                                                                                          \\
                          & $\bigcirc$ Store multiple variables \quad $\bigcirc$ Connect to a database \quad $\bigcirc$ Output data \quad $\bigcirc$ Repeat the same code                                                      \\
    \midrule
    Q2                    & What is the main purpose of functions in programming?                                                                                                                                              \\
                          & $\bigcirc$ Reuse code and avoid writing the same code repeatedly \quad $\bigcirc$ Increase code execution speed                                                                                    \\
                          & $\bigcirc$ Connect to the Internet \quad $\bigcirc$ Reduce program memory usage                                                                                                                    \\
    \midrule
    Q3                    & Which of the following is a constant?                                                                                                                                                              \\
                          & $\bigcirc$ A value entered by a user \quad $\bigcirc$ A counter updated each loop iteration                                                                                                        \\
                          & $\bigcirc$ A defined value of pi in the program (e.g., 3.14159) \quad $\bigcirc$ A variable storing database query results                                                                         \\
    \midrule
    Q4                    & Which case is most suitable for using an array (list)?                                                                                                                                             \\
                          & $\bigcirc$ Store one user's name \quad $\bigcirc$ Store names of 100 users \quad $\bigcirc$ Compute the sum of two numbers \quad $\bigcirc$ Print a message                                        \\
    \midrule
    Q5                    & Read the pseudocode below and answer:                                                                                                                                                              \\
                          & \begin{minipage}{0.7\textwidth}
                                \begin{verbatim}
FUNCTION MyFunc(n)
    IF n <= 1
        RETURN FALSE
    END IF
    FOR i FROM 2 TO sqrt(n)
        IF n MOD i == 0
            RETURN FALSE
        END IF
    END FOR
    RETURN TRUE
END FUNCTION
\end{verbatim}
                            \end{minipage}                                                                                                                                                                     \\
                          & What does this function check?                                                                                                                                                                     \\
                          & $\bigcirc$ Whether it is positive \quad $\bigcirc$ Whether it is even \quad $\bigcirc$ Whether it is prime \quad $\bigcirc$ Whether it is composite                                                \\
    \midrule
    Q6                    & Which pseudocode correctly computes the absolute difference between x and y?                                                                                                                       \\
                          & \begin{minipage}{0.7\textwidth}
                                \begin{verbatim}
A) 
X, Y = input two values
ABS_VALUE = X - Y
RETURN ABS_VALUE

B)
X, Y = input two values
IF X - Y < 0 THEN
    ABS_VALUE = Y - X
ELSE
    ABS_VALUE = X - Y
RETURN ABS_VALUE

C)
X, Y = input two values
ABS_VALUE = X + Y
RETURN ABS_VALUE

D)
X, Y = input two values
ABS_VALUE = (X - Y) * (X - Y)
RETURN ABS_VALUE
\end{verbatim}
                            \end{minipage}                                                                                                                                                                     \\
                          & $\bigcirc$ A \quad $\bigcirc$ B \quad $\bigcirc$ C \quad $\bigcirc$ D                                                                                                                              \\
    \midrule
    Q7                    & Which pseudocode correctly reverses an integer x (e.g., 12345 -> 54321)?                                                                                                                           \\
                          & \begin{minipage}{0.7\textwidth}
                                \begin{verbatim}
A) 
X = input integer
REVERSE = 0
WHILE X != 0:
    REVERSE = REVERSE * 10 + X % 10
    X = X / 10
RETURN REVERSE

B) 
X = input integer
REVERSE = -X
RETURN REVERSE

C) 
X = input integer
REVERSE = 0
WHILE X != 0:
     REVERSE = REVERSE + (X % 10) * (X % 10)
     X = X / 10
RETURN REVERSE

D)
X = input integer
REVERSE = 0
WHILE X != 0:
     REVERSE = REVERSE + X % 10
     X = X / 10
RETURN REVERSE
\end{verbatim}
                            \end{minipage}                                                                                                                                                        \\
                          & $\bigcirc$ A \quad $\bigcirc$ B \quad $\bigcirc$ C \quad $\bigcirc$ D                                                                                                                              \\
    \midrule
    Q8                    & Which pseudocode correctly computes the sum of the first N Fibonacci numbers?                                                                                                                      \\
                          & \begin{minipage}{0.7\textwidth}
                                \begin{verbatim}
A) 
N = input number of terms
SUM = 0
FOR I FROM 1 TO N:
    SUM = SUM + I
RETURN SUM

B) 
N = input number of terms
SUM = 1
FOR I FROM 2 TO N:
    SUM = SUM * I
RETURN SUM

C) 
N = input number of terms
A = 0, B = 1, SUM = 0
FOR I FROM 1 TO N:
    SUM = SUM + A
    NEXT = A + B
    A = B
    B = NEXT
RETURN SUM

D) 
N = input number of terms
FUNCTION FIB_SUM(N):
    IF N <= 1:
        RETURN N
    ELSE:
        RETURN FIB_SUM(N-1) + FIB_SUM(N-2)
RETURN FIB_SUM(N)
\end{verbatim}
                            \end{minipage}                                                                                                                                                                     \\
                          & $\bigcirc$ A \quad $\bigcirc$ B \quad $\bigcirc$ C \quad $\bigcirc$ D                                                                                                                              \\
    \midrule
    Q9                    & Read the pseudocode below and choose the possible logic error:                                                                                                                                     \\
                          & \begin{minipage}{0.7\textwidth}
                                \begin{verbatim}
FUNCTION InsertSorted(array, value)
    FOR i FROM 0 TO LENGTH(array) - 1
        IF value < array[i]
            INSERT value AT POSITION i IN array
            BREAK
        END IF
    END FOR
    IF value >= array[LENGTH(array) - 1]
        APPEND value TO array
    END IF
    RETURN array
END FUNCTION
\end{verbatim}
                            \end{minipage}                                                                                                                                                                     \\
                          & $\bigcirc$ If array is empty, accessing array[LENGTH(array) - 1] will cause an error                                                                                                               \\
                          & $\bigcirc$ Value is always appended to the end, regardless of correct order                                                                                                                        \\
                          & $\bigcirc$ The function does not update the array length after insertion, which may cause errors                                                                                                   \\
                          & $\bigcirc$ The value will be inserted multiple times into the array                                                                                                                                \\
    \midrule
    Q10                   & Read the pseudocode below and answer:                                                                                                                                                              \\
                          & \begin{minipage}{0.7\textwidth}
                                \begin{verbatim}
FUNCTION MyFunc(arr, left, right)
    IF left == right
        RETURN arr[left]
    ELSE
        mid = (left + right) / 2
        leftMax = MyFunc(arr, left, mid)
        rightMax = MyFunc(arr, mid + 1, right)
        IF leftMax > rightMax
            RETURN leftMax
        ELSE
            RETURN rightMax
        END IF
    END IF
END FUNCTION
\end{verbatim}
                            \end{minipage}                                                                                                                                                                     \\
                          & What is the purpose of this recursive function?                                                                                                                                                    \\
                          & $\bigcirc$ Sort the array \quad $\bigcirc$ Find the maximum value in the array \quad $\bigcirc$ Split the array into two equal parts \quad $\bigcirc$ Compute the sum of all elements in the array \\
\end{longtable}

\begin{longtable}{|l|p{0.80\textwidth}|}
    \caption{Prior Knowledge assessment questionnaire for Game Theory course.}
    \label{tab:foundation-game-theory}                                                                          \\
    \toprule
    \textbf{Question No.} & \textbf{Question Content}                                                           \\
    \midrule
    \endfirsthead
    \multicolumn{2}{c}%
    {{\tablename\ \thetable{} -- continued from previous page}}                                                 \\
    \toprule
    \textbf{Question No.} & \textbf{Question Content}                                                           \\
    \midrule
    \endhead
    \midrule \multicolumn{2}{r}{{Continued on next page}}                                                       \\
    \endfoot
    \bottomrule
    \endlastfoot
    Q1                    & Piecewise function extrema: For $f(x)$, find the maximum value on its domain.       \\
                          & $$
        f(x) = \begin{cases}
                   2x - 1 & \text{when } -2 \leq x \leq 1, \\
                   -x^2   & \text{when } x > 1
               \end{cases}
    $$                                                                                                          \\
    \midrule
    Q2                    & Solve the system: find $x$.                                                         \\
                          & $$
        \begin{cases}
            2x + 3y = 9 \\
            4x - 2y = 10
        \end{cases}
    $$                                                                                                          \\
    \midrule
    Q3                    & Derivative: For $f(x) = 3x^4 - 8x^3 + 6x^2 - 12x + 2$, find $f'(x)$ at $x = 1$.     \\
                          & $\bigcirc$ 2 \quad $\bigcirc$ -12 \quad $\bigcirc$ 12 \quad $\bigcirc$ 0            \\
    \midrule
    Q4                    & Function maximum: For $f(x) = x^3 - 6x^2 + 9x + 12$, find the maximum on $[-4, 4]$. \\
                          & \underline{\hspace{8cm}}                                                            \\
    \midrule
    Q5                    & Linear optimization: maximize $z = 2x + y$                                          \\
                          & Subject to constraints:                                                             \\
                          & $3x + y \leq 12$                                                                    \\
                          & $x + y \leq 6$                                                                      \\
                          & $x \geq 0, y \geq 0$                                                                \\
                          & Find the maximum value of $z$.                                                      \\
                          & \underline{\hspace{8cm}}                                                            \\
\end{longtable}

\subsection{Assignment questions}

After the course learning phase, participants had 20 minutes to complete the assignment. They could refer to the learning materials during this phase. The game theory assignment included 4 questions, and the Python assignment included 8 questions with point values reflecting item difficulty. The original participant-facing questions are reproduced in \autoref{tab:assignment-python} and \autoref{tab:assignment-game-theory}.

\begin{longtable}{|l|p{0.8\textwidth}|}
    \caption{Python Assignment questions.}
    \label{tab:assignment-python}                                                                                                                                                                                                                                                                                                                \\
    \toprule
    \textbf{Problem No.} & \textbf{Problem Description}                                                                                                                                                                                                                                                                                          \\
    \midrule
    \endfirsthead
    \multicolumn{2}{c}%
    {{\tablename\ \thetable{} -- continued from previous page}}                                                                                                                                                                                                                                                                                  \\
    \toprule
    \textbf{Problem No.} & \textbf{Problem Description}                                                                                                                                                                                                                                                                                          \\
    \midrule
    \endhead
    \midrule \multicolumn{2}{r}{{Continued on next page}}                                                                                                                                                                                                                                                                                        \\
    \endfoot
    \bottomrule
    \endlastfoot
    Q1                   & \textbf{Power}                                                                                                                                                                                                                                                                                                        \\
                         & Given an integer \texttt{x} (0, positive, or negative) and a non-negative integer \texttt{n}, write a function \texttt{power(x, n)} that returns \texttt{x} raised to the \texttt{n}th power. (1 pts)                                                                                                                 \\
    \midrule
    Q2                   & \textbf{Leap Year Check}                                                                                                                                                                                                                                                                                              \\
                         & Write a function \texttt{is\_leap\_year(year)} to check whether a given year is a leap year. Return \texttt{True} if it is, otherwise \texttt{False}. A leap year is:                                                                                                                                                 \\
                         & - Divisible by 4 but not by 100; or                                                                                                                                                                                                                                                                                   \\
                         & - Divisible by 400. (2 pts)                                                                                                                                                                                                                                                                                           \\
    \midrule
    Q3                   & \textbf{Insert List Element}                                                                                                                                                                                                                                                                                          \\
                         & Write a function \texttt{insert\_at\_index(lst, element, index)} that takes a list and two integers: an element and a positive index. Insert the element at the specified index. If the index exceeds the current length, append the element to the end. (2 pts)                                                      \\
    \midrule
    Q4                   & \textbf{Character Frequency}                                                                                                                                                                                                                                                                                          \\
                         & Write a function \texttt{char\_frequency(s)} that takes a string and returns a dictionary of character counts, preserving the order of first appearance. (3 pts)                                                                                                                                                      \\
    \midrule
    Q5                   & \textbf{Count Vowels in a String}                                                                                                                                                                                                                                                                                     \\
                         & Write a function \texttt{count\_vowels(s)} that returns the number of vowels \texttt{('a', 'e', 'i', 'o', 'u', 'A', 'E', 'I', 'O', 'U')} in a string \texttt{s}. Note that \texttt{s} may contain both upper- and lower-case letters. (3 pts)                                                                         \\
    \midrule
    Q6                   & \textbf{Greatest Common Divisor}                                                                                                                                                                                                                                                                                      \\
                         & Write a function \texttt{gcd(a, b)} that takes two positive integers and returns their greatest common divisor. (4 pts)                                                                                                                                                                                               \\
                         & Tip: A divisor is an integer that divides another integer exactly. For example, divisors of 12 include 1, 2, 3, 4, 6, 12. The GCD is the largest common divisor.                                                                                                                                                      \\
    \midrule
    Q7                   & \textbf{Compare Prices Across Shops}                                                                                                                                                                                                                                                                                  \\
                         & Given a list \texttt{prices} and a product name \texttt{product}. \texttt{prices[i]} is a dict for shop i, where keys are product names and values are prices. Write \texttt{find\_lowest\_price(prices, product)} to return the lowest price for \texttt{product}. Some shops may not have \texttt{product}. (4 pts) \\
    \midrule
    Q8                   & \textbf{Minimum Score Difference}                                                                                                                                                                                                                                                                                     \\
                         & Given a 0-indexed integer list \texttt{nums} where \texttt{nums[i]} is the score of student i, and an integer \texttt{k}, write \texttt{min\_difference(nums, k)} to pick any \texttt{k} scores so the max-min difference is minimized. Return this minimum difference. (5 pts)                                       \\
                         & Tip: Use \texttt{nums.sort()} to sort a list. For example,                                                                                                                                                                                                                                                            \\
                         & \begin{minipage}{0.7\textwidth}
                               \begin{verbatim}
nums = [1, 3, 2, 4]
nums.sort()
print(nums) # [1, 2, 3, 4]
\end{verbatim}
                           \end{minipage}                                                                                                                                                                                                                                                                                        \\
                         & Note: \texttt{sort} returns \texttt{None}.                                                                                                                                                                                                                                                                            \\
\end{longtable}

\begin{longtable}{|l|p{0.80\textwidth}|}
    \caption{Game Theory Assignment questions.}
    \label{tab:assignment-game-theory}                                                                                                                                                                                                                                                                                                                                                                                                                                                                        \\
    \toprule
    \textbf{Problem No.} & \textbf{Problem Description}                                                                                                                                                                                                                                                                                                                                                                                                                                                       \\
    \midrule
    \endfirsthead
    \multicolumn{2}{c}%
    {{\tablename\ \thetable{} -- continued from previous page}}                                                                                                                                                                                                                                                                                                                                                                                                                                               \\
    \toprule
    \textbf{Problem No.} & \textbf{Problem Description}                                                                                                                                                                                                                                                                                                                                                                                                                                                       \\
    \midrule
    \endhead
    \midrule \multicolumn{2}{r}{{Continued on next page}}                                                                                                                                                                                                                                                                                                                                                                                                                                                     \\
    \endfoot
    \bottomrule
    \endlastfoot
    Q1                   & \textbf{The Blonde Girl}                                                                                                                                                                                                                                                                                                                                                                                                                                                           \\
                         & In a movie, two men (A and B) go to a bar with several women. One is a beautiful blonde, and the rest are ordinary women. The men want to approach someone. This is modeled as a two-player game: matching with the blonde gives payoff 2, matching with an ordinary woman gives payoff 1, and getting no one gives payoff 0. (If both pursue the blonde, they interfere and nobody succeeds, so payoff is 0. If only one person approaches a woman, it is considered successful.) \\
                         & (1) The participants in this game are \texttt{Blank1} and \texttt{Blank2}. Each participant's strategies are \texttt{Blank3} and \texttt{Blank4}. (1 pts)                                                                                                                                                                                                                                                                                                                          \\
                         & (2) The payoff matrix is:                                                                                                                                                                                                                                                                                                                                                                                                                                                          \\
                         & \begin{tabular}{|c|c|c|} \hline
                               \texttt{Blank1}/\texttt{Blank2} & \texttt{Blank3}         & \texttt{Blank4}         \\ \hline
                               \texttt{Blank3}                 & \texttt{A1},\texttt{A2} & \texttt{B1},\texttt{B2} \\ \hline
                               \texttt{Blank4}                 & \texttt{C1},\texttt{C2} & \texttt{D1},\texttt{D2} \\ \hline
                           \end{tabular}                                                                                                                                                                                                                                                                                                                                                                                       \\
                         & Fill in \texttt{A1}, \texttt{A2}, \texttt{B1}, \texttt{B2}, \texttt{C1}, \texttt{C2}, \texttt{D1}, \texttt{D2}, and the Nash equilibria are \texttt{Blank5}. (2 pts)                                                                                                                                                                                                                                                                                                               \\
    \midrule
    Q2                   & \textbf{Yale Game Theory Classroom Game}                                                                                                                                                                                                                                                                                                                                                                                                                                           \\
                         & In class, the teacher gives students two choices, “A” and “B”. Each student writes one choice on a note. Students are randomly paired, and they do not know who their partner is in advance. Each student then receives a score:                                                                                                                                                                                                                                                   \\
                         & 1. If I choose “A” and the other chooses “B”, I get 3 and the other gets -1.                                                                                                                                                                                                                                                                                                                                                                                                       \\
                         & 2. If I choose “B” and the other chooses “A”, I get -1 and the other gets 3.                                                                                                                                                                                                                                                                                                                                                                                                       \\
                         & 3. If both choose “A”, both get 0.                                                                                                                                                                                                                                                                                                                                                                                                                                                 \\
                         & 4. If both choose “B”, both get 1.                                                                                                                                                                                                                                                                                                                                                                                                                                                 \\
                         & (1) The payoff matrix is below. Fill in each cell. (1 pts)                                                                                                                                                                                                                                                                                                                                                                                                                         \\
                         & \begin{tabular}{|c|c|c|} \hline
                               Me/Other & A                       & B                       \\ \hline
                               A        & \texttt{A1},\texttt{A2} & \texttt{B1},\texttt{B2} \\ \hline
                               B        & \texttt{C1},\texttt{C2} & \texttt{D1},\texttt{D2} \\ \hline
                           \end{tabular}                                                                                                                                                                                                                                                                                                                                                                                                              \\
                         & Fill in \texttt{A1}, \texttt{A2}, \texttt{B1}, \texttt{B2}, \texttt{C1}, \texttt{C2}, \texttt{D1}, \texttt{D2}.                                                                                                                                                                                                                                                                                                                                                                    \\
                         & (2) Using iterated strict dominance elimination, which strategy profiles are not eliminated? (1.5 pts)                                                                                                                                                                                                                                                                                                                                                                             \\
                         & A: (A,A) B: (A,B) C: (B,A) D: (B,B)                                                                                                                                                                                                                                                                                                                                                                                                                                                \\
                         & Your choice: \texttt{Blank1}.                                                                                                                                                                                                                                                                                                                                                                                                                                                      \\
                         & (3) Using the underline method, what is the pure-strategy Nash equilibrium? (1.5 pts)                                                                                                                                                                                                                                                                                                                                                                                              \\
                         & A: (A,A) B: (A,B) C: (B,A) D: (B,B)                                                                                                                                                                                                                                                                                                                                                                                                                                                \\
                         & Your choice: \texttt{Blank2}.                                                                                                                                                                                                                                                                                                                                                                                                                                                      \\
    \midrule
    Q3                   & \textbf{Splitting One Dollar}                                                                                                                                                                                                                                                                                                                                                                                                                                                      \\
                         & Two people negotiate over how to split one dollar. They simultaneously propose desired shares \(s_1\) and \(s_2\), with \(0 \leq s_1, s_2 \leq 1 \). If \(s_1 + s_2 \leq 1\), each gets their requested share; if \(s_1 + s_2 > 1\), both get nothing. Find the pure-strategy Nash equilibria. (5 pts)                                                                                                                                                                             \\
                         & $$u_i(s_i, s_j) = \begin{cases}
                                                     [[Blank1]], s_1 + s_2 \leq 1 \\
                                                     [[Blank2]], s_1 + s_2 > 1
                                                 \end{cases}$$                                                                                                                                                                                                                                                                                                                                                                                                  \\
                         & By definition of Nash equilibrium, \(s_i^\ast = \arg\max_{0 \leq s_i \leq 1} u_i(s_i, s_j^\ast)\).                                                                                                                                                                                                                                                                                                                                                                                 \\
                         & Solving yields: \texttt{Blank3}                                                                                                                                                                                                                                                                                                                                                                                                                                                    \\
                         & Therefore the pure-strategy Nash equilibria are \texttt{Blank4}.                                                                                                                                                                                                                                                                                                                                                                                                                   \\
    \midrule
    Q4                   & \textbf{Bertrand Duopoly Model}                                                                                                                                                                                                                                                                                                                                                                                                                                                    \\
                         & We consider two differentiated products. If firm 1 and firm 2 choose prices \(p_1\) and \(p_2\), demand for firm \(i\)'s product is:                                                                                                                                                                                                                                                                                                                                               \\
                         & $$q_i(p_i, p_j) = a - p_i + bp_j$$                                                                                                                                                                                                                                                                                                                                                                                                                                                 \\
                         & where \(b>0\), meaning firm \(i\)'s product is a differentiated substitute for firm \(j\)'s product. Assume no fixed costs and constant marginal cost \(c\). Find the pure-strategy Nash equilibrium. (7 pts)                                                                                                                                                                                                                                                                      \\
                         & Tip: Fixed costs do not change with output within a certain range (e.g., rent, management salaries, depreciation). Marginal cost is the additional cost of producing one more unit, i.e., the derivative of total cost with respect to output.                                                                                                                                                                                                                                     \\
                         & Fill in the blanks:                                                                                                                                                                                                                                                                                                                                                                                                                                                                \\
                         & Let firm i's payoff be \(u_i(p_i, p_j)\). From the description:                                                                                                                                                                                                                                                                                                                                                                                                                    \\
                         & $$u_i(p_i, p_j) = [[Blank1]]$$                                                                                                                                                                                                                                                                                                                                                                                                                                                     \\
                         & The Bertrand equilibrium is \(p_1^\ast=\)\texttt{Blank2}, \(p_2^\ast=\)\texttt{Blank3}.                                                                                                                                                                                                                                                                                                                                                                                            \\
\end{longtable}

\subsection{Review questions}

After the assignment phase, participants entered a 20-minute review phase. They could access the assignment questions, their answers, and the course learning materials. They also received additional review questions with viewable reference answers. The original 32 Python review questions and 10 game theory review questions are reproduced in \autoref{tab:review-python} and \autoref{tab:review-game-theory}, respectively.

\begin{longtable}{|l|p{0.80\textwidth}|}
    \caption{Review questions of Python course.}
    \label{tab:review-python}                                                                                                                                                                                                                                                                                                                               \\
    \toprule
    \textbf{Problem No.} & \textbf{Problem Description}                                                                                                                                                                                                                                                                                                     \\
    \midrule
    \endfirsthead
    \multicolumn{2}{c}%
    {{\tablename\ \thetable{} -- continued from previous page}}                                                                                                                                                                                                                                                                                             \\
    \toprule
    \textbf{Problem No.} & \textbf{Problem Description}                                                                                                                                                                                                                                                                                                     \\
    \midrule
    \endhead
    \midrule \multicolumn{2}{r}{{Continued on next page}}                                                                                                                                                                                                                                                                                                   \\
    \endfoot
    \bottomrule
    \endlastfoot
    Q1                   & \textbf{Check Positive Number}                                                                                                                                                                                                                                                                                                   \\
                         & Write a function \texttt{is\_positive(num)} that takes an integer and checks whether the number is positive. Return True if it is positive, otherwise return False. (1 pts)                                                                                                                                                      \\
    \midrule
    Q2                   & \textbf{Negate List}                                                                                                                                                                                                                                                                                                             \\
                         & Given an integer list \texttt{lst}, create a new list that contains the negation of each value. For example, given \texttt{[-1, 2, -3, 4]}, the result should be \texttt{[1, -2, 3, -4]}. (1 pts)                                                                                                                                \\
    \midrule
    Q3                   & \textbf{Count List Elements}                                                                                                                                                                                                                                                                                                     \\
                         & Write a function \texttt{count\_elements(lst)} that takes a list and returns the number of elements in the list. (1 pts)                                                                                                                                                                                                         \\
    \midrule
    Q4                   & \textbf{Is List Empty}                                                                                                                                                                                                                                                                                                           \\
                         & Write a function \texttt{is\_list\_empty(lst)} that takes a list and returns a boolean indicating whether the list is empty. (1 pts)                                                                                                                                                                                             \\
    \midrule
    Q5                   & \textbf{Merge Two Lists}                                                                                                                                                                                                                                                                                                         \\
                         & Write a function \texttt{merge\_lists(list1, list2)} that takes two lists and returns a new list containing all elements from both input lists. (1 pts)                                                                                                                                                                          \\
    \midrule
    Q6                   & \textbf{Pop and Return Value}                                                                                                                                                                                                                                                                                                    \\
                         & Write a function \texttt{pop\_and\_return\_value(my\_dict, key)} that takes a dictionary \texttt{my\_dict} and a key \texttt{key}. The function should remove the key-value pair and return the value. (1 pts)                                                                                                                   \\
    \midrule
    Q7                   & \textbf{Integer Sign}                                                                                                                                                                                                                                                                                                            \\
                         & Given an integer variable \texttt{x}, write a function \texttt{judge\_integer\_sign} that sets the variable \texttt{s} to the string "POSITIVE", "NEGATIVE", or "ZERO" depending on whether \texttt{x} is positive, negative, or zero. (2 pts)                                                                                   \\
    \midrule
    Q8                   & \textbf{Count Greater Than}                                                                                                                                                                                                                                                                                                      \\
                         & Write a function \texttt{count\_greater\_than} that takes an integer list \texttt{lst} and an integer \texttt{val}, and counts how many values in the list are strictly greater than \texttt{val}. (2 pts)                                                                                                                       \\
    \midrule
    Q9                   & \textbf{Calculate Average}                                                                                                                                                                                                                                                                                                       \\
                         & Write a function \texttt{calculate\_average(scores)} that takes a list of integer scores and returns the average score. (2 pts)                                                                                                                                                                                                  \\
    \midrule
    Q10                  & \textbf{Generate Star Strings}                                                                                                                                                                                                                                                                                                   \\
                         & Write a function \texttt{make\_star\_strings(nums)} that takes a list of non-negative integers \texttt{nums} and returns a list of strings made of asterisks (*). Each string should contain as many asterisks as the corresponding number. Note that 0 should produce an empty string. (2 pts)                                  \\
    \midrule
    Q11                  & \textbf{Sum of Even Numbers}                                                                                                                                                                                                                                                                                                     \\
                         & Write a function \texttt{sum\_even(nums)} that takes a list of numbers \texttt{nums} and returns the sum of all even numbers in the list. (2 pts)                                                                                                                                                                                \\
    \midrule
    Q12                  & \textbf{Count Positives in List}                                                                                                                                                                                                                                                                                                 \\
                         & Write a function \texttt{count\_positives(nums)} that takes a list of numbers \texttt{nums} and returns the count of positive numbers in the list. (2 pts)                                                                                                                                                                       \\
    \midrule
    Q13                  & \textbf{Are Students Passing}                                                                                                                                                                                                                                                                                                    \\
                         & Write a function \texttt{are\_students\_passing(grades)} that takes a dictionary \texttt{grades} of student names to scores. Return a new dictionary mapping each student to whether they pass (score >= 60 is True, otherwise False). (2 pts)                                                                                   \\
    \midrule
    Q14                  & \textbf{Count Until Break}                                                                                                                                                                                                                                                                                                       \\
                         & Write a function \texttt{count\_until\_break(nums)} that takes an integer list \texttt{nums}. The function should count elements until it encounters a \texttt{0}, then stop and return the count. (3 pts)                                                                                                                       \\
    \midrule
    Q15                  & \textbf{Total Salary}                                                                                                                                                                                                                                                                                                            \\
                         & Write a function \texttt{calculate\_total\_salary(salaries)} that takes a dictionary \texttt{salaries} mapping employee names to salaries (integers), and returns the total salary. (3 pts)                                                                                                                                      \\
    \midrule
    Q16                  & \textbf{Order Odds Before Evens}                                                                                                                                                                                                                                                                                                 \\
                         & Write a function \texttt{order\_odd\_even(nums)} that reorders the list \texttt{nums} so that all odd numbers come before even numbers, while preserving the original relative order. (3 pts)                                                                                                                                    \\
    \midrule
    Q17                  & \textbf{Remove List Elements}                                                                                                                                                                                                                                                                                                    \\
                         & Given a list \texttt{nums} and a value \texttt{val}, remove all elements equal to \texttt{val} and return the resulting list. The order of elements should not change. (3 pts)                                                                                                                                                   \\
    \midrule
    Q18                  & \textbf{Product of Even Indices}                                                                                                                                                                                                                                                                                                 \\
                         & Write a function \texttt{product\_even\_index(nums)} that takes a list of numbers and returns the product of elements at even indices. (3 pts)                                                                                                                                                                                   \\
    \midrule
    Q19                  & \textbf{Generate Username}                                                                                                                                                                                                                                                                                                       \\
                         & Write a function \texttt{generate\_username(name, birth\_year)} that takes a name and a birth year, and returns a username made of the first letter of the name and the last two digits of the birth year. (3 pts)                                                                                                               \\
    \midrule
    Q20                  & \textbf{Chinese Zodiac}                                                                                                                                                                                                                                                                                                          \\
                         & Write a function \texttt{chinese\_zodiac(year)} that takes a year and returns its Chinese zodiac animal. The zodiac order is fixed, starting from "鼠", with 12 animals: ["鼠", "牛", "虎", "兔", "龙", "蛇", "马", "羊", "猴", "鸡", "狗", "猪"]. The zodiac cycles every 12 years. For example, 2020 is 鼠, so 2021 is 牛, and so on. (3 pts) \\
    \midrule
    Q21                  & \textbf{Sum of Integer Divisions}                                                                                                                                                                                                                                                                                                \\
                         & Write a function \texttt{sum\_of\_divisions(nums, divisor)} that takes a list of numbers \texttt{nums} and an integer \texttt{divisor}. Return the sum of each element divided by \texttt{divisor}, taking only the integer part. (3 pts)                                                                                        \\
    \midrule
    Q22                  & \textbf{Factorial}                                                                                                                                                                                                                                                                                                               \\
                         & Write a function \texttt{factorial(n)} that takes an integer and returns its factorial. The factorial of \texttt{n} is the product of all positive integers less than or equal to \texttt{n}. If the input is negative, return \texttt{None}. (4 pts)                                                                            \\
    \midrule
    Q23                  & \textbf{Find Missing Number}                                                                                                                                                                                                                                                                                                     \\
                         & Given a list of n distinct numbers from 0 to n with one missing, write \texttt{find\_missing\_number(nums)} to return the missing number. (4 pts)                                                                                                                                                                                \\
    \midrule
    Q24                  & \textbf{Longest Common Prefix}                                                                                                                                                                                                                                                                                                   \\
                         & Write a function \texttt{longest\_common\_prefix(str1, str2)} that takes two strings and returns their longest common prefix. (4 pts)                                                                                                                                                                                            \\
    \midrule
    Q25                  & \textbf{Majority Element}                                                                                                                                                                                                                                                                                                        \\
                         & Given an array of size \texttt{n}, write \texttt{majority\_element(nums)} to find the element that appears more than [n/2] times. If no element appears more than [n/2], return \texttt{None}. (4 pts)                                                                                                                           \\
    \midrule
    Q26                  & \textbf{Calculate Taxed Prices}                                                                                                                                                                                                                                                                                                  \\
                         & Write a function \texttt{calculate\_taxed\_prices(prices)} that takes a dictionary \texttt{prices} of item names to original prices. Compute the taxed price using: if price <= 10, tax rate is 10\%; if price > 10, tax rate is 12\%. Return a new dictionary with taxed prices. (4 pts)                                        \\
    \midrule
    Q27                  & \textbf{Merge Strings}                                                                                                                                                                                                                                                                                                           \\
                         & Write a function \texttt{merge\_strings(word1, word2)} that takes two strings and merges them by alternating letters, starting with \texttt{word1}. If one string is longer, append the remaining letters to the end. Return the merged string. (4 pts)                                                                          \\
    \midrule
    Q28                  & \textbf{Replace Digits With Count}                                                                                                                                                                                                                                                                                               \\
                         & Write a function \texttt{replace\_digits\_with\_count(s)} that takes a string \texttt{s}. Replace each digit character with that many repeated copies of itself. For example, "a3b2c4d" becomes "a333b22c4444d". (4 pts)                                                                                                         \\
    \midrule
    Q29                  & \textbf{ATM Transactions}                                                                                                                                                                                                                                                                                                        \\
                         & Write a function \texttt{atm\_transactions(balance, transactions)} that takes an initial balance and a transaction dictionary. The dictionary has keys "D" (deposit) and "W" (withdraw), each mapped to a list of amounts. Compute and return the final balance after all transactions. (4 pts)                                  \\
    \midrule
    Q30                  & \textbf{Find All Primes}                                                                                                                                                                                                                                                                                                         \\
                         & Write a function \texttt{find\_primes(n)} that takes an integer \texttt{n} and returns a list of all prime numbers from 2 to \texttt{n} (inclusive). (5 pts)                                                                                                                                                                     \\
    \midrule
    Q31                  & \textbf{Perfect Number Check}                                                                                                                                                                                                                                                                                                    \\
                         & Given an integer \texttt{num}, write \texttt{is\_perfect\_number(num)} to check whether it is a perfect number. A perfect number equals the sum of its positive divisors excluding itself (e.g., 28). (5 pts)                                                                                                                    \\
    \midrule
    Q32                  & \textbf{Taxi Fare}                                                                                                                                                                                                                                                                                                               \\
                         & Write a function \texttt{calculate\_taxi\_fare(distance, waiting\_time, time\_of\_day)} that takes distance (km), waiting time (minutes), and time of day ("day" or "night") and computes the fare using the rules below. (5 pts)                                                                                                \\
                         & Pricing rules:                                                                                                                                                                                                                                                                                                                   \\
                         & - In "day", the base fare is 10, including the first 3 km.                                                                                                                                                                                                                                                                       \\
                         & - In "night", the base fare is 13, including the first 3 km.                                                                                                                                                                                                                                                                     \\
                         & - After the base distance, daytime costs 2 per km; nighttime costs 2.5 per km.                                                                                                                                                                                                                                                   \\
                         & - Waiting time costs 0.5 per minute in both day and night.                                                                                                                                                                                                                                                                       \\
\end{longtable}

\begin{longtable}{|l|p{0.80\textwidth}|}
    \caption{Review questions of Game Theory course.}
    \label{tab:review-game-theory}                                                                                                                                                                                                                                                                                                                                                                                                                                                                                      \\
    \toprule
    \textbf{Problem No.} & \textbf{Problem Description}                                                                                                                                                                                                                                                                                                                                                                                                                                                                 \\
    \midrule
    \endfirsthead
    \multicolumn{2}{c}%
    {{\tablename\ \thetable{} -- continued from previous page}}                                                                                                                                                                                                                                                                                                                                                                                                                                                         \\
    \toprule
    \textbf{Problem No.} & \textbf{Problem Description}                                                                                                                                                                                                                                                                                                                                                                                                                                                                 \\
    \midrule
    \endhead
    \midrule \multicolumn{2}{r}{{Continued on next page}}                                                                                                                                                                                                                                                                                                                                                                                                                                                               \\
    \endfoot
    \bottomrule
    \endlastfoot
    Q1                   & \textbf{Stag Hunt Game}                                                                                                                                                                                                                                                                                                                                                                                                                                                                      \\
                         & Two hunters go hunting. They can hunt a deer or a hare. A deer requires cooperation by two people, while a hare can be hunted alone; the payoff from deer is higher than that from hare. Let the payoff from deer be x and from hare be y.                                                                                                                                                                                                                                                   \\
                         & (1) The players in this stag hunt game are \texttt{Blank1} and \texttt{Blank2}. Each player can choose the strategies \texttt{Blank3} and \texttt{Blank4}. (1 pts)                                                                                                                                                                                                                                                                                                                           \\
                         & (2) The payoff matrix is:                                                                                                                                                                                                                                                                                                                                                                                                                                                                    \\
                         & \begin{tabular}{|c|c|c|} \hline
                               \texttt{Blank1}/\texttt{Blank2} & \texttt{Blank3}         & \texttt{Blank4}         \\ \hline
                               \texttt{Blank3}                 & \texttt{A1},\texttt{A2} & \texttt{B1},\texttt{B2} \\ \hline
                               \texttt{Blank4}                 & \texttt{C1},\texttt{C2} & \texttt{D1},\texttt{D2} \\ \hline
                           \end{tabular}                                                                                                                                                                                                                                                                                                                                                                                                 \\
                         & Fill in \texttt{A1}, \texttt{A2}, \texttt{B1}, \texttt{B2}, \texttt{C1}, \texttt{C2}, \texttt{D1}, \texttt{D2}. The pure-strategy Nash equilibria are \texttt{Blank5}. (2 pts)                                                                                                                                                                                                                                                                                                               \\
    \midrule
    Q2                   & \textbf{Hawk-Dove Game}                                                                                                                                                                                                                                                                                                                                                                                                                                                                      \\
                         & In nature, interactions between animals can be viewed as games. For example, when two hawks discover the same food at the same time, they fight and may both be injured, so each hawk's payoff is -2. When two doves meet, they share peacefully and each gets 1. When a hawk meets a dove, the dove retreats and the hawk gets all the food, so the hawk's payoff is 2 and the dove's payoff is 0. Now suppose two birds meet.                                                              \\
                         & (1) The players in the hawk-dove game are \texttt{Blank1} and \texttt{Blank2}. Each player can choose the strategies \texttt{Blank3} and \texttt{Blank4}. (1 pts)                                                                                                                                                                                                                                                                                                                            \\
                         & (2) The payoff matrix is:                                                                                                                                                                                                                                                                                                                                                                                                                                                                    \\
                         & \begin{tabular}{|c|c|c|} \hline
                               \texttt{Blank1}/\texttt{Blank2} & \texttt{Blank3}         & \texttt{Blank4}         \\ \hline
                               \texttt{Blank3}                 & \texttt{A1},\texttt{A2} & \texttt{B1},\texttt{B2} \\ \hline
                               \texttt{Blank4}                 & \texttt{C1},\texttt{C2} & \texttt{D1},\texttt{D2} \\ \hline
                           \end{tabular}                                                                                                                                                                                                                                                                                                                                                                                                 \\
                         & Fill in \texttt{A1}, \texttt{A2}, \texttt{B1}, \texttt{B2}, \texttt{C1}, \texttt{C2}, \texttt{D1}, \texttt{D2}. The pure-strategy Nash equilibria are \texttt{Blank5}. (2 pts)                                                                                                                                                                                                                                                                                                               \\
    \midrule
    Q3                   & \textbf{Found Document Game}                                                                                                                                                                                                                                                                                                                                                                                                                                                                 \\
                         & Company representative A is going to sign a contract expected to bring 100,000 in profit. Unfortunately, a key document is lost and picked up by passerby B. If A cannot retrieve it in time, the deal fails, causing a potential reputation loss of 30,000; the document has no value to B.                                                                                                                                                                                                 \\
                         & A contacts B, and both know the document is important to A and worthless to B. B has two choices:                                                                                                                                                                                                                                                                                                                                                                                            \\
                         & * Ask for a reward of 20,000 to return the document. B gains 20,000 but suffers moral pressure equivalent to a 15,000 loss.                                                                                                                                                                                                                                                                                                                                                                  \\
                         & * Return it for free: no monetary gain, but moral satisfaction worth 15,000.                                                                                                                                                                                                                                                                                                                                                                                                                 \\
                         & A's choices:                                                                                                                                                                                                                                                                                                                                                                                                                                                                                 \\
                         & * Pay the 20,000 reward to retrieve the document. Total loss is 20,000 (paying avoids the 30,000 reputation loss and 100,000 profit loss).                                                                                                                                                                                                                                                                                                                                                   \\
                         & * Give up the document and bear a total loss of 130,000.                                                                                                                                                                                                                                                                                                                                                                                                                                     \\
                         & (1) The payoff matrix is:                                                                                                                                                                                                                                                                                                                                                                                                                                                                    \\
                         & \begin{tabular}{|c|c|c|} \hline
                               A/B          & Ask for Reward          & No Reward               \\ \hline
                               Retrieve     & \texttt{A1},\texttt{A2} & \texttt{B1},\texttt{B2} \\ \hline
                               Not Retrieve & \texttt{C1},\texttt{C2} & \texttt{D1},\texttt{D2} \\ \hline
                           \end{tabular}                                                                                                                                                                                                                                                                                                                                                                                                                    \\
                         & Fill in \texttt{A1}, \texttt{A2}, \texttt{B1}, \texttt{B2}, \texttt{C1}, \texttt{C2}, \texttt{D1}, \texttt{D2}. (1 pts)                                                                                                                                                                                                                                                                                                                                                                      \\
                         & (2) The pure-strategy Nash equilibrium is \texttt{Blank1}. (1 pts)                                                                                                                                                                                                                                                                                                                                                                                                                           \\
    \midrule
    Q4                   & \textbf{Market Barrier}                                                                                                                                                                                                                                                                                                                                                                                                                                                                      \\
                         & In a market there are two companies (A and B). Each considers whether to raise market barriers to deter new entrants. If one company raises barriers and the other does not, the barrier-raising company gains a higher market share (payoff 3) and the other loses some share (payoff 1). If neither raises barriers, they keep current shares (payoff 2). If both raise barriers, competition drives costs too high and both get 0.                                                        \\
                         & (1) The players are \texttt{Blank1} and \texttt{Blank2}. Each can choose strategies \texttt{Blank3} and \texttt{Blank4}. (1 pts)                                                                                                                                                                                                                                                                                                                                                             \\
                         & (2) The payoff matrix is:                                                                                                                                                                                                                                                                                                                                                                                                                                                                    \\
                         & \begin{tabular}{|c|c|c|} \hline
                               \texttt{Blank1}/\texttt{Blank2} & \texttt{Blank3}         & \texttt{Blank4}         \\ \hline
                               \texttt{Blank3}                 & \texttt{A1},\texttt{A2} & \texttt{B1},\texttt{B2} \\ \hline
                               \texttt{Blank4}                 & \texttt{C1},\texttt{C2} & \texttt{D1},\texttt{D2} \\ \hline
                           \end{tabular}                                                                                                                                                                                                                                                                                                                                                                                                 \\
                         & Fill in \texttt{A1}, \texttt{A2}, \texttt{B1}, \texttt{B2}, \texttt{C1}, \texttt{C2}, \texttt{D1}, \texttt{D2}. The pure-strategy Nash equilibria are \texttt{Blank5}. (2 pts)                                                                                                                                                                                                                                                                                                               \\
    \midrule
    Q5                   & \textbf{Trust Game}                                                                                                                                                                                                                                                                                                                                                                                                                                                                          \\
                         & Two players play a trust game. The trustor (player 1) has 10 units of resources. They can transfer some portion \((0 \leq x \leq 10)\) to the trustee (player 2), and the transferred amount is tripled. The trustee can then return some amount \((y \leq 3x)\) to the trustor.                                                                                                                                                                                                             \\
                         & (1) Player 1's payoff function \(u_1(x, y)\) is \texttt{Blank1}, and player 2's payoff function \(u_2(x, y)\) is \texttt{Blank2}. (1 pts)                                                                                                                                                                                                                                                                                                                                                    \\
                         & (2) The pure-strategy Nash equilibrium is \texttt{Blank3}. (3 pts)                                                                                                                                                                                                                                                                                                                                                                                                                           \\
    \midrule
    Q6                   & \textbf{Two-Thirds of the Average Guess}                                                                                                                                                                                                                                                                                                                                                                                                                                                     \\
                         & In a game, n participants simultaneously choose a number in [0, 360]. The winner is the person whose number is closest to two-thirds of the average. Participants do not know others' choices. Assuming all players are rational, what number will they choose?                                                                                                                                                                                                                              \\
                         & (1) Let player i's choice be \(x_i\). The payoff function is \(u_i(x_1, ..., x_n)\) = \texttt{Blank1}. (1 pts)                                                                                                                                                                                                                                                                                                                                                                               \\
                         & (2) The pure-strategy Nash equilibrium is \texttt{Blank2}. (3 pts)                                                                                                                                                                                                                                                                                                                                                                                                                           \\
    \midrule
    Q7                   & \textbf{Divider-Chooser Game}                                                                                                                                                                                                                                                                                                                                                                                                                                                                \\
                         & A and B share a cake. The cake can be divided into two portions. A first splits the cake into two pieces, and then B chooses one. Both want as much cake as possible. Suppose A's split is \(x\) and \(1-x\) (\(x\leq0.5\)).                                                                                                                                                                                                                                                                 \\
                         & (1) A's payoff \(u_A\) is \texttt{Blank1}, and B's payoff \(u_B\) is \texttt{Blank2}. (3 pts)                                                                                                                                                                                                                                                                                                                                                                                                \\
                         & (2) The pure-strategy Nash equilibrium is \texttt{Blank3}. (2 pts)                                                                                                                                                                                                                                                                                                                                                                                                                           \\
    \midrule
    Q8                   & \textbf{Bertrand Model 2}                                                                                                                                                                                                                                                                                                                                                                                                                                                                    \\
                         & Suppose two firms (1 and 2) sell a homogeneous product. Each sets its price \(p_1\) and \(p_2\). Consumers buy from the lower-priced firm. If prices are equal, consumers split equally. Each firm has constant marginal cost \(c\), and market demand is \(D(p)=D_0-k \cdot p\), where \(p\) is the lowest market price, and \(D_0\), \(k\) are constants.                                                                                                                                  \\
                         & (1) The payoff function is \(u_i(p_i, p_j)\) = \texttt{Blank1}. (2 pts)                                                                                                                                                                                                                                                                                                                                                                                                                      \\
                         & (2) The pure-strategy Nash equilibrium is \texttt{Blank2}. (4 pts)                                                                                                                                                                                                                                                                                                                                                                                                                           \\
    \midrule
    Q9                   & \textbf{Public Goods Problem}                                                                                                                                                                                                                                                                                                                                                                                                                                                                \\
                         & A village has 3 herders who graze sheep on common land. The numbers of sheep are \(g_1, g_2, g_3\). The land can support at most \(G_{max} = 10\) sheep; beyond this, the land is overgrazed and each sheep's value becomes 0. The cost per sheep is \(c=4\). The value per sheep \(v(G)\) decreases with total grazing \(G\) as \(v(G) = 20 - 2G\) for \(G < G_{max}\); otherwise \(v(G)=0\).                                                                                               \\
                         & (1) Villager i's payoff function \(u_i(g_i, g_j, g_k)\) = \texttt{Blank1}. (2 pts)                                                                                                                                                                                                                                                                                                                                                                                                           \\
                         & (2) Given \(g_j=g_j^*\) and \(g_k=g_k^*\), villager i's optimal grazing is \texttt{Blank2}. (2 pts)                                                                                                                                                                                                                                                                                                                                                                                          \\
                         & (3) The pure-strategy Nash equilibrium is \texttt{Blank3}. (2 pts)                                                                                                                                                                                                                                                                                                                                                                                                                           \\
    \midrule
    Q10                  & \textbf{Final-Offer Arbitration}                                                                                                                                                                                                                                                                                                                                                                                                                                                             \\
                         & A firm and a union dispute wages. Both submit wage demands to a neutral arbitrator. The firm and union simultaneously submit their desired wages, denoted \(w_1\) and \(w_2\) (\(w_1 < w_2\)). The arbitrator chooses one of the two demands as the outcome. The arbitrator has an ideal wage \(x\) and simply picks the offer closest to \(x\). The arbitrator knows \(x\), while the other two parties do not; they believe \(x\) is a random variable with CDF \(F(x)\) and PDF \(f(x)\). \\
                         & (1) The expected wage is \texttt{Blank1}. (2 pts)                                                                                                                                                                                                                                                                                                                                                                                                                                            \\
                         & (2) Let \(w_1^*, w_2^*\) be the Nash equilibrium offers. The equilibrium condition is \texttt{Blank2} (expressed using the CDF). (6 pts)                                                                                                                                                                                                                                                                                                                                                     \\
\end{longtable}

\subsection{Exam questions}

After the review phase, participants proceeded to a 20-minute exam without GPT access. They could refer to the learning materials, assignment questions, and their assignment answers. The game theory exam included 4 questions, and the Python exam included 8 questions. These questions differed from those in the assignment and review phases and had point values reflecting item difficulty. The original participant-facing exam questions are reproduced in \autoref{tab:exam-python} and \autoref{tab:exam-game-theory}.

\begin{longtable}{|l|p{0.80\textwidth}|}
    \caption{Python Exam questions.}
    \label{tab:exam-python}                                                                                                                                                                                                                                                                                                                                                                                                                                 \\
    \toprule
    \textbf{Problem No.} & \textbf{Problem Description}                                                                                                                                                                                                                                                                                                                                                                                                     \\
    \midrule
    \endfirsthead
    \multicolumn{2}{c}%
    {{\tablename\ \thetable{} -- continued from previous page}}                                                                                                                                                                                                                                                                                                                                                                                             \\
    \toprule
    \textbf{Problem No.} & \textbf{Problem Description}                                                                                                                                                                                                                                                                                                                                                                                                     \\
    \midrule
    \endhead
    \midrule \multicolumn{2}{r}{{Continued on next page}}                                                                                                                                                                                                                                                                                                                                                                                                   \\
    \endfoot
    \bottomrule
    \endlastfoot
    Q1                   & \textbf{String Length}                                                                                                                                                                                                                                                                                                                                                                                                           \\
                         & Write a function \texttt{string\_length(s)} that returns the length of the input string.(1 pts)                                                                                                                                                                                                                                                                                                                                  \\
    \midrule
    Q2                   & \textbf{Absolute Difference}                                                                                                                                                                                                                                                                                                                                                                                                     \\
                         & Given two values \texttt{x} and \texttt{y}, write a function \texttt{absolute\_difference(x, y)} to compute the absolute difference.(2 pts)                                                                                                                                                                                                                                                                                      \\
    \midrule
    Q3                   & \textbf{Square Dictionary}                                                                                                                                                                                                                                                                                                                                                                                                       \\
                         & Write a function \texttt{square\_dict(n)} that returns a dictionary with \texttt{n} elements. Keys are 0 to \texttt{n-1}, and each value is the square of its key. Order should be from 0 to \texttt{n-1}. (2 pts)                                                                                                                                                                                                               \\
    \midrule
    Q4                   & \textbf{Palindrome Number}                                                                                                                                                                                                                                                                                                                                                                                                       \\
                         & Write a function \texttt{is\_palindrome(x)}. Given an integer \texttt{x}, return \texttt{True} if it is a palindrome integer; otherwise return \texttt{False}. (3 pts)                                                                                                                                                                                                                                                           \\
                         & Tip: Use \texttt{str} to convert an integer to a string, e.g., \texttt{str(12345) = "12345"}.                                                                                                                                                                                                                                                                                                                                    \\
    \midrule
    Q5                   & \textbf{Largest Elements in a List}                                                                                                                                                                                                                                                                                                                                                                                              \\
                         & Write a function \texttt{find\_two\_largest\_numbers(lst)} that returns the largest and second-largest values in the list. The input list length is at least 2. (3 pts)                                                                                                                                                                                                                                                          \\
    \midrule
    Q6                   & \textbf{Missing Employee Info}                                                                                                                                                                                                                                                                                                                                                                                                   \\
                         & Given two dictionaries \texttt{employee\_name} and \texttt{employee\_salary}, where keys are employee IDs. Values are names in \texttt{employee\_name} and salaries in \texttt{employee\_salary}. Write a function \texttt{find\_missing\_info(employee\_name, employee\_salary)} to find employee IDs with missing info (has a name but no salary, or has a salary but no name). Return a list of IDs sorted ascending. (4 pts) \\
                         & Tip: Use \texttt{nums.sort()} to sort a list. For example,                                                                                                                                                                                                                                                                                                                                                                       \\
                         & \begin{minipage}{0.7\textwidth}
                               \begin{verbatim}
nums = [1, 3, 2, 4]
nums.sort()
print(nums) # [1, 2, 3, 4]
\end{verbatim}
                           \end{minipage}                                                                                                                                                                                                                                                                                                                                                                                                   \\
                         & Note: \texttt{sort} returns \texttt{None}.                                                                                                                                                                                                                                                                                                                                                                                       \\
    \midrule
    Q7                   & \textbf{Fibonacci Sequence}                                                                                                                                                                                                                                                                                                                                                                                                      \\
                         & Write a function \texttt{fibonacci\_sequence(n)} that generates a Fibonacci sequence of length \texttt{n} (\texttt{n} is a non-negative integer) and returns it as a list. (4 pts)                                                                                                                                                                                                                                               \\
                         & Tip: In the Fibonacci sequence, each number is the sum of the previous two, starting with 0 and 1.                                                                                                                                                                                                                                                                                                                               \\
    \midrule
    Q8                   & \textbf{Are Two Halves Similar}                                                                                                                                                                                                                                                                                                                                                                                                  \\
                         & Write a function \texttt{are\_halves\_similar(s)} that takes an even-length string \texttt{s}. Split it into two equal halves, \texttt{a} and \texttt{b}.                                                                                                                                                                                                                                                                        \\
                         & Two strings are \textbf{similar} if they contain the same number of vowels \texttt{('a', 'e', 'i', 'o', 'u', 'A', 'E', 'I', 'O', 'U')}. Note that \texttt{s} may contain both upper- and lower-case letters.                                                                                                                                                                                                                     \\
                         & If \texttt{a} and \texttt{b} are similar, return \texttt{True}; otherwise return \texttt{False}. (5 pts)                                                                                                                                                                                                                                                                                                                         \\
\end{longtable}

\begin{longtable}{|l|p{0.80\textwidth}|}
    \caption{Game Theory Exam questions.}
    \label{tab:exam-game-theory}                                                                                                                                                                                                                                                                                                                                                                                                                                                                                                             \\
    \toprule
    \textbf{Problem No.} & \textbf{Problem Description}                                                                                                                                                                                                                                                                                                                                                                                                                                                                                      \\
    \midrule
    \endfirsthead
    \multicolumn{2}{c}%
    {{\tablename\ \thetable{} -- continued from previous page}}                                                                                                                                                                                                                                                                                                                                                                                                                                                                              \\
    \toprule
    \textbf{Problem No.} & \textbf{Problem Description}                                                                                                                                                                                                                                                                                                                                                                                                                                                                                      \\
    \midrule
    \endhead
    \midrule \multicolumn{2}{r}{{Continued on next page}}                                                                                                                                                                                                                                                                                                                                                                                                                                                                                    \\
    \endfoot
    \bottomrule
    \endlastfoot
    Q1                   & \textbf{Abstract Game 3}                                                                                                                                                                                                                                                                                                                                                                                                                                                                                          \\
                         & Given the following normal-form game.                                                                                                                                                                                                                                                                                                                                                                                                                                                                             \\
                         & (1) Using iterated elimination, the strategies that are not eliminated are \texttt{Blank1}. (1.5 pts)                                                                                                                                                                                                                                                                                                                                                                                                             \\
                         & (2) Using the underline method, the pure-strategy Nash equilibria are \texttt{Blank2}. (1.5 pts)                                                                                                                                                                                                                                                                                                                                                                                                                  \\
                         & \begin{tabular}{|c|c|c|c|} \hline
                               Player 1 / Player 2 & L    & C    & R    \\ \hline
                               T                   & 2, 0 & 1, 1 & 4, 2 \\ \hline
                               M                   & 3, 4 & 1, 2 & 2, 3 \\ \hline
                               B                   & 1, 3 & 0, 2 & 3, 0 \\ \hline
                           \end{tabular}                                                                                                                                                                                                                                                                                                                                                                                                                                                                 \\
    \midrule
    Q2                   & \textbf{Smart Pig Game}                                                                                                                                                                                                                                                                                                                                                                                                                                                                                           \\
                         & One big pig and one small pig are kept in the same pen. There is a trough on one side and a pedal on the other (the pig that steps the pedal is farther from the trough). Each press puts 10 units of food into the trough, and the pig that presses pays a cost of 2 units. If the big pig reaches the trough first, the big and small pigs eat 9 and 1 units respectively; if the small pig reaches first, they eat 6 and 4; if both pigs press the pedal at the same time, the big and small pigs eat 7 and 3. \\
                         & (1) The players in the Smart Pig Game are \texttt{Blank1} and \texttt{Blank2}. Each player can choose the strategies \texttt{Blank3} and \texttt{Blank4}. (1 pts)                                                                                                                                                                                                                                                                                                                                                 \\
                         & (2) The payoff matrix is: (1.5 pts)                                                                                                                                                                                                                                                                                                                                                                                                                                                                               \\
                         & \begin{tabular}{|c|c|c|} \hline
                               \texttt{Blank1}/\texttt{Blank2} & \texttt{Blank3}         & \texttt{Blank4}         \\ \hline
                               \texttt{Blank3}                 & \texttt{A1},\texttt{A2} & \texttt{B1},\texttt{B2} \\ \hline
                               \texttt{Blank4}                 & \texttt{C1},\texttt{C2} & \texttt{D1},\texttt{D2} \\ \hline
                           \end{tabular}                                                                                                                                                                                                                                                                                                                                                                                                                      \\
                         & Fill in \texttt{A1}, \texttt{A2}, \texttt{B1}, \texttt{B2}, \texttt{C1}, \texttt{C2}, \texttt{D1}, \texttt{D2}.                                                                                                                                                                                                                                                                                                                                                                                                   \\
                         & (3) Using the underline method, the pure-strategy Nash equilibrium is \texttt{Blank5}. (1.5 pts)                                                                                                                                                                                                                                                                                                                                                                                                                  \\
    \midrule
    Q3                   & \textbf{n-Player Cournot Game}                                                                                                                                                                                                                                                                                                                                                                                                                                                                                    \\
                         & Consider a Cournot model with \(n\) firms. Let \(q_i\) be firm i's output, and \(Q=\sum_{i=1}^{n} q_i\) be total market output. The market-clearing price is \(P\), with \(P=P(Q)=a-Q\) (when \(Q<a\); otherwise \(P=0\)). Assume there is no fixed cost and marginal cost is a constant \(c\).                                                                                                                                                                                                                   \\
                         & Note: Fixed cost does not change with output within a range (e.g., rent, management salaries, depreciation). Marginal cost is the additional cost of producing one more unit; it is the derivative of total cost with respect to output.                                                                                                                                                                                                                                                                          \\
                         & (1) Under these assumptions, each firm's profit function is: (2 pts)                                                                                                                                                                                                                                                                                                                                                                                                                                              \\
                         & $$\pi_i(q_1,\cdots,q_n)=[[Blank1]]$$                                                                                                                                                                                                                                                                                                                                                                                                                                                                              \\
                         & where \(i=1,2,...,n\).                                                                                                                                                                                                                                                                                                                                                                                                                                                                                            \\
                         & (2) In the Nash equilibrium of the n-player Cournot game, \(q_i^*\) equals: \texttt{Blank2}. (3 pts)                                                                                                                                                                                                                                                                                                                                                                                                              \\
    \midrule
    Q4                   & \textbf{Hotelling Model}                                                                                                                                                                                                                                                                                                                                                                                                                                                                                          \\
                         & Suppose there is a 1 km beach with uniformly distributed visitors. Two ice-cream stands (stand 1 and stand 2) plan to open on the beach and choose locations \(a\) and \((1-b)\), with \(a + b < 1\). The ice cream is identical in quality. Prices are \(p_1\) and \(p_2\). Assume there is no fixed cost and marginal cost is a constant \(c\).                                                                                                                                                                 \\
                         & Note: Fixed cost does not change with output within a range (e.g., rent, management salaries, depreciation). Marginal cost is the additional cost of producing one more unit; it is the derivative of total cost with respect to output.                                                                                                                                                                                                                                                                          \\
                         & \begin{minipage}{0.7\textwidth}
                               \centering
                               \includegraphics[width=0.5\linewidth]{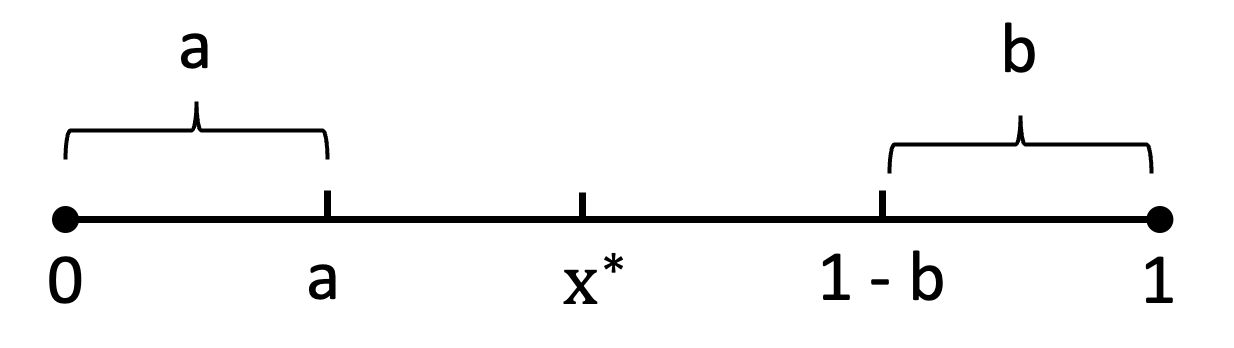}
                           \end{minipage}                                                                                                                                                                                                                                                                                                                                                                                                                                                        \\
                         & (1) Assume consumers know both prices. The travel cost per unit distance is \(t\), and consumers choose the stand with the lowest total cost (ice-cream price plus travel cost). Let \(x^* \in [0, 1]\) be the indifferent consumer: if \(x< x^*\), choose stand 1; if \(x>x^*\), choose stand 2. Then \(x^*\) is \texttt{Blank1}. (3 pts)                                                                                                                                                                        \\
                         & (2) The Nash equilibrium locations \(a\) and \(b\) are \texttt{Blank2}, \texttt{Blank3}. (4 pts)                                                                                                                                                                                                                                                                                                                                                                                                                  \\
                         & (3) Now suppose locations are fixed at \(a\) and \((1-b)\), with \(a + b < 1\). The firms choose prices \(p_1\) and \(p_2\). The Nash equilibrium prices are \texttt{Blank4}, \texttt{Blank5}. (4 pts)                                                                                                                                                                                                                                                                                                            \\
\end{longtable}

\subsection{Post-task survey}

After the exam phase, all participants completed a post-task survey. Experimental participants evaluated the assistant across study phases and reported how and why they used it. All participants reported their motivations, views of AI assistance in education, and general study feedback. The original participant-facing wording is reproduced in \autoref{tab:post-task}.

\begin{longtable}{|l|p{0.80\textwidth}|}
    \caption{Post-task survey questions for Python and Game Theory courses.}
    \label{tab:post-task}                                                                                                                                                                                                                             \\
    \toprule
    \textbf{Question No.} & \textbf{Question Content}                                                                                                                                                                                                 \\
    \midrule
    \endfirsthead
    \multicolumn{2}{c}%
    {{\tablename\ \thetable{} -- continued from previous page}}                                                                                                                                                                                       \\
    \toprule
    \textbf{Question No.} & \textbf{Question Content}                                                                                                                                                                                                 \\
    \midrule
    \endhead
    \midrule \multicolumn{2}{r}{{Continued on next page}}                                                                                                                                                                                             \\
    \endfoot
    \bottomrule
    \endlastfoot
    Q1                    & How often did you use the chat assistant during the learning stage?                                                                                                                                                       \\
                          & $\bigcirc$ Not used \quad $\bigcirc$ Almost never \quad $\bigcirc$ Occasionally \quad $\bigcirc$ Often \quad $\bigcirc$ Very often \quad $\bigcirc$ I cannot use the chat assistant \quad $\bigcirc$ Other                \\
    \midrule
    Q2                    & How often did you use the chat assistant during Homework 1?                                                                                                                                                               \\
                          & $\bigcirc$ Not used \quad $\bigcirc$ Almost never \quad $\bigcirc$ Occasionally \quad $\bigcirc$ Often \quad $\bigcirc$ Very often \quad $\bigcirc$ I cannot use the chat assistant \quad $\bigcirc$ Other                \\
    \midrule
    Q3                    & How often did you use the chat assistant during the review stage?                                                                                                                                                         \\
                          & $\bigcirc$ Not used \quad $\bigcirc$ Almost never \quad $\bigcirc$ Occasionally \quad $\bigcirc$ Often \quad $\bigcirc$ Very often \quad $\bigcirc$ I cannot use the chat assistant \quad $\bigcirc$ Other                \\
    \midrule
    Q4                    & In which stage was the chat assistant most helpful?                                                                                                                                                                       \\
                          & $\bigcirc$ Learning stage \quad $\bigcirc$ Homework stage (HW1) \quad $\bigcirc$ Review stage \quad $\bigcirc$ Not sure \quad $\bigcirc$ About equally helpful                                                            \\
                          & $\bigcirc$ Not helpful in any stage \quad $\bigcirc$ I cannot use the chat assistant \quad $\bigcirc$ Other                                                                                                               \\
    \midrule
    Q5                    & How accurate were the assistant's answers?                                                                                                                                                                                \\
                          & $\bigcirc$ Very accurate \quad $\bigcirc$ Somewhat accurate \quad $\bigcirc$ Neutral \quad $\bigcirc$ Somewhat inaccurate \quad $\bigcirc$ Very inaccurate \quad $\bigcirc$ I cannot use the chat assistant               \\
    \midrule
    Q6                    & How well did the assistant understand your questions?                                                                                                                                                                     \\
                          & $\bigcirc$ Very accurate \quad $\bigcirc$ Somewhat accurate \quad $\bigcirc$ Neutral \quad $\bigcirc$ Somewhat inaccurate \quad $\bigcirc$ Very inaccurate \quad $\bigcirc$ I cannot use the chat assistant               \\
    \midrule
    Q7                    & Response personalization: Did it adjust guidance based on your replies within a conversation?                                                                                                                             \\
                          & $\bigcirc$ Always adjusts \quad $\bigcirc$ Often adjusts \quad $\bigcirc$ Sometimes adjusts \quad $\bigcirc$ Rarely adjusts \quad $\bigcirc$ Never adjusts \quad $\bigcirc$ I cannot use the chat assistant               \\
    \midrule
    Q8                    & Learning-level personalization: Did it consider your prior progress and knowledge in this course?                                                                                                                         \\
                          & $\bigcirc$ Always considers my progress \quad $\bigcirc$ Often considers my progress \quad $\bigcirc$ Sometimes considers my progress                                                                                     \\
                          & $\bigcirc$ Rarely considers my progress \quad $\bigcirc$ Never considers my progress \quad $\bigcirc$ I cannot use the chat assistant                                                                                     \\
    \midrule
    Q9                    & Effect on course learning: Compared to doing it alone (no assistant), did the assistant improve your efficiency and outcomes?                                                                                             \\
                          & $\bigcirc$ Significantly improved \quad $\bigcirc$ Slightly improved \quad $\bigcirc$ No change \quad $\bigcirc$ Slightly reduced \quad $\bigcirc$ Significantly reduced \quad $\bigcirc$ I cannot use the chat assistant \\
    \midrule
    Q10                   & Effect on Homework 1: Compared to doing it alone (no assistant), did the assistant improve your efficiency and outcomes?                                                                                                  \\
                          & $\bigcirc$ Significantly improved \quad $\bigcirc$ Slightly improved \quad $\bigcirc$ No change \quad $\bigcirc$ Slightly reduced \quad $\bigcirc$ Significantly reduced \quad $\bigcirc$ I cannot use the chat assistant \\
    \midrule
    Q11                   & Effect on review: Compared to doing it alone (no assistant), did the assistant improve your efficiency and outcomes?                                                                                                      \\
                          & $\bigcirc$ Significantly improved \quad $\bigcirc$ Slightly improved \quad $\bigcirc$ No change \quad $\bigcirc$ Slightly reduced \quad $\bigcirc$ Significantly reduced \quad $\bigcirc$ I cannot use the chat assistant \\
    \midrule
    Q12                   & Effect on final assessment (HW2): Compared to not using the assistant, did it improve your efficiency and outcomes?                                                                                                       \\
                          & $\bigcirc$ Significantly improved \quad $\bigcirc$ Slightly improved \quad $\bigcirc$ No change \quad $\bigcirc$ Slightly reduced \quad $\bigcirc$ Significantly reduced \quad $\bigcirc$ I cannot use the chat assistant \\
    \midrule
    Q13                   & How polite was the assistant in its interactions?                                                                                                                                                                         \\
                          & $\bigcirc$ Too polite, sometimes unnecessary \quad $\bigcirc$ Appropriately polite \quad $\bigcirc$ Cannot evaluate / not sure                                                                                            \\
                          & $\bigcirc$ Not polite enough \quad $\bigcirc$ Very impolite \quad $\bigcirc$ I cannot use the chat assistant                                                                                                              \\
    \midrule
    Q14                   & How patient was the assistant in its interactions?                                                                                                                                                                        \\
                          & $\bigcirc$ Too patient, sometimes unnecessary \quad $\bigcirc$ Appropriately patient \quad $\bigcirc$ Cannot evaluate / not sure                                                                                          \\
                          & $\bigcirc$ Not patient enough \quad $\bigcirc$ Very impatient \quad $\bigcirc$ I cannot use the chat assistant                                                                                                            \\
    \midrule
    Q15                   & Compared to traditional learning, how interesting was learning with the assistant?                                                                                                                                        \\
                          & $\bigcirc$ Much more interesting \quad $\bigcirc$ Slightly more interesting \quad $\bigcirc$ About the same                                                                                                               \\
                          & $\bigcirc$ Slightly more boring \quad $\bigcirc$ Much more boring \quad $\bigcirc$ I cannot use the chat assistant                                                                                                        \\
    \midrule
    Q16                   & Did the assistant encourage you to explore and learn new knowledge independently?                                                                                                                                         \\
                          & $\bigcirc$ Strongly encouraged \quad $\bigcirc$ Somewhat encouraged \quad $\bigcirc$ Neutral \quad $\bigcirc$ Not very encouraged \quad $\bigcirc$ Not encouraged at all \quad $\bigcirc$ I cannot use the chat assistant \\
    \midrule
    Q17                   & Motivation: What were your main reasons for participating? (Select all that apply)                                                                                                                                        \\
                          & $\square$ Learn related knowledge (game theory or programming) \quad $\square$ Understand how LLMs help learning                                                                                                          \\
                          & $\square$ Curiosity and interest in the experiment \quad $\square$ Support the researchers \quad $\square$ Receive compensation \quad $\square$ Other                                                                     \\
    \midrule
    Q18                   & During the experiment, what were your main purposes for using the assistant? (Select all that apply)                                                                                                                      \\
                          & $\square$ Help summarize reference materials \quad $\square$ Concept explanations \quad $\square$ Get answers to specific questions                                                                                       \\
                          & $\square$ Get explanations for specific answers \quad $\square$ Get solution approaches \quad $\square$ Help recommend problems (if applicable)                                                                           \\
                          & $\square$ Help check answer correctness \quad $\square$ I cannot use the chat assistant \quad $\square$ Other                                                                                                             \\
    \midrule
    Q19                   & Which aspects of the assistant helped you understand the learning content? (Select all that apply)                                                                                                                        \\
                          & $\square$ Help summarize reference materials \quad $\square$ Concept explanations \quad $\square$ Get answers to problems                                                                                                 \\
                          & $\square$ Get explanations for problem answers \quad $\square$ Get solution approaches \quad $\square$ Validate my answers                                                                                                \\
                          & $\square$ Not helpful \quad $\square$ I cannot use the chat assistant \quad $\square$ Other                                                                                                                               \\
    \midrule
    Q20                   & How did you integrate the assistant into your learning? (Select all that apply)                                                                                                                                           \\
                          & $\square$ I tried independently first and used the assistant only when stuck                                                                                                                                              \\
                          & $\square$ I used the assistant throughout the learning process                                                                                                                                                            \\
                          & $\square$ I used the assistant mainly at the beginning to get a rough understanding, then studied independently                                                                                                           \\
                          & $\square$ I used the assistant later to review and consolidate                                                                                                                                                            \\
                          & $\square$ I did not use the assistant or used it very little \quad $\square$ I cannot use the chat assistant \quad $\square$ Other                                                                                        \\
    \midrule
    Q21                   & Overall attitude: How do you view applying chat assistants or similar AI tools in education?                                                                                                                              \\
                          & $\bigcirc$ Strongly support; it can greatly improve education experience and outcomes                                                                                                                                     \\
                          & $\bigcirc$ Somewhat support, but risks should be handled cautiously                                                                                                                                                       \\
                          & $\bigcirc$ Neutral; I see both pros and cons                                                                                                                                                                              \\
                          & $\bigcirc$ Reserved; I worry about possible negative effects                                                                                                                                                              \\
                          & $\bigcirc$ Strongly oppose; using such tools in education may cause serious problems                                                                                                                                      \\
                          & $\bigcirc$ Other                                                                                                                                                                                                          \\
    \midrule
    Q22                   & Willingness: How willing are you to use GPT or similar AI tools for learning in the future?                                                                                                                               \\
                          & $\bigcirc$ Very willing in the future \quad $\bigcirc$ Somewhat willing in the future \quad $\bigcirc$ Neutral / unsure                                                                                                   \\
                          & $\bigcirc$ Not very willing in the future \quad $\bigcirc$ Not willing at all in the future \quad $\bigcirc$ Other                                                                                                        \\
    \midrule
    Q23                   & Potential risks: If using GPT-like AI tools long-term for learning, which issues concern you? (Select all that apply)                                                                                                     \\
                          & $\square$ I worry about over-reliance, which may weaken critical thinking and problem-solving.                                                                                                                            \\
                          & $\square$ I worry the information from these tools may be inaccurate or unreliable.                                                                                                                                       \\
                          & $\square$ I worry these tools may lead to plagiarism or academic dishonesty.                                                                                                                                              \\
                          & $\square$ I worry this may reduce interaction and collaboration with peers/teachers. \quad $\square$ Other                                                                                                                \\
    \midrule
    Q24                   & Difficulty: How would you rate the overall difficulty of Homework 1?                                                                                                                                                      \\
                          & $\bigcirc$ Very challenging \quad $\bigcirc$ Somewhat challenging \quad $\bigcirc$ Moderate \quad $\bigcirc$ Relatively easy \quad $\bigcirc$ Very easy                                                                   \\
    \midrule
    Q25                   & Difficulty: How would you rate the overall difficulty of Homework 2?                                                                                                                                                      \\
                          & $\bigcirc$ Very challenging \quad $\bigcirc$ Somewhat challenging \quad $\bigcirc$ Moderate \quad $\bigcirc$ Relatively easy \quad $\bigcirc$ Very easy                                                                   \\
    \midrule
    Q26                   & Time adequacy: Was the time for Homework 1 sufficient?                                                                                                                                                                    \\
                          & $\bigcirc$ More than sufficient \quad $\bigcirc$ Sufficient \quad $\bigcirc$ Moderate \quad $\bigcirc$ Tight \quad $\bigcirc$ Very tight                                                                                  \\
    \midrule
    Q27                   & Time adequacy: Was the time for Homework 2 sufficient?                                                                                                                                                                    \\
                          & $\bigcirc$ More than sufficient \quad $\bigcirc$ Sufficient \quad $\bigcirc$ Moderate \quad $\bigcirc$ Tight \quad $\bigcirc$ Very tight                                                                                  \\
    \midrule
    Q28                   & Time adequacy: Was the time for learning sufficient?                                                                                                                                                                      \\
                          & $\bigcirc$ More than sufficient \quad $\bigcirc$ Sufficient \quad $\bigcirc$ Moderate \quad $\bigcirc$ Tight \quad $\bigcirc$ Very tight                                                                                  \\
    \midrule
    Q29                   & Review usefulness: Did the review stage help improve your mastery of the material?                                                                                                                                        \\
                          & $\bigcirc$ Very helpful \quad $\bigcirc$ Somewhat helpful \quad $\bigcirc$ Moderate \quad $\bigcirc$ A little helpful \quad $\bigcirc$ Not helpful                                                                        \\
    \midrule
    Q30                   & During this study, what difficulties or challenges did you encounter? (Select all that apply)                                                                                                                             \\
                          & $\square$ Difficulty understanding the materials \quad $\square$ Lack of effective learning materials or resources                                                                                                        \\
                          & $\square$ Time management and scheduling \quad $\square$ Maintaining motivation and interest \quad $\square$ Homework pressure                                                                                            \\
                          & $\square$ Other technical issues (lag, platform response, etc.) \quad $\square$ None \quad $\square$ Other                                                                                                                \\
    \midrule
    Q31                   & How easy was it to operate the chat assistant?                                                                                                                                                                            \\
                          & $\bigcirc$ Very easy \quad $\bigcirc$ Somewhat easy \quad $\bigcirc$ Neutral \quad $\bigcirc$ Somewhat complex \quad $\bigcirc$ Very complex \quad $\bigcirc$ I cannot use the chat assistant                             \\
    \midrule
    Q32                   & Please describe your overall experience during the test. You may mention ease of use, usefulness of features, any issues, or notable experiences. If you have suggestions for improvement, please describe them below.    \\
                          & \framebox[0.75\textwidth]{\rule{0pt}{3\baselineskip}}
\end{longtable}





\end{CJK*}
\end{document}